\documentclass[twocolumn]{aa} 
\usepackage{natbib}
\usepackage{amsmath}
\usepackage{xspace}
\usepackage[utf8]{inputenc}
\usepackage{caption}
\usepackage{subcaption}
\usepackage[varg]{txfonts}
\usepackage{multirow}
\usepackage{xcolor}

\bibpunct{(}{)}{;}{a}{}{,} 
\usepackage{xspace} 

\newcommand{\iso}[2]{\ensuremath{\mathrm{{^{#2}}#1}}} 
\newcommand{\electron}{\ensuremath{\mathrm{e^{-}}}}

\newcommand{\neutrinoe}{\ensuremath{\mathrm{\nu_{e}}}}
\newcommand{\msun}{\ensuremath{\mathrm{\,M_\odot}}} 
\newcommand{\rsun}{\ensuremath{\mathrm{\,R_\odot}}} 
\newcommand{\mdotsun}{\ensuremath{\mathrm{\,M_{\odot}\,yr^{-1}}}} 
\newcommand{\denu}{\ensuremath{\mathrm{\,g\,cm^{-3}}}} 

\newcommand{\rhoc}{\ensuremath{\rho_{\mathrm{c}}}} 
\newcommand{\logrhoc}{\ensuremath{\log_{10} (\rho_{\mathrm{c}}/ \mathrm{g\,cm^{-3}})}} 
\newcommand{\logrho}{\ensuremath{\log_{10} (\rho/ \mathrm{g\,cm^{-3}})}} 
\newcommand{\mch}{\ensuremath{M_{\mathrm{Ch}}}} 
\newcommand{\mdot}{\ensuremath{\dot{M}}} 

\newcommand{\mesa}{{\textsc{mesa}}\xspace} 
\newcommand{\ia}{SN\,Ia\xspace} 
\newcommand{\ias}{SNe\,Ia\xspace} 

\newcommand{\seriesone}{\textsc{Series\,I}\xspace} 
\newcommand{\seriestwo}{\textsc{Series\,II}\xspace} 

\usepackage[detect-none]{siunitx}
\usepackage[breaklinks,colorlinks,urlcolor=blue,citecolor=blue]{hyperref}
\graphicspath{{./}{figures/}}

\definecolor{ochre}{rgb}{0.8, 0.47, 0.13}

\definecolor{johncolor}{rgb}{0.6, 0.4, 0.8}

\definecolor{savvascolor}{rgb}{1.0, 0.0, 0.0}

\begin{document}

\title{Thermonuclear and electron-capture supernovae from stripped-envelope stars}

\author{S. Chanlaridis\inst{1,2,3,4}
   \and J. Antoniadis\inst{1,5,4}
   \and D. R. Aguilera-Dena\inst{1}
   \and G. Gr\"{a}fener\inst{4}
   \and N. Langer\inst{4,5}
   \and N. Stergioulas\inst{3}
          }

   \institute{Institute of Astrophysics, Foundation for Research \& Technology -- Hellas (FORTH), GR-70013 Heraklion, Greece 
            \and
            Department of Physics, University of Crete, University Campus, GR-70013 Heraklion, Greece 
            \and 
            Department of Physics, Aristotle University of Thessaloniki, University Campus, GR-54124 Thessaloniki, Greece 
            \and
            Argelander-Institut f\"{u}r Astronomie, Auf dem H\"{u}gel 71, DE-53121 Bonn, Germany 
            \and
            Max-Planck-Institut f\"{u}r Radioastronomie, Auf dem H\"{u}gel 69, DE-53121 Bonn, Germany  \\
             \email{schanlaridis@physics.uoc.gr}
             }

   \date{Received; accepted}

\abstract
{When stripped from their hydrogen-rich envelopes, stars with initial masses between $\sim$7 and 11\msun\ may develop massive degenerate cores and collapse. Depending on the final structure and composition, the  outcome can range from a thermonuclear explosion, to the formation of a neutron star in an electron-capture supernova (ECSN). It has  recently been  demonstrated  that stars in this mass range may be more prone to disruption than previously thought: they may initiate explosive oxygen burning when their central densities are still below $\rhoc \lesssim 10^{9.6} $\denu. At the same time, their envelopes expand significantly, leading to the complete depletion of helium. This combination makes them  interesting candidates for type\,Ia supernovae---which we call (C)ONe \ias---and this might have broader implications for the formation of neutron stars via  ECSNe.}
{To constrain the observational counterparts of (C)ONe \ias and the key properties that enable them, it is crucial to constrain the evolution, composition, and precollapse structure of their progenitors, as well as the evolution of these quantities with cosmic time. In turn, this requires a detailed 
investigation of the final evolutionary stages preceding the collapse, and their sensitivity to input physics.}
{Here, we modeled the evolution of  252 single, 
nonrotating helium stars  covering the 
initial mass range $\numrange{0.8}{3.5}\msun$, 
with metallicities between $Z=10^{-4}$ and $0.02$, and overshoot efficiency factors from $\rm f_{OV} = 0.0$ to $0.016$ 
across all convective boundaries. We used these models to constrain several properties of these stars, including  their central densities, compositions, envelope masses, and radii at the onset explosive oxygen ignition, as well as  the final outcome as a function of initial helium star mass. We further investigate the sensitivity of these properties to mass loss rate assumptions using an additional grid of 110 models with varying wind efficiencies. }
{We find that helium star models with masses between $\sim$1.8  and 2.7\msun\ are able to evolve onto $\numrange{1.35}{1.37}\msun$ (C)ONe cores that initiate explosive burning at central densities between $\logrhoc \sim 9.3$ and 9.6. We constrained the amount of residual carbon retained after core carbon burning as a function of initial conditions, and conclude that it plays a critical role in determining the final outcome: Chandrasekhar-mass degenerate cores that retain more than approximately $0.005\msun$ of carbon result in (C)ONe \ias, while those with lower carbon mass become ECSNe. 
We find that (C)ONe \ias are more likely to occur at high metallicities, whereas at low metallicities ECSNe dominate. However,  both \ia and ECSN progenitors expand significantly during the final evolutionary stages, so that for the most extended models, a further binary interaction may occur.
We constrain the relative ratio between (C)ONe \ias and SNe Ib/c to be 0.17--0.30 at $Z=0.02$, and 0.03--0.13 at $Z\leq 10^{-3}$.}
{We  conclude with a discussion on potential observational properties of (C)ONe \ias and their progenitors. In the few thousand years leading to the explosion, at least some progenitors should  be  identifiable as luminous metal-rich super-giants, embedded in hydrogen-free circumstellar nebulae. }

\keywords{stars: evolution --- binaries: general --- supernovae: general --- nuclear reactions, nucleosynthesis, abundances}

\authorrunning {Chanlaridis et.al.}
\titlerunning {Non-accreting progenitors of thermonuclear supernovae}
\maketitle

\defcitealias{antoniadis2020}{ACGL20} 
\section{Introduction} \label{sec:intro}

The relation between a star's initial properties and the type of remnant it produces remains equivocal.  In isolation, stars with zero-age main sequence 
(ZAMS)  masses above $\sim $11\msun\ are  thought to produce 
neutron stars (NSs) or black holes (BHs), while those with 
masses below $\sim$7\msun\ leave behind white dwarfs (WDs). The 
fate of stars with intermediate masses is less certain, as 
predictions for their final size and composition remain  
challenging \citep[][and references therein]{dec07:siess, Poelarends:2007ip, doherty2015, Farmer:2015afs}. 

It is believed that most stars in this intermediate-mass regime evolve through a super-asymptotic giant branch (SAGB) phase, during which they burn carbon nonexplosively \citep[e.g.,][and references therein]{Ritossa:1999ApJ, siess2006, Poelarends:2007ip,doherty2015,Farmer:2015afs,Poelarends:2017dua}.
Prior to carbon ignition,  the mass of the helium core can be substantially reduced due to 
the  second dredge-up (2DU), in which processed helium-rich matter is mixed throughout the stellar envelope. Subsequent shell burning is highly unstable and proceeds in  a series  of thermal pulses \citep{Iben:1997abc,Vassiliadis:1993zz,Poelarends:2007ip}.
If the envelope is lost before the degenerate core reaches the Chandrasekhar-mass limit  ($\mch \simeq \numrange{1.35}{1.37}\msun$, for the compositions investigated in this paper), then the final outcome is a WD. Conversely, if the core mass exceeds \mch, the star produces  a supernova (SN), the properties
of which depend on the stellar structure and ignition conditions\footnote{Here, ignition refers to the time that the energy liberation rate per unit mass exceeds the local neutrino cooling rate, $\epsilon_{\rm nuc} - \epsilon_{\rm \nu}\ge 0$} at the onset of the thermonuclear runaway \citep{rose1969,wheeler1978}.
Most hydrogen-rich SAGBs are believed to lose their envelope before their core becomes sufficiently massive to collapse. Hence, an explosive ending is likely rare for  noninteracting stars in this mass range \citep{Iben:1983ts,Poelarends:2007ip}. However, if the hydrogen-rich envelope is lost before core helium ignition, both unstable shell burning and the 2DU are avoided, leading to more efficient core growth. This significantly broadens the mass range in which  near-\mch\ SAGB cores may form \citep[e.g.,][and references therein]{Podsiadlowski:2003py,Poelarends:2007ip,Woosley:2019sdf}.   
The most important formation path for naked helium  stars in this mass range is thought to be envelope stripping in binaries either through stable mass transfer, or in a common-envelope event \citep{Tauris:2015xra,Tauris2017ApJ,Siess2018leb,Laplace2020aa}. 

At the high densities expected inside near-\mch\ SAGB cores, the 
 ashes of carbon fusion participate in weak interactions, for instance Urca cycles and electron captures (henceforth $e$-captures), that can drastically influence the final evolutionary stages. 
Owing to the large abundance of \iso{Ne}{20}, the most 
important of these is the $e$-capture sequence \iso{Ne}{20}$(\electron,\neutrinoe)$\iso{F}{20}$(\electron,\neutrinoe)$\iso{O}{20},  
which---especially at densities above $10^{10}\denu$---rapidly consumes free electrons, leading to an $e$-capture supernova (ECSN). The result of the collapse is determined by the competition between 
deleptonization and nuclear energy release, and can be either a NS formed in a low-energy core-collapse ECSN (ccECSN), or a thermonuclear explosion (tECSN), in which at least part of the star is disrupted \citep{miyaji1980,nomoto1984,canal1992,kitaura2006,gutierrez1996a,jones2014}.
Whether one outcome prevails over the other depends on several uncertain factors, such as the ignition location and density \citep{Leung:2019phz}, the structure and chemical makeup of the star, the  speed and size of the nuclear flame \citep{timmes1992a,schwab2020}, the reaction rates of 
the various $e$-capture interactions \citep{kirsebom2019a}, and the efficiency of convective energy transport \citep{Schwab:2018cnb}. 

Multidimensional hydrodynamic simulations exploring some of these parameters
\citep[e.g.,][]{Jones:2016asr,Jones:2018ule,Leung:2019phz}
generally favor tECSNe for ignition densities around $10^{10}\denu$ or smaller, even if the thermonuclear flame never transitions into a supersonic shock \citep[e.g.,][]{Jones:2018ule}. 
These models suggest that tECSNe may eject more than $\sim$1\msun\ of material, leaving behind only small bound remnants. Because of the small electron-to-baryon ratio ($Y_{\rm e}$) of the material  
processed in nuclear statistical equilibrium, tECSNe have been proposed as candidate sources for neutron-rich elements such as \iso{Ca}{48} and \iso{Ti}{50} that are typically
underproduced in normal type\,Ia SNe \citep[\ias;][]{Jones:2018ule}. Intriguingly, the chemical composition of some presolar meteoric grains compares well with  the  predicted nucleosynthetic signature of these events \citep{Jones:2018ule}. Some compact objects also seem consistent with the expectations for tECSN remnants \citep{raddi2019a,tauris2019}. 

Recent advances in our understanding of SAGB stars suggest that they might be even more prone to disruption than previously thought. Firstly, the second forbidden 2$^{+}$\,$\rightarrow$\,0$^{+}$ 
ground state transition in the $\beta$-decay of \iso{F}{20} has now been measured \citep{kirsebom2019b}, and found to be several orders of magnitude stronger than previously predicted. This makes $e$-captures on \iso{Ne}{20} a key energy source even at densities below  $10^{10}\denu$ \citep{kirsebom2019a,kirsebom2019b,Leung:2019phz}. Secondly, state-of-the-art simulations consistently show that some carbon is retained after the carbon burning phase \citep[][see also Sect.~\ref{sec:results}]{Iben:1983ts,Garcia1997,siess2006,Denissenkov:2013qaa,Farmer:2015afs,doherty2015,Lecoanet:2016abca,Jones:2018ule}. Due to its lower ignition temperature compared to oxygen, residual carbon can significantly 
influence  the explosion conditions. In some cases, during the main carbon burning episode, the flame, which ignites off-center, may  never reach the core, thereby creating a hybrid structure in which a CO-dominated core is surrounded by an 
ONe-dominated envelope \citep[][and references therein]{Denissenkov:2013qaa,Farmer:2015afs}. 

In a previous paper \citep[][henceforth \citetalias{antoniadis2020}]{antoniadis2020}, we examin the evolution of two stripped helium star models with masses of 1.8 and 2.5\msun. We show that these stars initiate explosive
oxygen burning at central densities $\lesssim 6\times 10^9\denu$, thereby completely avoiding  $e$-captures on \iso{Ne}{20}. At the same time, their envelopes expand significantly, leading to intense mass loss and the complete depletion of helium. The combination of low ignition densities, and helium-free composition, make these stars atypical progenitor candidates for thermonuclear  \ias \citep[henceforth (C)ONe \ias; see also][]{waldman2006a, waldman2008}. In addition, this behavior may also reduce significantly the mass range for which thermonuclear and core-collapse ECSNe may occur. To gauge  which flavor of \ia this channel may be relevant to, as well as the broader consequences for ECSNe, it is important to constrain the range of masses for which this outcome is possible, the  stellar structure and composition at the onset of collapse, as well as the sensitivity of these properties to underlying physical uncertainties. 

In this work we present detailed one-dimensional stellar evolution models that cover a broad range of initial helium star 
masses, compositions, and physical assumptions. The text is organized as follows: in Sect.~\ref{sec:methods} we describe our methodology and input physics. In Sect.~\ref{sec:results} we present the results of our calculations and in Sect.~\ref{sec:physical_uncertainties} we discuss various physical uncertainties in the input physics. Finally, in Sect.~\ref{sec:summary} we conclude with a concise summary and a discussion on the ramifications of our models for ECSNe and \ias.


\section{Simulations and physical assumptions} \label{sec:methods}

We perform our numerical simulations using the implicit 
one-dimensional code \textbf{M}odules 
for \textbf{E}xperiments in \textbf{S}tellar 
\textbf{A}strophysics \citep[\mesa\,v10398;][]{Paxton:2010ji,Paxton:2013pj,Paxton:2015jva,Paxton:2017eie}. Our main grid (henceforth \seriesone) consists of 252 
helium star models ($X=0$), covering the $\numrange{0.8}{3.5}\msun$  mass 
range with a resolution of 0.1\msun, at three metallicities 
($Z = 10^{-4}, 10^{-3}$, and 0.02).
Other physical assumptions in our models  are identical to those described in \citetalias{antoniadis2020} and are 
only briefly summarized here for completeness\footnote{Our \mesa configuration 
files (inlists) are publicly available at \url{https://zenodo.org/record/3580243}}.
To set the initial composition, we considered the  abundances of \cite{grevesse1998} as a reference for the solar isotopic composition. Models of different metallicities have abundances that are scaled from the aforementioned reference for the solar composition. To better account for an enhanced carbon-oxygen mixture as a result of helium burning, we used the type-2 OPAL Rosseland mean opacity tables \citep{OPAL}. 
 We considered ion and electron screening corrections as described in \cite{PCR2009} and \cite{Itoh2002},  respectively.
We used the  \texttt{HELM} \citep{HELM:eos} and \texttt{PC} \citep{PC:eos} equations-of-state. For their blending,  we followed the options suggested by \cite{Schwab:2017epw} so that the core is always treated using the \texttt{PC}  equation-of-state.

Convection was treated according to the modified mixing-length theory (MLT) prescription of \cite{MLT_Henyey} with a mixing-length parameter of $\alpha_{\text{ML}} = 2.0$. Contrary to the default MLT option in \mesa, this treatment allows for the convective efficiency to vary with the opaqueness of the convective element \citep{Paxton:2010ji}. 
Stability against convection was evaluated using the Ledoux criterion. 
Convective energy transport was computed using the \texttt{MLT++} variation of standard MLT \citep{Paxton:2015jva}. This artificially reduces large  superadiabatic temperature gradients to improve numerical stability (for a discussion on the consequences of this choice we refer to Sect.~\ref{sec:uncertainties}). 
We also adopted an  efficiency parameter of $\alpha_{\text{SEM}} = 1.0$ for semiconvection \citep{Langer1991}, and a diffusion coefficient of  $D_{\text{TH}} = 1.0$ for thermohaline mixing \citep{Brown_2013}. Both semiconvection and thermohaline mixing are treated by \mesa\ as diffusive processes \citep{kippenhahn1980,Langer1983}. We did not employ the predictive mixing implementation  described in \cite{Paxton:2017eie}. Lastly, we considered the effect of  overshooting across convective boundaries. Even though overshooting is likely not particularly strong  in this mass range, especially between the CO-core and envelope interface \citep{Lecoanet:2016abca}, it might have a significant influence on the propagation of the carbon burning flame \citep{Farmer:2015afs}. Motivated by this, we computed models with small variations in overshooting efficiency. More specifically, following \cite{Jones2013apj}, we considered three scaling factors of $f_{\rm OV}$ = 0.0, 0.014 and 0.016, where $f_{\rm OV}$ is defined  as a fraction of the local pressure scale height, $H_p$, and relates to the diffusion coefficient, $D_{\text{OV}}$ via,

\begin{align}
	\label{eq:OV}
	D_{\text{OV}} = D_0 \exp \left( - \frac{2z}{H_v} \right); \hspace{0.5cm} H_v = f_{\text{OV}} H_p,
\end{align}
where $z$ is the geometric distance from the edge of the convective zone \citep{Herwig2000}.

Our nuclear network consisted of 43 
isotopes and considered all  relevant weak reactions, using the 
weak rates from \cite{Suzuki2016}. We ensured that all these isotopes 
were  included in the \texttt{PC} equation-of-state calculation \citep[see][for details]{Schwab:2017epw}.
Mass loss rates due to stellar winds were implemented using the \texttt{Dutch} wind scheme \citep{Dutch}, and assuming no further binary interactions take place following the formation of the helium star. In our case, this implied two different rates, depending on the effective temperature of the star: for $T_{\text{eff}} > 10^4$\,K and a surface abundance of hydrogen of $X < 0.4$ by mass fraction (which is always satisfied in our models), we applied the prescription of \cite{Nugis2000} with a scaling factor of $\eta = 1$ (default value). For $T_{\text{eff}} < 10^4$\,K the mass loss rate followed the prescription of \cite{deJager1988}. 

To probe the influence of stellar wind on core growth, we  evolved additional stellar models with varying wind scaling factors within the \texttt{Dutch} scheme. This secondary grid (henceforth \seriestwo) consists of 110 models with $(Z,f_{\rm OV})=(0.02,0.0)$, covering the $\numrange{1.4}{3.5}\msun$  mass range with a step size of 0.1\msun. 

We did not employ any physical condition as a termination
criterion in our calculations. Instead, we let our models evolve until 
no further numerical solution could be found. In practice, this meant 
that calculations stopped either because the timestep became too small 
($\delta t <10^{-6}$\,s), the number of \mesa ``backups'' exceeded 50, 
or due to the core reaching an extremely high temperature 
($T_{c}>10^{9.5}$\,K), outside the validity regime of our 
equation-of-state. As we discuss in the next section, in terms of evolutionary stages, the majority of ONe-core models stopped shortly after the carbon burning phase, when 
their central density was $\rhoc \simeq  10^{9.0}$\,g\,cm$^{-3}$, and the mass of the degenerate core was $\sim 1.0-1.3\msun$. However, a handful of models  managed to reach the Chandrasekhar-mass limit ($\numrange{1.35}{1.37}\msun$) and stopped only after the onset of 
explosive oxygen burning and shortly before achieving the conditions for nuclear statistical equilibrium. A summary of the parameters discussed above is given in Table~\ref{tab:parameters}.   

\begin{table}[t]
    \centering
        \begin{tabular*}{\linewidth}{@{\extracolsep{0.16\textwidth}}p{0.3\linewidth}p{0.3\linewidth}@{}}
        \hline \hline \\
        Parameter & Value(s) \\\\
        \hline \\
        Convection ($\alpha_{\text{\tiny{ML}}}$) & $2.0$ \\\\
        Semiconvection ($\alpha_{\text{\tiny{SC}}}$) & $1.0$ \\\\
        Thermohaline ($D_{\text{\tiny{TH}}}$) & $1.0$ \\\\
        Wind scaling factor ($\eta$) & $0.1$, $0.25$, $0.5$, $0.8$, $1.0$, $1.6$ \\\\
        Overshooting ($f_{\text{OV}}$) & $0.0$, $0.014$, $0.016$ \\\\
        Metallicity ($Z$) & $10^{-4}$, $10^{-3}$, $0.02$ \\
        \hline
        \end{tabular*}
    \caption{Summary of mixing coefficients and variable parameters used in our calculations.}
     \label{tab:parameters}
\end{table}


\section{Results} \label{sec:results}

\begin{figure}
    \centering
    \includegraphics[width=\columnwidth]{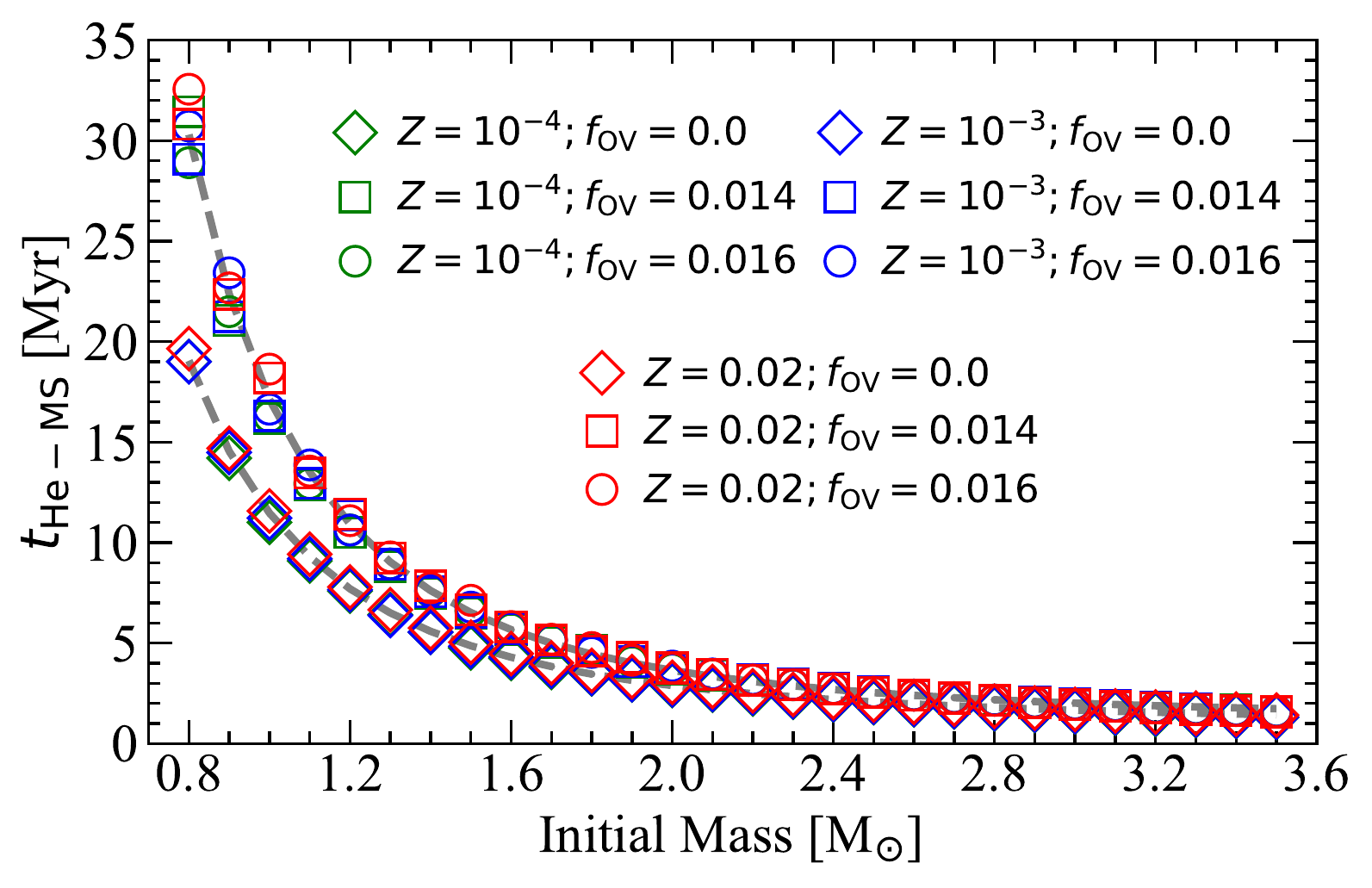}
    \caption{Helium main sequence lifetime as a function of initial mass. Dashed gray curves show the fitting function: $ax^b + c$, with ($a$, $b$, $c$) = ($10.54$, $-2.43$, $0.931$) and ($15.93$, $-2.7$, $1.193$) for models without overshooting and models with $f_{\rm OV} = 0.014, 0.016$ respectively.}
    \label{fig:tams_vs_mass}
\end{figure}

Figure~\ref{fig:tams_vs_mass} shows the duration of the core helium burning phase as a function of initial mass, while Fig.~\ref{fig:hrd_series1} illustrates the Hertzsprung-Russell diagram for five representative models from \seriesone (see below). Figure~\ref{fig:RhoT} shows the core temperature and density evolution for the same stars. 

All models undergo a  period of stable core helium burning that lasts between 1.5 and 32.5\,Myr. The duration of the core helium burning phase (henceforth helium main sequence; He-MS) in our models can be approximated by a power-law of the form,

\begin{equation}\label{eq:tams_power_law}
    t_{\text{He-MS}}/{\rm Myr} = 
    \begin{cases}
        10.54 \left(\frac{M_{\text{He, i}}}{\msun}\right)^{-2.43} + 0.931; \hspace{0.25cm} f_{\text{OV}} = 0.0, \\ \\
        15.93 \left(\frac{M_{\text{He, i}}}{\msun}\right)^{-2.70} + 1.193; \hspace{0.25cm} f_{\text{OV}} > 0.0, 
    \end{cases}
\end{equation}
where $M_{\rm{He, i}}$ is the initial mass of the model. In agreement with \cite{Yan_2016}, we find that models in which overshooting is included result in larger convective core masses and longer lifetimes. 
The difference in lifetimes between models with $f_{\rm OV}=0$ and $f_{\rm OV}>0$ is more pronounced at lower 
initial masses. This is due to an increase in the amount of fuel added to the core which does not result in a 
strong luminosity increase. The helium burning lifetime of helium star models in the metallicity range covered by our grids is largely insensitive to variations of this parameter\footnote{We note that metallicity variations would affect the mass of \iso{He}{4}-rich regions (and thus their burning lifetime) if we were to start our calculations with a normal \iso{H}{1}-rich star. This however does not affect the results presented here.}. This is in contrast to what has been found for larger-mass helium stars ($M_{\rm{He, i}} \gtrsim 4\msun$) which are subject to stronger mass loss rates during core helium burning \citep{Aguilera-Dena:2021abc}. The influence of overshooting and initial composition is further discussed in Sect.~\ref{sec:physical_uncertainties}.

In \mesa, the outer boundary of the CO and ONe cores is by default defined based on the abundance of \iso{He}{4} and other relevant isotopes such as \iso{C}{12} and \iso{O}{16}. We find this to be problematic because some of our models lose their entire \iso{He}{4} envelope. For this reason, we choose to define the core boundary as the location where the ratio of the ideal gas pressure to degenerate 
pressure is of  order  unity ($P_{\text{gas}} / P_{\text{deg}} = 1$).
Following  core helium burning, the newly formed CO cores lie within the mass range $0.21 
\lesssim M_c/\rm M_\odot \lesssim 1.86$. 
At this stage, the main source of energy moves into a stable radiative helium burning shell, while 
the core contracts and cools as a result of neutrino emission. The subsequent evolution varies 
strongly with initial mass, allowing us to distinguish several distinct evolutionary paths. In what follows, 
we explore in detail the subsequent evolution of our \seriesone stellar models, grouping them 
according to their core composition and final fate.

\subsection{CO \& hybrid CONe white dwarfs}\label{sec:wd_evolution}
For $M_{\rm{He, i}}\lesssim 1.4\msun$ 
($\lesssim1.2\msun$ for $Z=10^{-4}$;  see 
Figs.~\ref{fig:parameterSpace} and \ref{fig:final_fates}  for 
exact mass boundaries as a function of $Z$ and $f_{\rm OV}$), CO cores do not become massive enough 
to initiate carbon burning. Instead, stable shell helium burning gradually increases the mass of the core, while the envelope is expelled over a timescale of a few Myr, if no mass transfer takes place. These stars ultimately become CO WDs. The difference in the mass threshold most likely stem from the metallicity-depended wind although other effects, such as a different mean molecular weight, can contribute to the cause.

\begin{figure}
    \centering
    \includegraphics[width=\columnwidth]{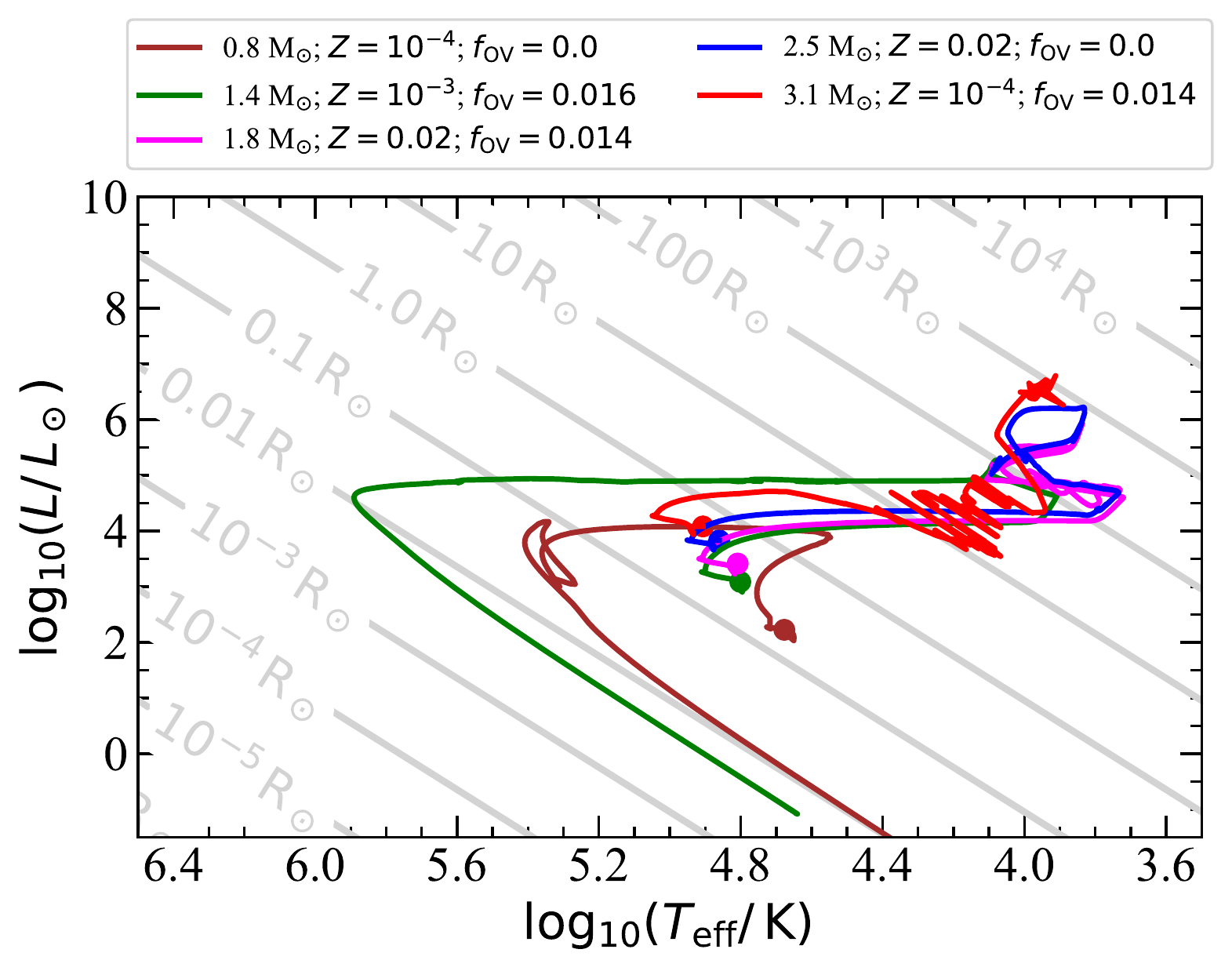}
    \caption{Hertzsprung-Russell diagram for five \seriesone models starting from the core helium burning phase (indicated by the circular mark). Lines of constant radius were calculated using the Stefan-Boltzmann law; $R= \left( L/4\pi \sigma T_{\rm eff}^4 \right)^{1/2}$.}
    \label{fig:hrd_series1}
\end{figure}

\begin{figure*}[t]
    \centering
    \includegraphics[width=\textwidth]{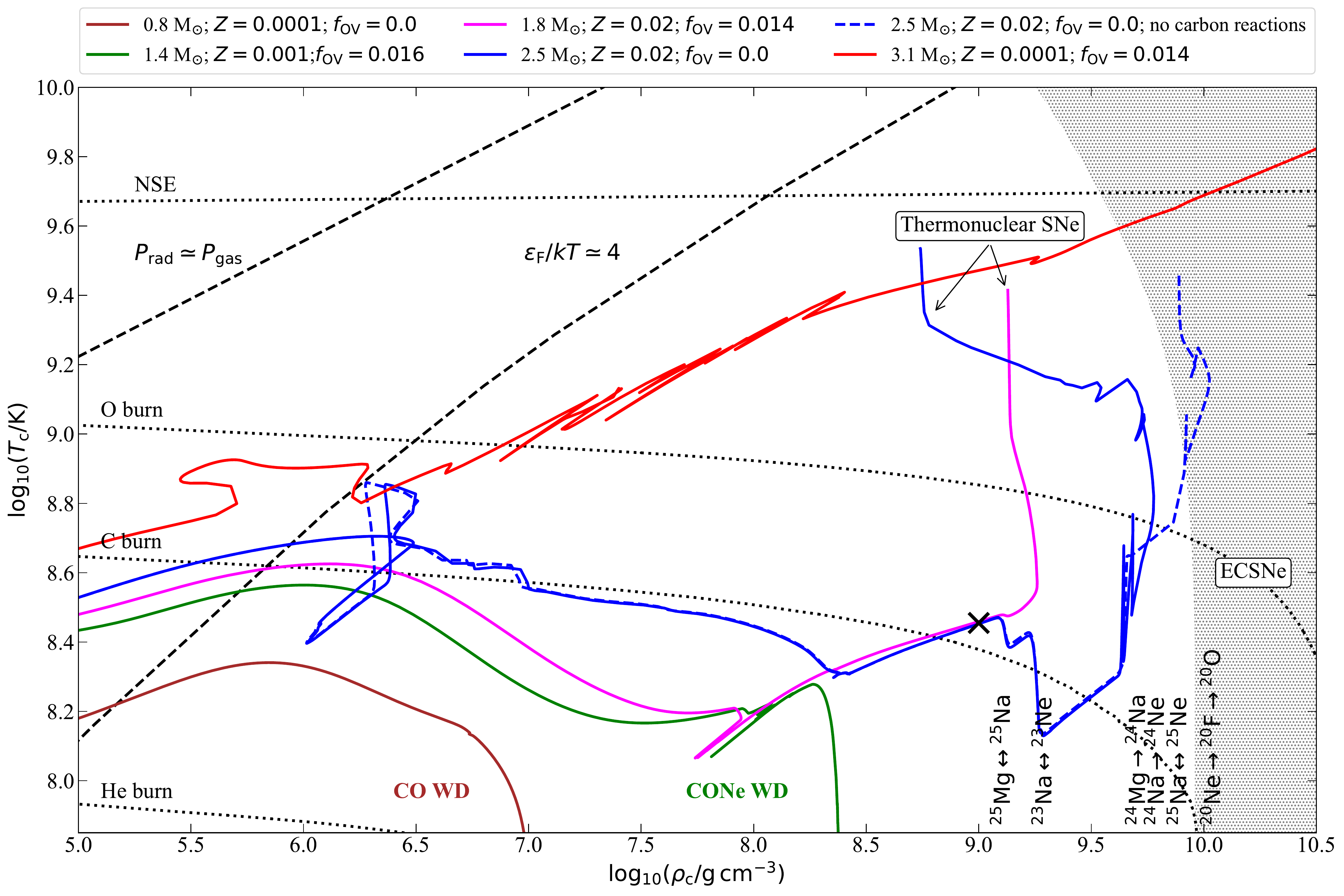}
    \caption{Examples of the evolution of five \seriesone models with different initial masses in the $\log(\rho_{\rm c})-\log(T_{\rm c})$ plane. Black dotted lines show approximate ignition curves taken from \mesa. Black dashed lines indicate different pressure regimes while the hatched area shows the approximate region for $e$-captures on $\rm ^{20}Ne$ nuclei. 
    The blue dashed line refers to the same stellar model as the one with the solid blue line; the only difference is that for the former, all carbon-participating reaction have been switched off prior to Urca cooling leading most likely to an ECSN. The density at which carbon reactions were turned off is marked by a black cross-mark.}
    \label{fig:RhoT}
\end{figure*}

\begin{figure*}
    \centering
    \includegraphics[scale=0.6]{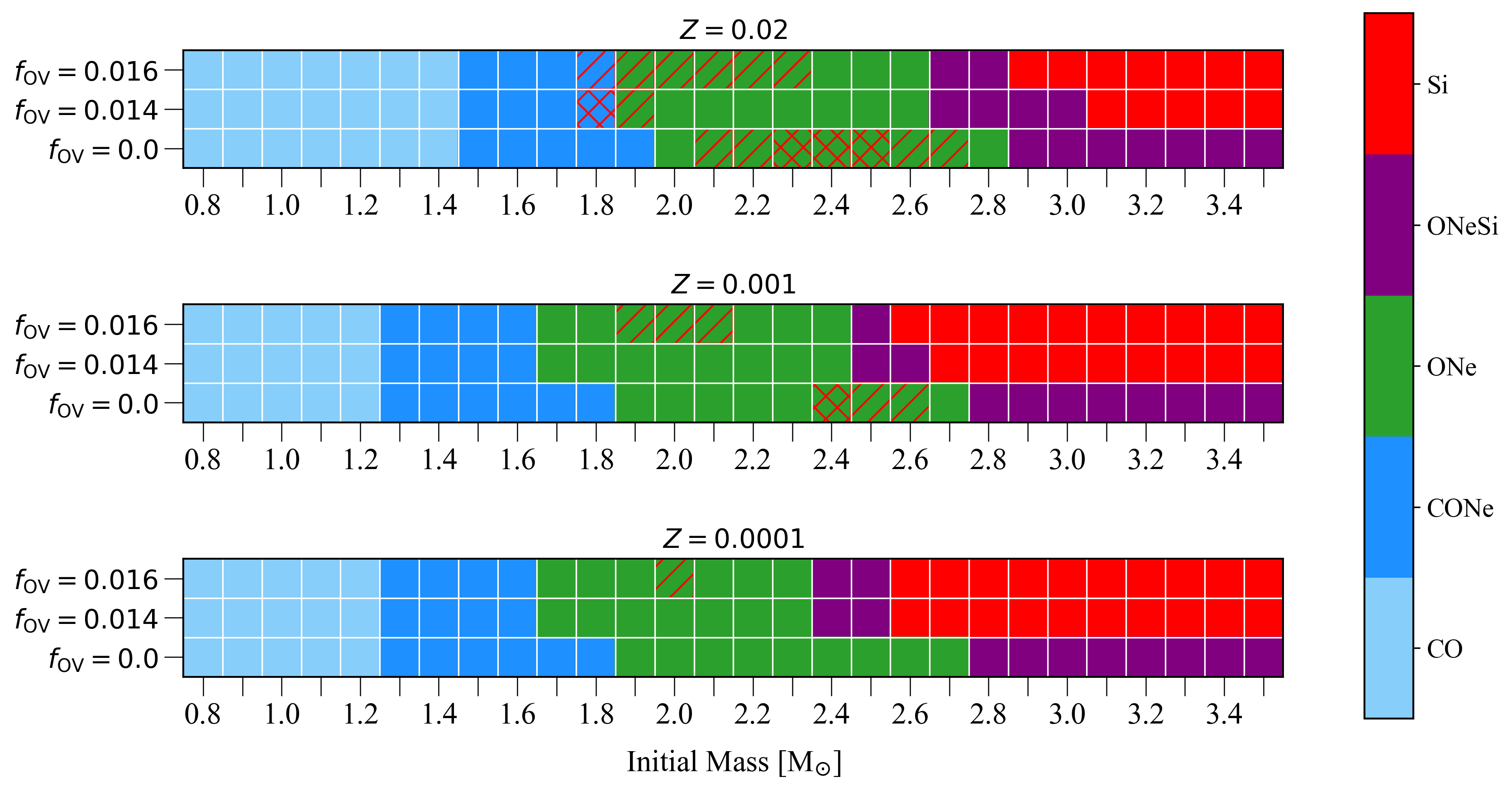}
    \caption{Compositions and evolutionary stages of our \seriesone models when calculations were terminated, in raster format. Metallicity for the bottom, middle, and top panel is $Z=10^{-4}$, $Z=10^{-3}$, and $Z=0.02$ respectively, while the colorbar indicates the composition of the core at the endpoint of the simulation (see text). Semihatched regions show models that stopped only after developing a near-$\mch$ core within the range $\numrange{1.35}{1.37}\msun$, shortly before oxygen ignition. All these models also have no envelopes and contain no helium. Fully hatched regions show models with near-$\mch$ cores that reached the explosive oxygen burning stage. All nonhatched ONe models stopped before their entire envelope was lost, and have cores with masses  between  ($1.0 < M_c/\rm M_\odot \leq 1.36$).}  
        \label{fig:parameterSpace}
\end{figure*}

\begin{figure*}
    \centering
    \includegraphics[scale=0.6]{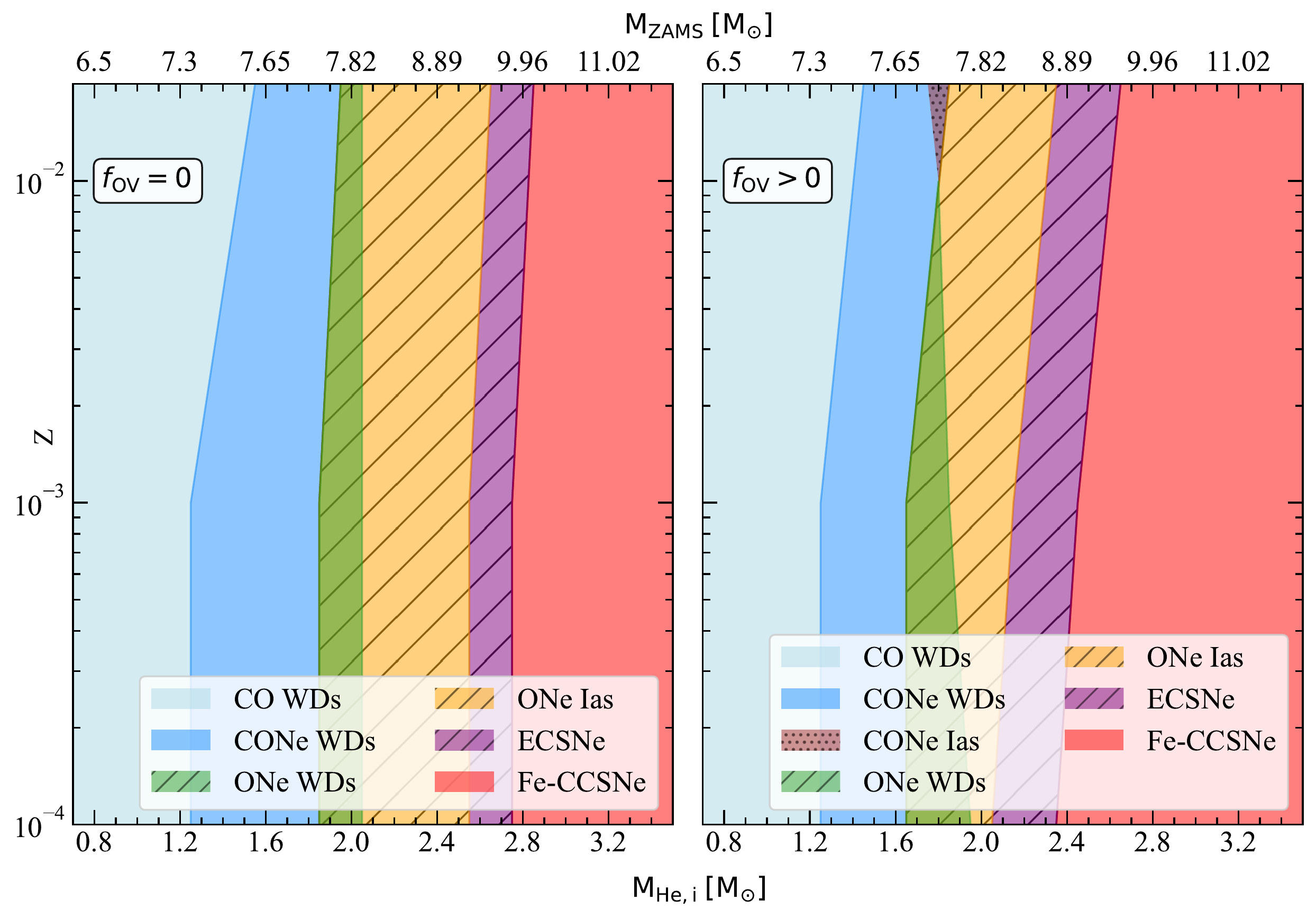}
    \caption{Final fates as a function of initial helium star mass and initial metallicity. The left panel corresponds to models without overshooting, while the right panel corresponds to models with overshooting across convective boundaries ($\rm f_{OV} = 0.014, 0.016$). Hatched regions show models with ONe cores featuring residual carbon, which evolve either toward an ONe WD, a thermal runaway, or an ECSN.
    The boundary between ONe \ias and ECSNe corresponds to models with average residual carbon mass fractions of $X(\iso{C}{12}) \simeq 0.004$ (see Sect.~\ref{sec:physical_uncertainties}). 
    The ZAMS masses (of solar metallicity models) were mapped based on Fig. 4 of  \cite{Farmer:2015afs}. For $M_{\rm ZAMS} < 7.0\msun$ we extrapolated the mass using the fitting function $M_{\rm ZAMS} = 2(M_{\rm He, core} + 2.45)$. For more details, see Sect.~\ref{sec:results}.}
    \label{fig:final_fates}
\end{figure*}

The evolution of a star representative of this class is 
illustrated in Figs.~\ref{fig:hrd_series1} and \ref{fig:RhoT} with brown color.  
This particular model experiences a late thermal pulse  
while it descends from the post-AGB 
track toward its terminal WD cooling track. This 
flash lasts $\sim 470\,\text{yr}$, resulting in a peak \iso{He}{4}-burning luminosity of 
$\log_{10}(L_{\rm He, peak}/{\rm L}_\odot) \simeq 5.22$, and does not lead to significant mass loss. The final outcome is a $0.78\msun$ CO WD  with a mean abundance of $X(\iso{C}{12}) = 0.47$, $X(\iso{O}{16}) = 0.52$ and $Y = 0.01$\footnote{Unless otherwise stated, this notation refers to a mass averaged mass fraction of an isotope for the rest of the paper.}

At the end of the simulation, our CO WDs have masses between 0.7 and 1.16\msun\ and possess a \iso{He}{4}-rich envelope of $\lesssim 0.07\msun$. For the lowest-mass model, the shell burning and 
envelope ejection phase lasts $\sim 6.4\,\text{Myr}$. During this time, the envelope expands to $R \simeq 0.93$\,R$_\odot$ and reaches an effective temperature of 
$\log_{10} (T_{\rm{eff}}/{\rm K}) \simeq 4.75$. For the highest-mass model, the 
envelope expands substantially more, to $R \simeq 50$\,R$_\odot$, reaching a surface 
temperature of $\log_{10} (T_{\rm{eff}}/{\rm K}) \simeq 3.85$, and is expelled in $\sim 1.5$\,Myr.

As we move to higher initial masses ($M_{\rm{He, i}}\gtrsim 1.3\msun$), our models experience off-center carbon ignition, due to thermal neutrino losses leading to a temperature inversion in the inner core. The mass of the core during the off-center ignition is on average $\rm M_{CO} \simeq 1.05$\msun, consistent  with  typical values found in literature \citep[e.g.,][]{doherty2015}. Similarly to what has been found for \iso{H}{1}-rich SAGB stars \citep[e.g.,][]{Garcia1997,Gil-Pons:2003sep,siess2006,Poelarends:2007ip,Doherty:2010slg,Farmer:2015afs}, the ignition location 
varies with initial mass, with increasing \iso{He}{4}-ZAMS masses resulting in ignition locations closer to the center. 
As long as the temperature at the base of the carbon burning region remains higher than the carbon ignition threshold, the 
burning front advances inward.  For stars that exhibit this behavior and have relatively low masses, the flame quenches before reaching the center.
In this case, the core exhibits a hybrid-like structure, where the inner layers are primarily composed of carbon and oxygen, while the outer layers have an oxygen-neon composition \citep[for a detailed discussion on the flame quenching mechanisms we refer to the work of][]{Siess2009, Denissenkov:2013qaa, chen2014b, Farmer:2015afs}. 
In our \seriesone models, the lowest-mass helium star that ignites carbon off-center is a $1.3\msun$ model with $(Z,f_{\rm OV})=(10^{-4},0.0)$. The ignition happens at a mass coordinate of $\sim 1.2\msun$ and the flame stops at a mass coordinate of $\sim 1.1\msun$. The highest-mass model that forms such a hybrid structure is a $1.8\msun$ helium star with $Z=0.02,f_{\text{OV}} = (0.02,0.014)$, that ignites carbon at a  mass coordinate of $\sim 1.1\msun$. The flame quenching occurs at $\sim 0.9\msun$. 

The evolution of a representative model from this category is illustrated with green in Figs.~\ref{fig:hrd_series1} and \ref{fig:RhoT}. An overview of mass boundaries for hybrid WDs, and their dependence on metallicity and overshooting can be seen in Figs.~\ref{fig:parameterSpace} and \ref{fig:final_fates}. 
In the latter, we map helium core masses to ZAMS masses based on Fig. 4 of \cite{Farmer:2015afs}. This does not take into account the role of metallicity in setting the relation between initial mass and helium core mass, but we expect the difference to be small. In the absence of further binary interactions after the hydrogen envelope removal, the final hybrid CONe WDs in our models have masses between 1.2 and 1.33\msun. This means that they only need to accrete very small amounts of matter to reach the Chandrasekhar-mass limit (see Sect.~\ref{sec:summary} for a discussion).

\subsection{ONe cores}\label{sec:one_core_evolution}
\begin{figure}
    \centering
    \includegraphics[width=\columnwidth]{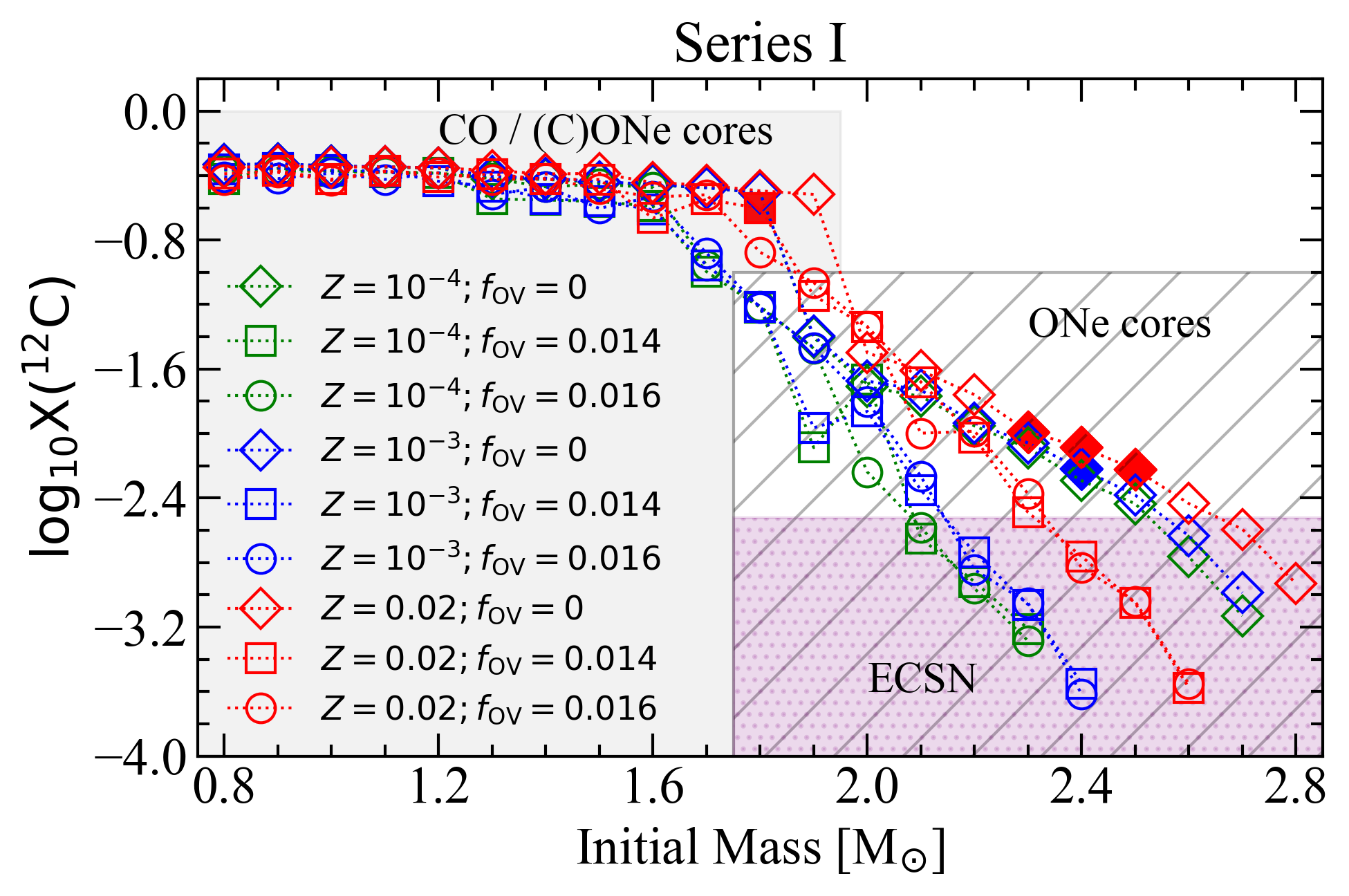}
    \includegraphics[width=\columnwidth]{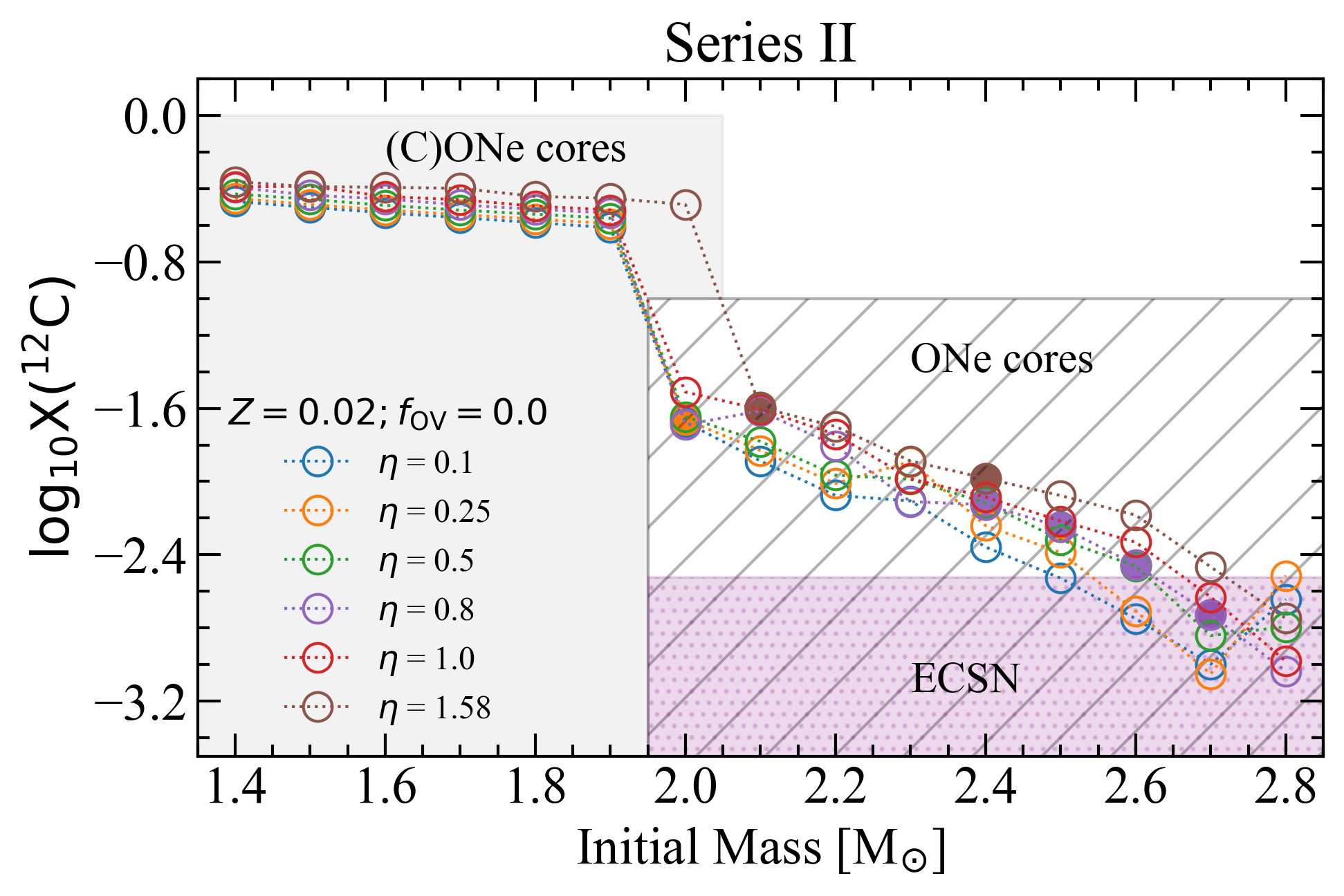}
    \caption{Average mass fraction of residual carbon in the degenerate core, as a function of initial mass. For models that undergo carbon burning, the mass fractions were calculated shortly before the Urca cooling stage, that is when the density reached a value of $\log_{10} (\rho_c / \text{g cm}^{-3}) \approx 9.0$. Models that ignite residual carbon are represented with filled data points. All \seriestwo models assume $Z=0.02$, $f_{\rm OV}=0.0$, and only vary in terms of the wind scaling factor $\eta$.}
    \label{fig:carbon_mass_fractions}
\end{figure}

Stars with initial masses between $\sim 1.9\msun$ and $\sim  2.8\msun$ (see Figs.~\ref{fig:parameterSpace} and \ref{fig:final_fates}),  stand out due to their ability to develop highly degenerate ONe cores. As discussed in \citetalias{antoniadis2020}, one key property that  plays a critical role in determining the final fate of these stars is the amount of residual carbon distributed throughout the core  after the main carbon burning phase. 
With the exception of a few models with $f_{\rm OV}> 0$ (see Sect.~\ref{sec:cone_core_evolution}), for  most stars in this mass range, carbon still ignites  in a shell but the carbon flame
is able to propagate all the way to the center, thereby forming an ONe-dominated core. 
However, our simulations show that in all cases, some residual carbon remains
distributed throughout the degenerate core, with average mass fractions of $0.0003 \lesssim X(\iso{C}{12}) \lesssim 0.11$ (see Appendix~\ref{apx:composition} for representative carbon abundance profiles). 
Figure~\ref{fig:carbon_mass_fractions} shows variations in  the average residual carbon abundance of the degenerate core as a function of initial mass.
These abundances were calculated using the \mesa profile nearest to $\logrhoc=9$, that is after the main carbon burning phase and before the onset of Urca reactions. As can be seen, the amount of residual carbon varies strongly both with initial mass and metallicity, especially for $f_{\rm OV}>0$.  
 The residual-carbon profile remains practically unchanged until the  onset of exothermic weak reactions at $\logrhoc\simeq 9.6$ (see Fig.~\ref{fig:RhoT}).

\begin{figure*}
    \centering
    \begin{subfigure}{.5\textwidth}
        \centering
        \includegraphics[width=\columnwidth]{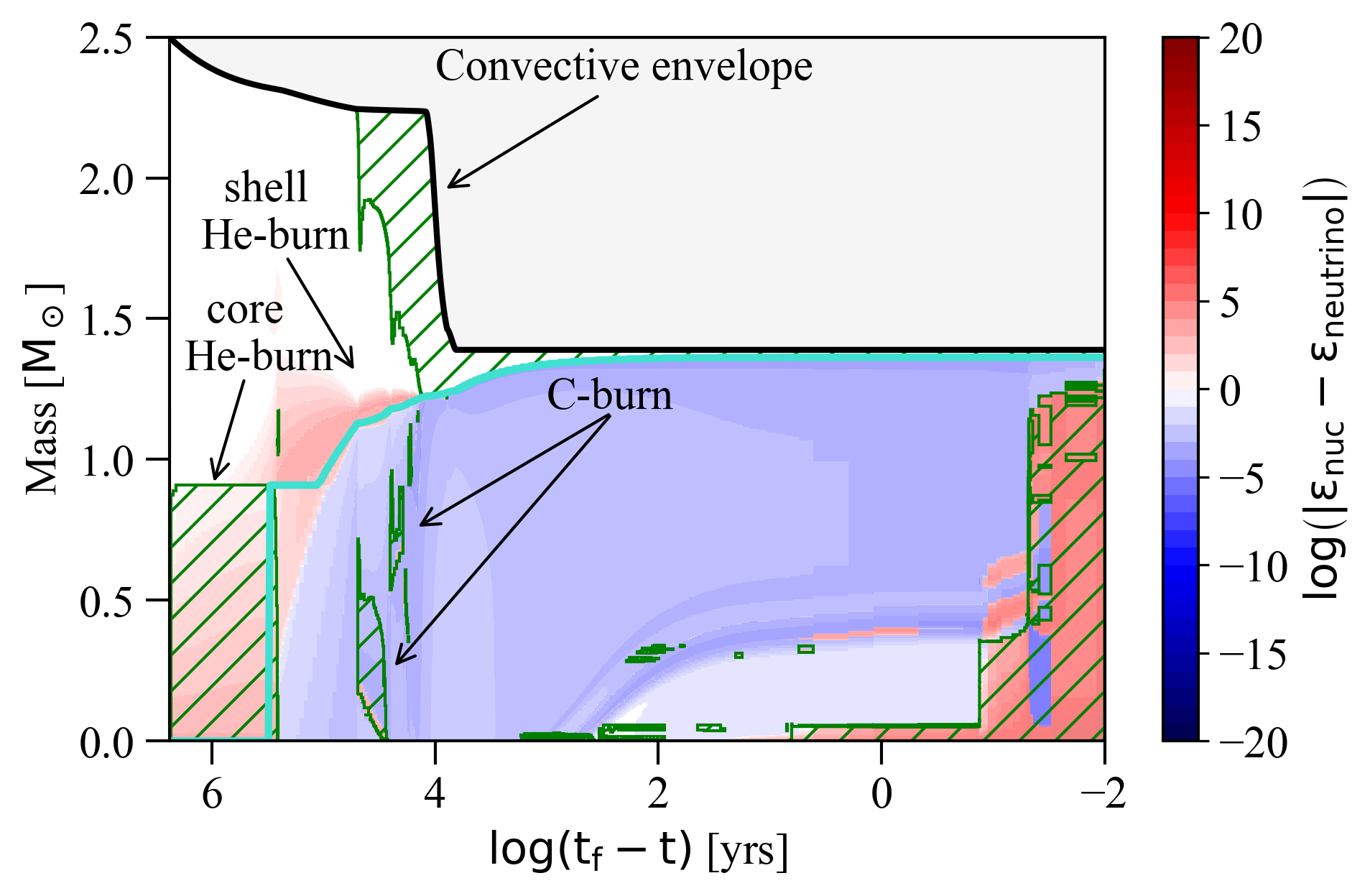}
    \end{subfigure}\hfill
    \begin{subfigure}{.5\textwidth}
        \centering
        \includegraphics[width=\columnwidth]{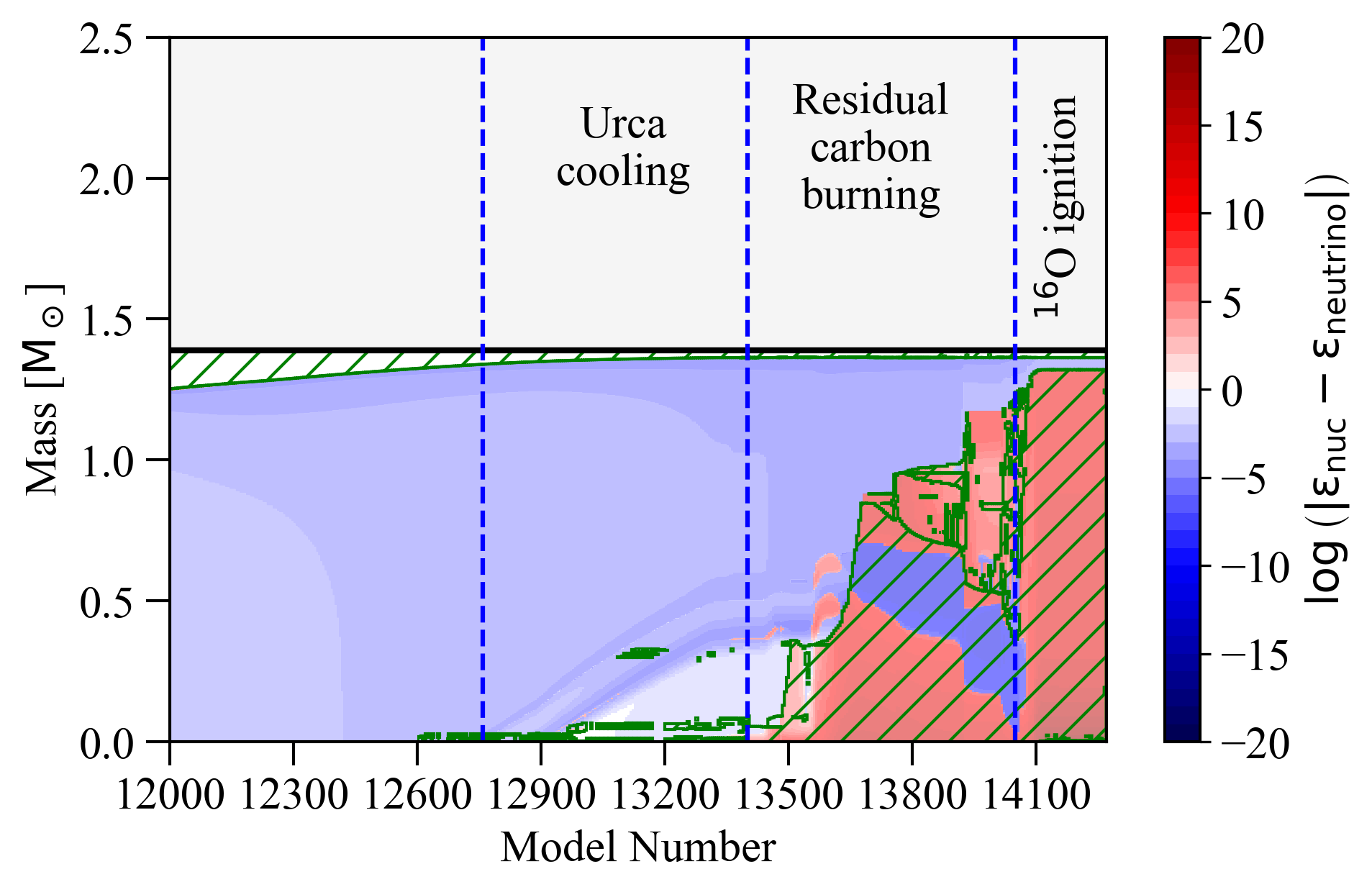}
    \end{subfigure}\hfill
    
     \begin{subfigure}{.5\textwidth}
        \centering
        \includegraphics[width=\columnwidth]{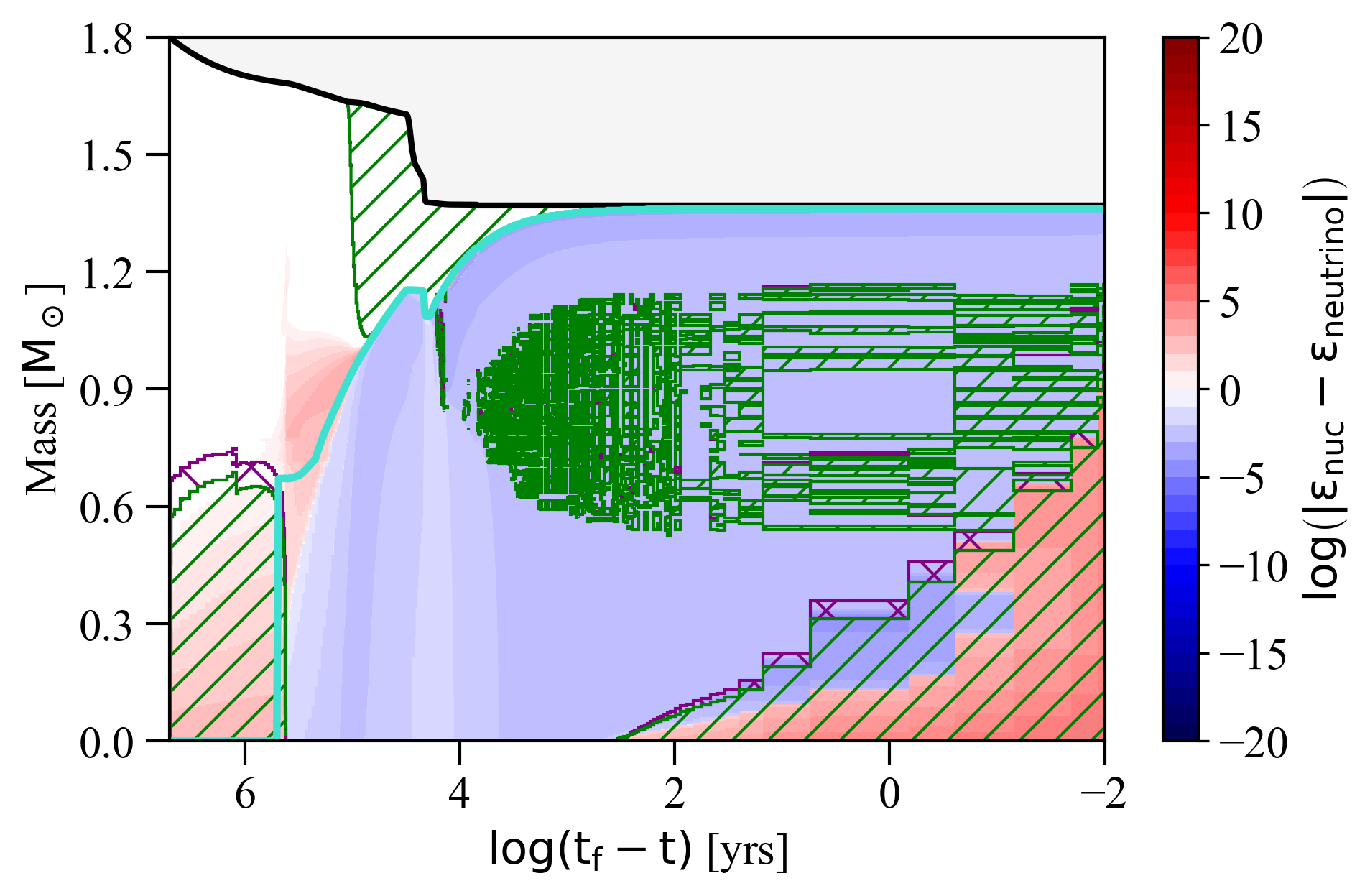}
    \end{subfigure}\hfill
    \begin{subfigure}{.5\textwidth}
        \centering
        \includegraphics[width=\columnwidth]{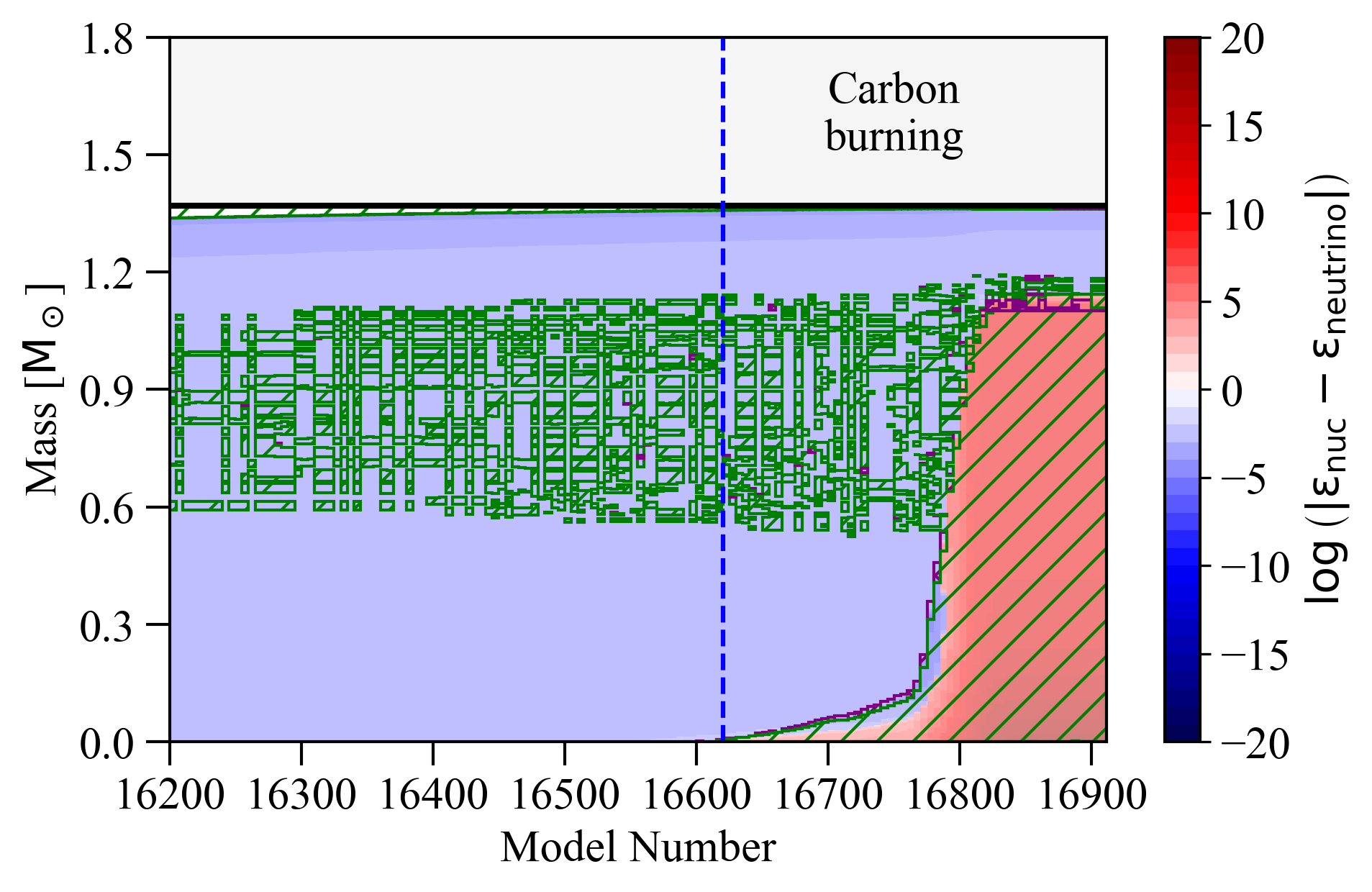}
    \end{subfigure}

    \caption{On the left, we present a Kippenhahn diagram of two \seriesone stellar models as a function of remaining evolution time. On the right, we present a section of the same diagrams but as a function of model number in which the location of some important stages can be seen more accurately. Green hatched areas denote regions of convection. The intensity shown in the colorbar, indicates the net energy rate. Turquoise line shows the CO core. \textbf{Top panel}: Structure of a $2.5 \rm M_{\odot}$ helium star with $Z = 0.02$ and no overshooting. During $e$-captures on $\rm ^{24}Mg$ and subsequent carbon burning, the core becomes convective. \textbf{Bottom panel}: Structure of a hybrid $1.8 \rm M_{\odot}$ helium star with $Z = 0.02$ and $f_{\rm OV} = 0.014$. Convective overshooting (purple hatched area) leads to larger core during the helium main sequence phase. Compressional heating ignites the carbon resulting in a thermal runaway.}
    \label{fig:Kipp}
\end{figure*}

\begin{figure}
    \centering
    \includegraphics[width=\columnwidth]{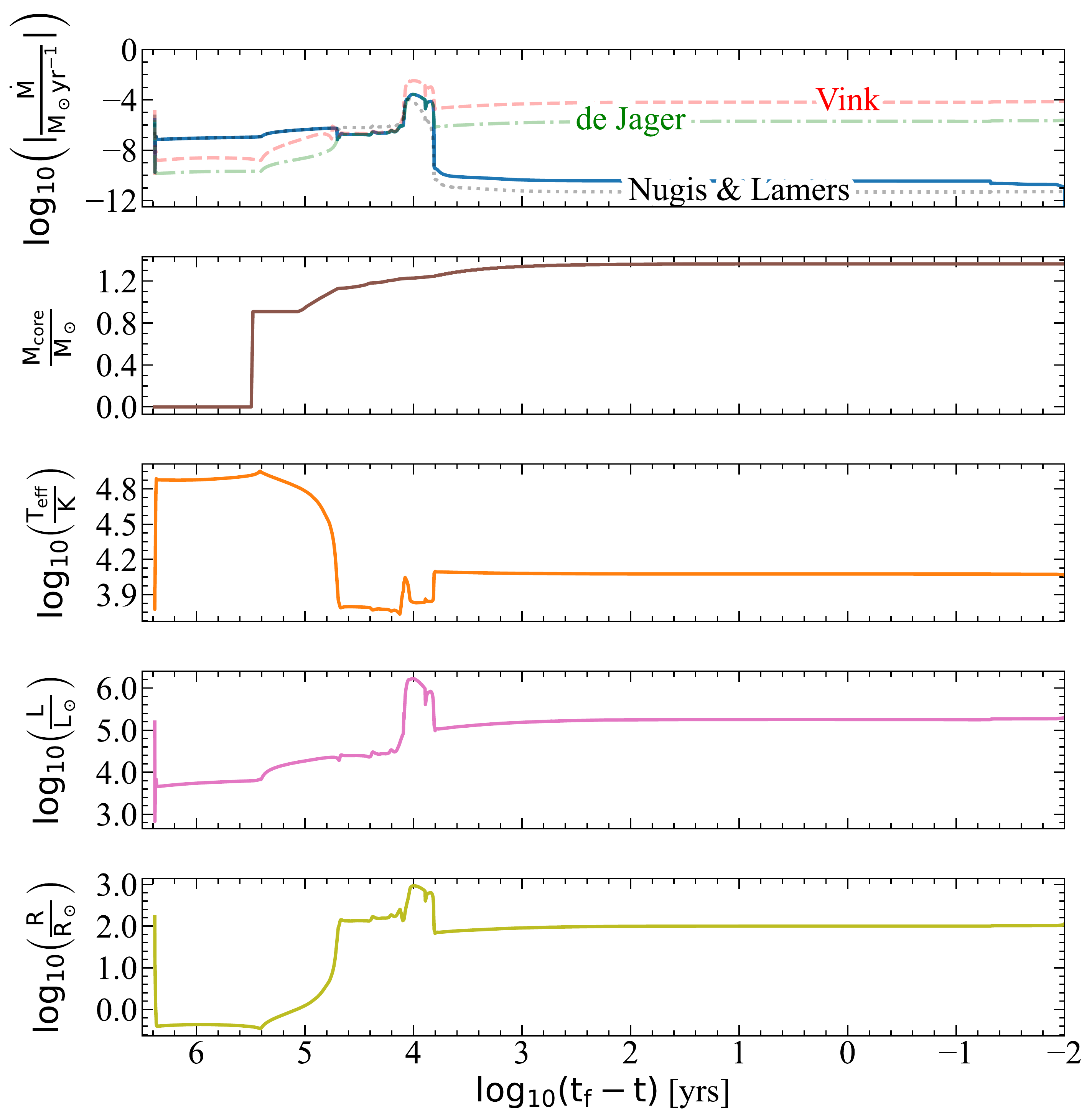}
    \caption{Mass loss rate, core mass growth, $T_{\text{eff}}$, luminosity, and radius vs remaining time for the $2.5\msun$ fiducial model of \seriesone. The translucent dashed, dotted, and dash-dotted lines represent the mass loss rates based on \cite{Vink:2017ujd}, \cite{Nugis2000}, and \cite{deJager1988} respectively.}
    \label{fig:lum_teff_mdot_rad}
\end{figure}

\begin{figure}
    \centering
    \includegraphics[width=\columnwidth]{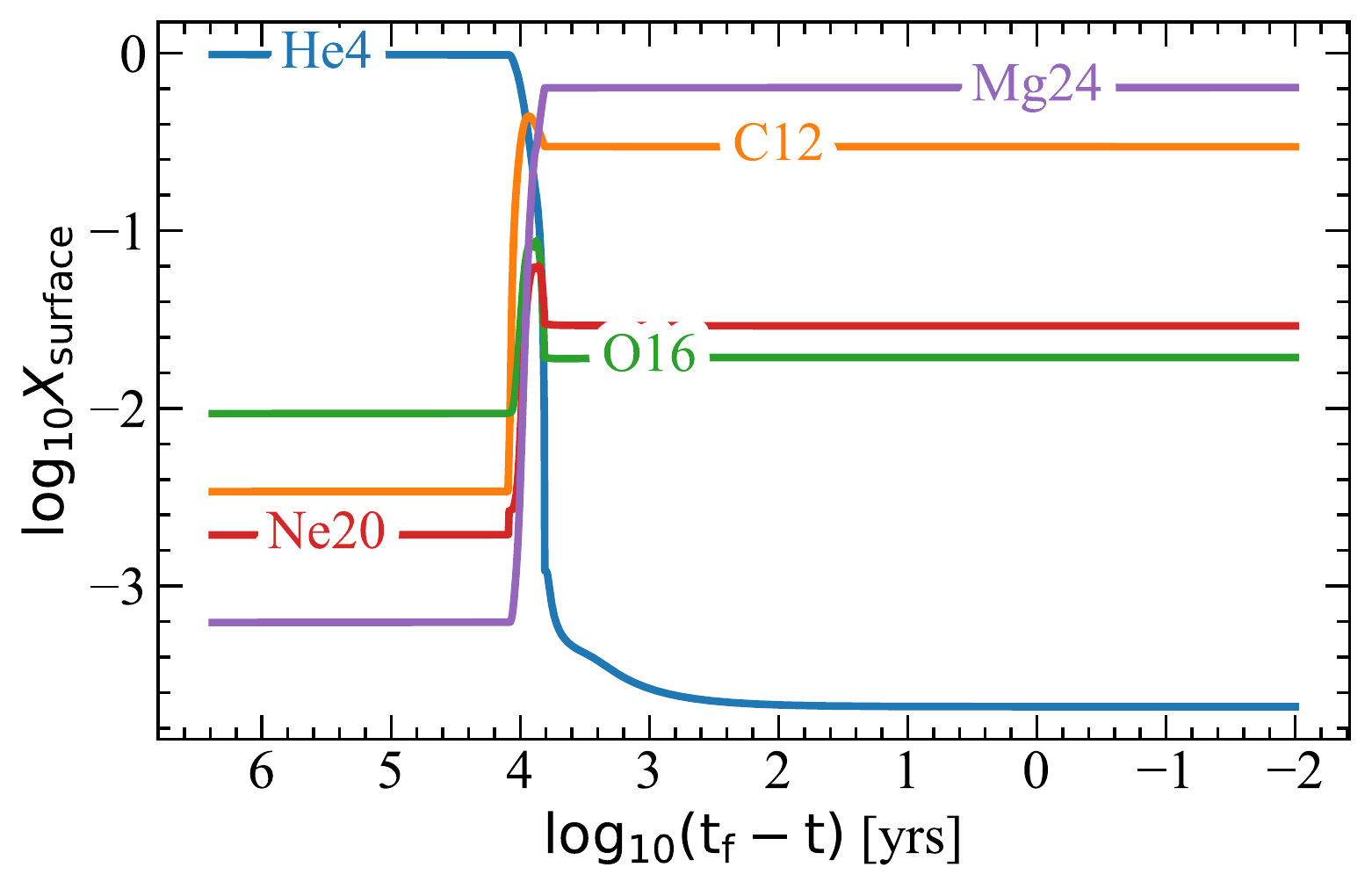}
    \caption{Surface abundance evolution for the most dominant species in our $2.5\msun$ fiducial model of \seriesone.}
    \label{fig:surface_abun_evol}
\end{figure}

To better illustrate the properties of these stars during their late evolutionary stages,  we now focus on a fiducial model with an initial mass of $\rm 2.5\msun$ and $(Z,f_{\rm OV}) = (0.02,0.0)$, as representative for this mass range. 
The evolution of this model is illustrated by the blue curves in Figs.~\ref{fig:hrd_series1} and \ref{fig:RhoT}, while Fig.~\ref{fig:Kipp} 
shows the corresponding Kippenhahn diagram. Following the depletion of helium in the center, the model initially ignites carbon at a mass coordinate $\sim 0.3\msun$ when the CO core has a mass of $\sim 1.2\msun$, and the total mass of the star is $\sim 2.3\msun$. The carbon flame advances in both directions processing the material in $\sim 40\,\text{kyr}$. During this period, the star resembles a 
red giant with an extended envelope that has  $\log_{10} \left( T_{\text{eff}}\,/ \text{K} \right) \simeq 3.76$, $\log_{10} \left( L\,/ {\rm L}_\odot \right) \simeq 4.3$, and $\log_{10} \left( R\,/ {\rm R}_\odot \right) \simeq 2.1$.
Its mass loss rate during this phase is $\dot{M} \simeq 10^{-6}\msun\,\text{yr}^{-1}$, as  estimated using the \cite{deJager1988} wind scheme that has been empirically calibrated for red super giants (see top panel of Fig.~\ref{fig:lum_teff_mdot_rad}). 
The surface composition is dominated by helium  (see Fig.~\ref{fig:surface_abun_evol}). 

\begin{figure}
    \centering
    \includegraphics[width=\columnwidth]{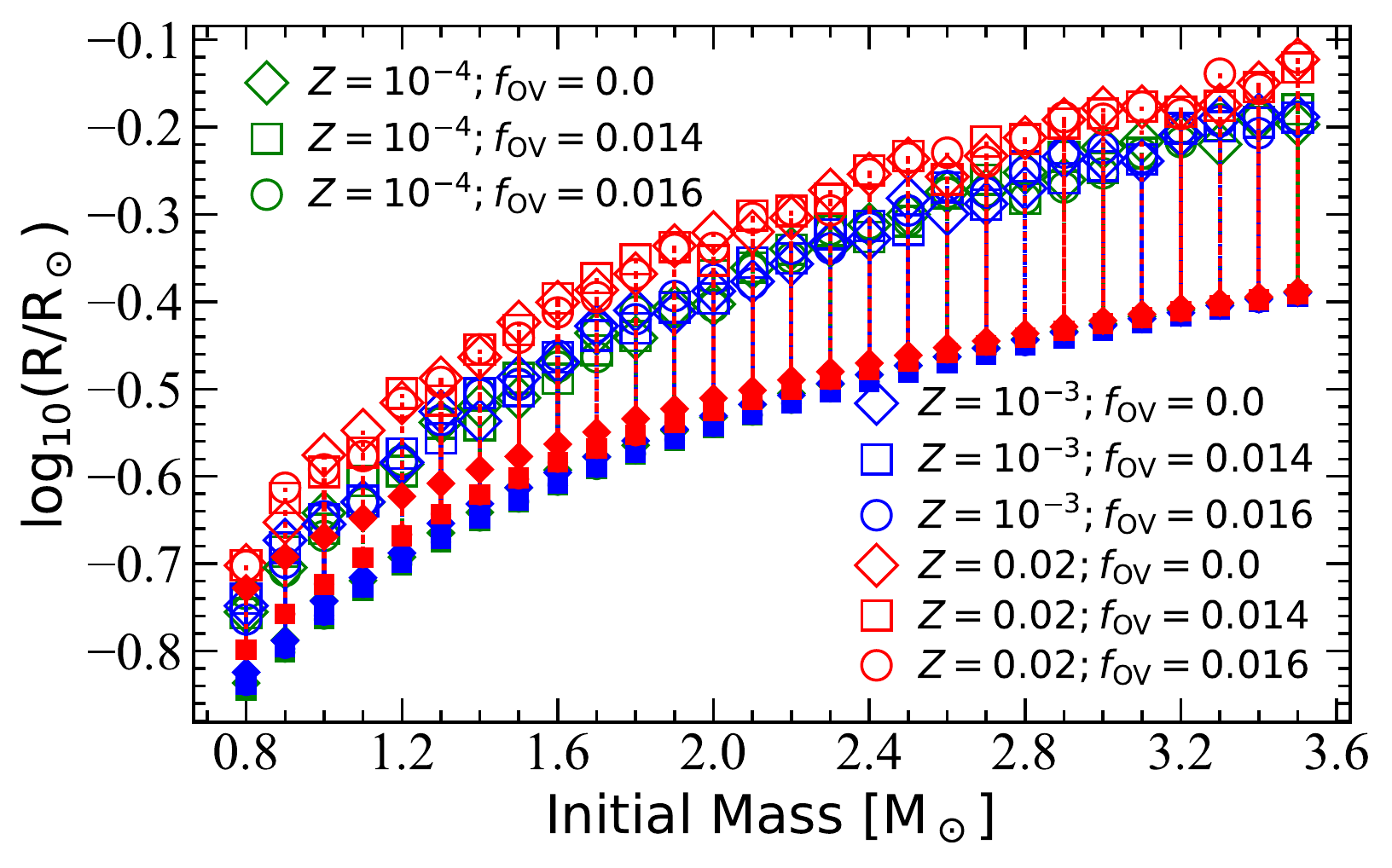}
    \includegraphics[width=\columnwidth]{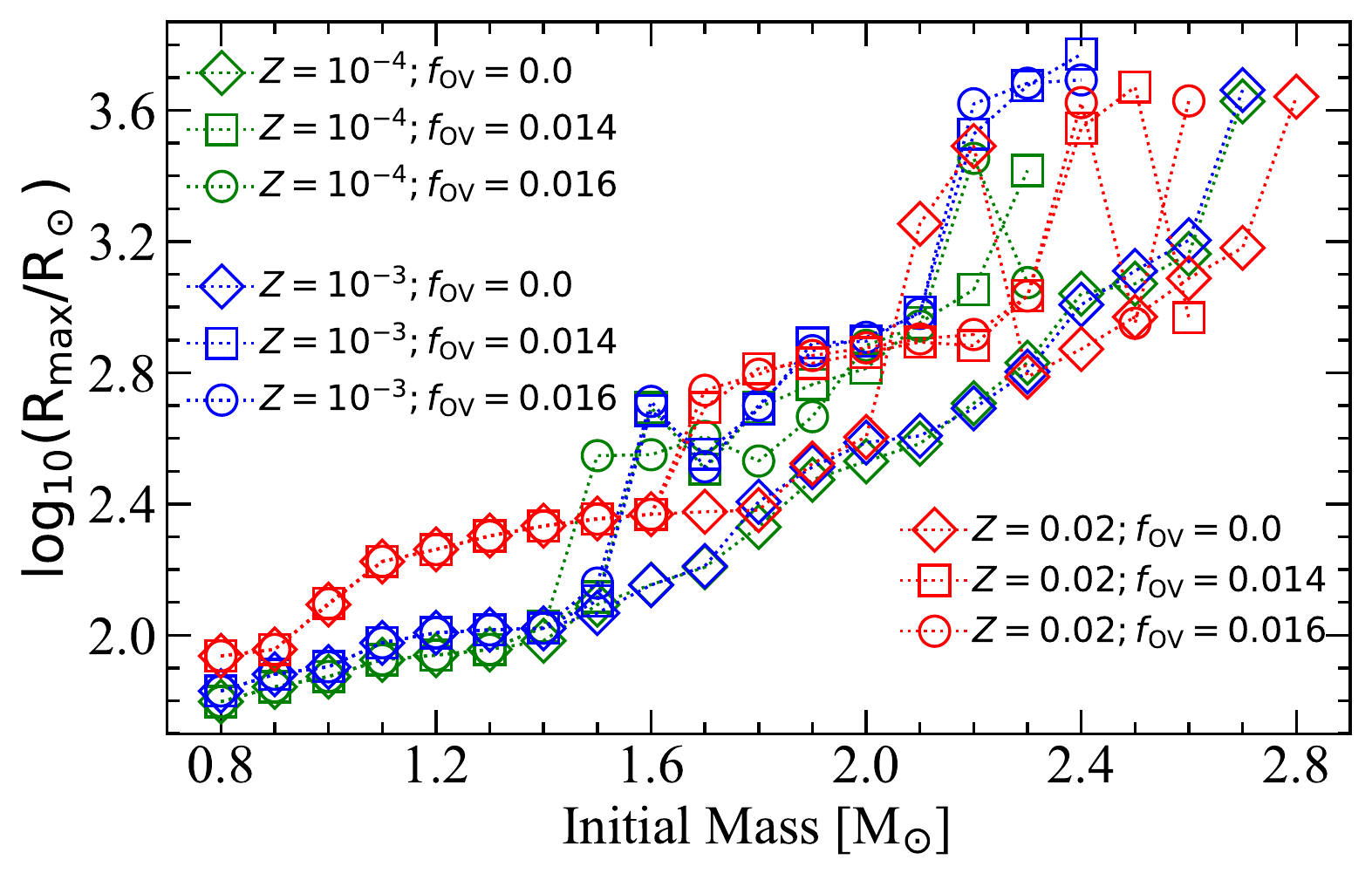}
    \caption{Radius as a function of initial mass for our \seriesone models. The upper panel shows the evolution of the radius during helium main sequence phase. The commencement and finish of core helium burning are shown by nonfilled and filled data points, respectively. The maximum radius reached is depicted in the lower panel. These radii might be underestimated since  \texttt{MLT++}  suppresses envelope inflation (see Sect.~\ref{sec:uncertainties}).}
    \label{fig:mass_radius}
\end{figure}

Following carbon burning, the degenerate core is composed primarily of oxygen and neon; $X(\iso{O}{16}) \simeq 0.37$, $X(\iso{Ne}{20}) \simeq 0.41$, as well as traces of residual carbon; $X(\iso{C}{12}) \simeq 0.006$, or $ \rm M_c (\iso{C}{12}) \simeq 0.008\msun$, where $\rm M_c (\iso{C}{12})$ refers to the mass of the degenerate core that is composed of carbon. The top panel of Fig.~\ref{fig:Kipp} shows the Kippenhahn diagram for this model as a function of remaining time (left), and as a function of model number, focusing on the evolution following carbon burning (right). 
For the last $\sim$10$^4$\,yr of the evolution, the center contracts and cools due to thermal neutrino losses, while the envelope 
expands and develops a deep convective region that penetrates into the thin nuclear burning shell. Using the standard MLT formalism, during this stage the envelope becomes dynamically unstable, resulting in numerical convergence issues. This is a known 
problem in radiation-dominated envelopes of massive stars where convection is not strong enough to carry away  energy. With the 
\texttt{MLT++} that we employed in our calculations  (see Sect.~\ref{sec:uncertainties}),  the envelope is allowed to expand to $R \simeq 
930$\,R$_\odot$, as can be seen in the bottom panel of Fig.~\ref{fig:lum_teff_mdot_rad}, and Fig.~\ref{fig:mass_radius} where
we display the maximum radius of our \seriesone models as a function of the initial helium star mass. The star reaches high luminosities ($L \simeq 10^{6.25}$\,L$_\odot$) 
resulting in a very strong wind of $\dot{M} \simeq 
10^{-3.7}$\,M$_\odot$\,yr$^{-1}$ that removes the 
helium envelope on a timescale of $\simeq 3$\,kyr. 
At the same time, a large amount of metals is produced in the 
envelope as a result of helium-rich material being processed in a very hot burning shell. 
This is yet another unique feature of models in this mass range which, 
for the final $\sim 10^{4}$\,yr of their evolution, possess an 
extended atmosphere composed entirely of metals, as can be seen 
in Figs.~\ref{fig:lum_teff_mdot_rad} and \ref{fig:surface_abun_evol}.

While the evolution and exact composition of the envelope are somewhat 
uncertain due to the dependence on the energy-transport mechanism and 
the limited nuclear network that we employed,  the estimates for the wind mass loss rate 
are consistent with  the theoretical upper limit for 
super-Eddington winds  \citep{Owocki:2004zz, Smith2006}. The 
aforementioned behavior, also appears to be indifferent 
to variations in wind efficiency (see Sect.~\ref{sec:physical_uncertainties}).

The  mass of the star following this phase is $1.39\msun$, 
while the mass of the degenerate core has reached  $\sim 1.26\msun$, due to shell accretion 
at an average rate of $\rm \dot{M}_{\rm acc} \simeq 4.8 \times 10^{-6} \msun yr^{-1}$. 
Subsequently, the core continues to grow in mass at a rate 
of $\rm \dot{M}_{\rm acc} \simeq 2.2 \times 10^{-5} \msun yr^{-1}$ 
until it reaches a  mass  of $\simeq 1.37\msun$ (see Fig.~\ref{fig:lum_teff_mdot_rad}). 
The late evolution of the core is represented by the blue curve in Fig.~\ref{fig:RhoT}. 
For the last 5\,kyr, compressional heating is balanced by thermal 
neutrino emission, until the onset of Urca 
reactions. The ($\rm ^{25}Mg$,$\rm ^{25}Na$), and ($\rm ^{23}Na$,$\rm ^{23}Ne$) Urca pairs are of utmost importance, at 
densities $\rm \logrhoc \approx 9.1$, and 
$\rm \logrhoc \approx 9.2$, respectively. 
Both aforementioned reactions cool the core, effectively delaying the onset of runaway thermonuclear burning. 
Following the Urca cooling phase, there is no significant neutrino emission and the 
core evolves along the adiabatic curve 
between $9.2 \lesssim \logrhoc \lesssim 9.6$. 
Further compression leads to exothermic $e$-captures on $\rm ^{24}Mg$ nuclei at $\rm \logrhoc \approx 9.6$. 
Following the onset of these reactions, the temperature rises sufficiently for residual carbon to ignite, as the star enters a brief carbon burning phase ($\sim$3\,yr),
as seen in the top right panel of Fig.~\ref{fig:Kipp}.
The central temperature at the time the core has reached its maximum compactness is $\rm \logrhoc \approx 8.82$. 
The energy yield from carbon burning is sufficient to raise the temperature further for 
explosive oxygen burning  to occur. As a result, a thermonuclear runaway ensues at 
$\logrhoc \simeq 9.77$, when the average electron-to-baryon ratio of the core is 
$Y_e = 0.496$, and  the total binding energy is $E_{\text{bin}} = - 5.76 \times 10^{50}\,\text{erg}$. 
This ignition density is much lower than what is expected for typical ECSNe; $\logrhoc \simeq 10$. 
Our simulation indicates that the energy released at this stage, causes the core to expand.
In turn, this means that $e$-captures on \iso{Ne}{20} are likely completely avoided.  
As discussed in \citetalias{antoniadis2020} and in Sect.~\ref{sec:explosion_properties}, 
the available nuclear energy in principle suffices to completely unbind the star which, 
by virtue of vigorous shell burning, contains no helium at the time of the explosion. 
Such a thermonuclear runaway would therefore be observable as a SN\,Ia subtype. 
Further details on the chemical makeup of this model, at various 
evolutionary points, can be found in Appendix~\ref{apx:composition}.

In total, only four ONe models from \seriesone evolved past the runaway oxygen burning phase  
(green fully hatched squares in Fig.~\ref{fig:parameterSpace}) before terminating due to 
numerical difficulties related to  high  temperatures ($T_{\rm c}\gtrsim 10^{9.4}$\,K).  
An additional seven models from \seriestwo (Section~\ref{sec:physical_uncertainties}) 
also initiated oxygen burning before stopping for the same reason ($M_{\rm He, i}=2.0-2.7\msun$). 
Furthermore, sixteen models terminated during the Urca cooling phase, 
when they already had a near-\mch\ core---within the $\numrange{1.35}{1.37}\msun$ range---and no helium in their envelopes (hatched squares in Fig.~\ref{fig:parameterSpace}).
All remaining \seriesone ONe models terminated during the envelope ejection phase (i.e., following carbon burning), due to numerical convergence issues, when their core mass was in 
the $\numrange{1.0}{1.36}\msun$ range (green, non-hatched regions). Consequently their final fate is not directly constrained by our simulations. We encountered similar numerical issues for the majority of models that we label as CONe WDs. We note that it might be the case that some of these CONe cores could still end up as ONe cores (given that the C-flame will not be quenched).

\begin{figure*}[t]
    \centering
    \begin{subfigure}{.5\textwidth}
        \centering
        \includegraphics[scale=0.4]{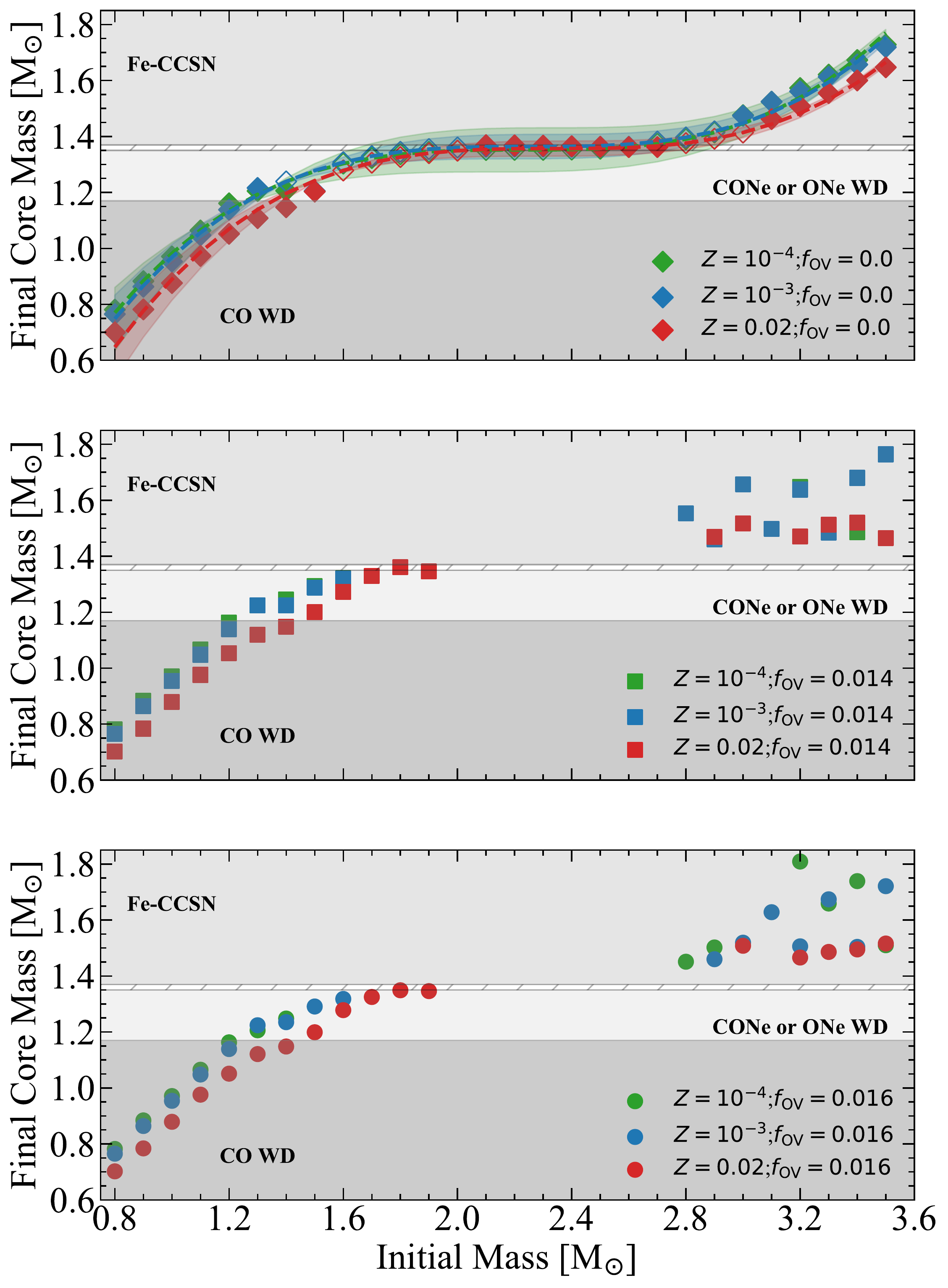}
    \end{subfigure}%
    \begin{subfigure}{.5\textwidth}
        \centering
        \includegraphics[scale=0.4]{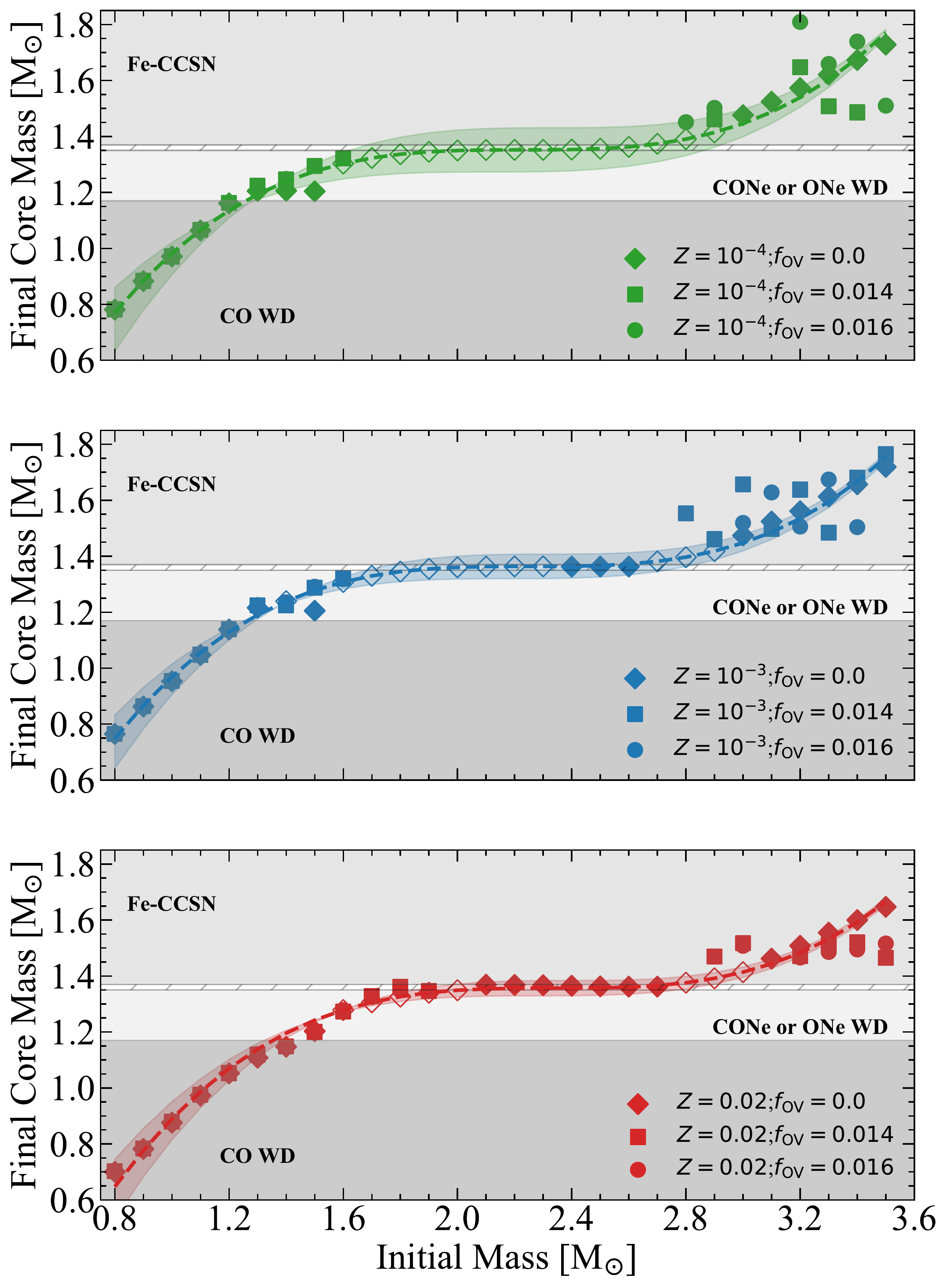}
    \end{subfigure}
    \caption{Final core masses for different initial metallicity (left), and overshoot factors (right).
    Models that experience thermal runaway develop a core between $M_c \simeq \numrange{1.35}{1.37}\msun$ 
    (thin, hatched region) due to helium shell burning. Missing data points correspond to models which retain an 
    envelope of more than $\sim 0.1\msun$ at the end of the simulation run, hence their final core mass could not be 
    accurately estimated. Nonfilled data points correspond to the interpolated core mass based on a fitting function
    (dashed curves) that was applied to models without overshooting and assumes a form: $a(x-b)^3 + c$. 
    These coefficients are $(a,b,c) = (0.201 \pm 0.007, 2.224 \pm 0.036, 1.352 \pm 0.026)$, $(0.201 \pm 0.006, 2.252 \pm 0.022, 1.364 \pm 0.015)$, and $(0.196 \pm 0.006, 2.336 \pm 0.018, 1.356 \pm 0.009)$ 
    for models with $Z = 10^{-4}$, $Z=10^{-3}$, and $Z=0.02$ respectively. Green, blue, and red colored regions represent $3\sigma$ confidence levels based on the standard deviation error estimates of the fitting parameters.}
    \label{fig:coreGrowth}
\end{figure*}

The lowest-mass model with an ONe core that reached \mch,  had   
$(M_{\rm He, i},Z,f_{\rm OV})=(1.9,0.02,0.014)$. Hence, it is 
reasonable to assume that all stars with $f_{\rm OV}>0$, $Z=0.02$ and  higher masses ($M_{\rm He, i}\geq 1.9\msun$)  
are also capable of developing a near-\mch\ core and initiating explosive oxygen burning.
This conclusion is further supported by the core growth rates of these models when the simulations stopped; \citep[$\mdot \simeq 10^{-5}\mdotsun$; see also][]{Poelarends:2017dua}. 
For $(Z,f_{\rm OV}) = (0.02,0.0)$, the least massive model that developed a \mch\ core had an initial mass of  $M_{\rm He,i}=2.1\msun$. 
As can be seen in Fig.~\ref{fig:parameterSpace}, for lower metallicities, terminal core masses and final outcomes are even more uncertain. 
However, based on the overall trend of the initial-to-final mass relation shown in Fig.~\ref{fig:coreGrowth}, we conclude that at least  models with  $M_{\rm He, i}\gtrsim 2.1\msun$ reach the Chandrasekhar-mass limit,  independently of metallicity and overshooting.
ONe cores that  do not manage to reach the Chandrasekhar-mass limit  become massive ONe WDs instead (we recall that our most massive CONe WDs have masses of $\sim 1.33\msun$). However, none of our models successfully settled on the WD cooling track. 

To further constrain the initial-to-final mass relation, we  attempted to interpolate between our models,  by fitting a power-law function of the form $a(x-b)^3 + c$, as can be seen in 
Fig.~\ref{fig:coreGrowth}. Here, we only used models that terminated after ejecting most of their envelope (residual envelope mass $M_{\rm env} < 0.1\msun$), as well as those that evolved toward a core-collapse SN. We found reasonable fits only for  $f_{\rm OV} = 0$ models, as those terminated at similar evolutionary stages. These fits similarly suggest that the Chandrasekhar-mass limit can be reached for $M_{\rm He, i}\gtrsim \numrange{2.0}{2.1}\msun$.

\subsection{Hybrid CONe cores with near-\mch\ masses}\label{sec:cone_core_evolution}
Two of our \seriesone models developed a hybrid 
structure in their core, but still managed to reach 
the Chandrasekhar-mass limit. Both stars had an 
initial mass of $1.8\msun$,  $Z = 0.02$, and 
$f_\text{OV} = 0.014, 0.016$. The explosive 
ignition of carbon and oxygen was initiated at even
lower densities  compared to the case of 
ONe models, due to the significantly larger amount
of residual carbon. The degenerate core for the model with $f_{\text{OV}} = 0.014$
is represented by the magenta track in Fig.~\ref{fig:RhoT}. The core composition is dominated by  carbon and oxygen, 
with $X(\iso{C}{12}) \approx 0.25$ and 
$X(\iso{O}{16}) \approx 0.46$ respectively (average values for the degenerate core). Since the carbon flame does not reach the center, the Urca pairs ($\rm ^{25}Mg$,$\rm ^{25}Na$), and ($\rm ^{23}Na$,$\rm ^{23}Ne$) do not play a major role in the temperature evolution. 
The lower panel of Fig.~\ref{fig:Kipp} shows the Kippenhahn diagram for 
the same model. In the absence of significant Urca cooling, once the 
compressional heating timescale is equal to that  of thermal neutrino 
emission, the density continues to rise until the onset of explosive 
carbon- and oxygen burning at a density of $\rm \log_{10} 
\left(\rho_c^{\rm ign} / \rm g \ cm^{-3} \right) \approx 9.26$. The 
substantially larger amount of carbon compared to the 
ONe cores discussed in the previous section, could in principle lead to more energetic 
explosions. The central temperature, when density reaches its maximum 
value, is $\rm \log(T_c / \rm K) \approx 8.58$.
As in the case of ONe cores, the helium envelope of those hybrid models 
is lost only after the core has reached a near-Chandrasekhar mass value. 
If the envelope is removed at an earlier stage, for instance as a result of mass transfer or a common envelope event, then a sub-M$_{\text{Ch}}$ 
hybrid CONe WD would have been formed, similarly to the models   
discussed in Sect.~\ref{sec:wd_evolution}.

\subsection{Pre-explosion properties of (C)ONe cores: \ias and ECSNe}\label{sec:explosion_properties} 
Based on the results presented  above, stars with a degenerate near-\mch\ core in the $\numrange{1.35}{1.37}\msun$ range may experience one of the following outcomes: firstly, for ONe cores, if oxygen ignites at low densities ($\logrhoc \lesssim 9.8$) and the core immediately expands as a response to the thermonuclear runaway, then $e$-captures on \iso{Ne}{20} may be completely avoided. In this case, the star may  undergo a thermonuclear runaway that resembles a \ia explosion. The kinetic energy of the ejecta and the composition of the material  depend on the explosion conditions, for instance $Y_{\rm e}$ and $E_{\rm bin}$.

Secondly, hybrid near-\mch\  cores ignite oxygen at even lower densities ($\logrhoc \lesssim 9.3$). These objects do not experience $e$-captures and may yield \ia-like explosions with higher kinetic energies compared to ONe cores. 

Lastly, if an ONe core ignites oxygen at high densities ($\logrhoc \gtrsim 10$) then $e$-captures on \iso{Ne}{20} lead to rapid deleptonization, and the star  evolves toward an ECSN. Depending on the explosion conditions, the latter can be either a tECSN, in which part of the star is disrupted, or a ccECSN that forms a neutron star.

In the remainder of this section, we describe the pre-explosion properties of models that successfully initiated explosive oxygen burning. A more in-depth discussion on the possible final fates of near-\mch\ models that terminated before reaching this stage follows in Sect.~\ref{sec:physical_uncertainties}. 

\begin{figure}
    \centering
    \includegraphics[width=\columnwidth]{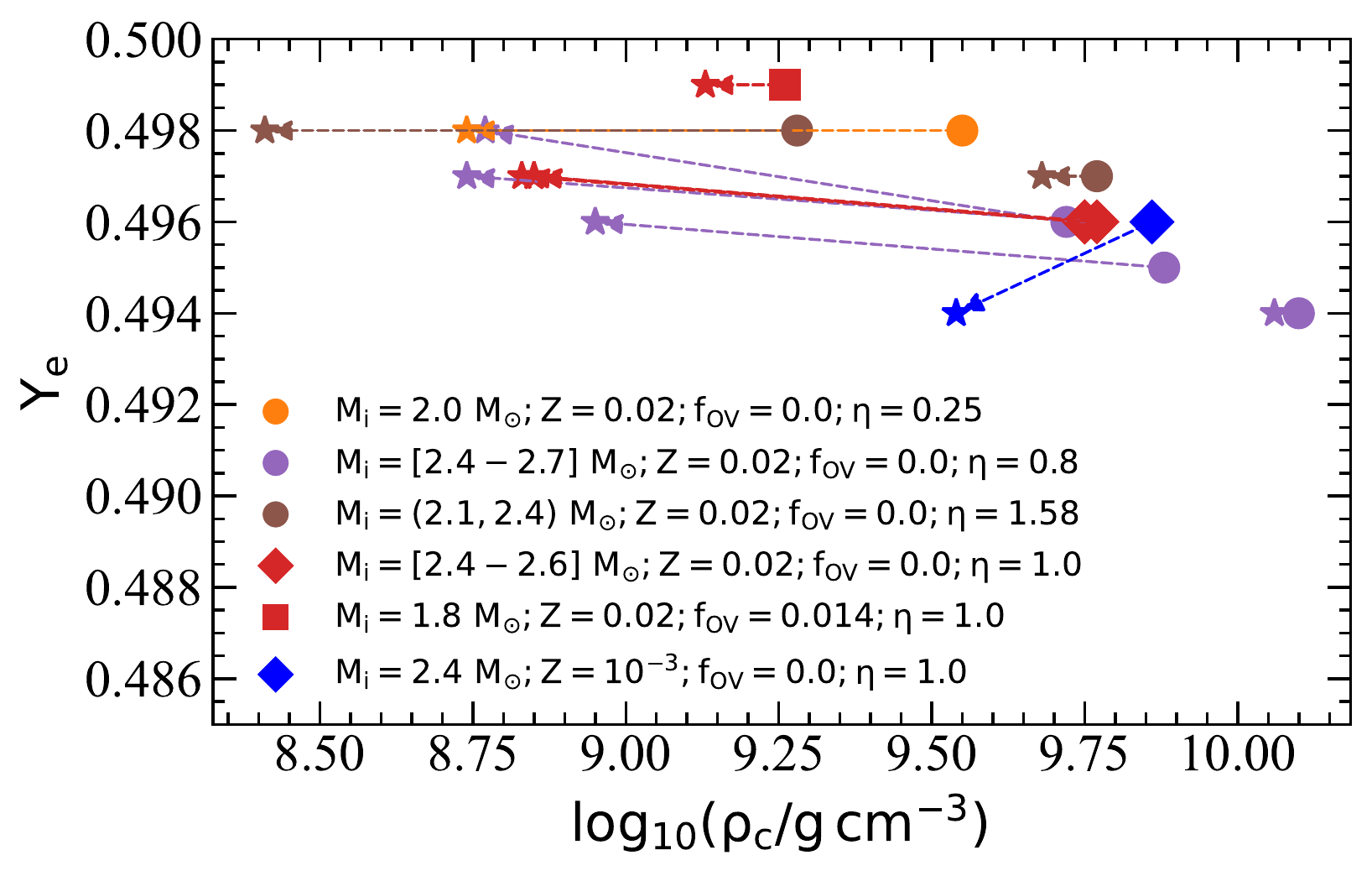}
    \includegraphics[width=\columnwidth]{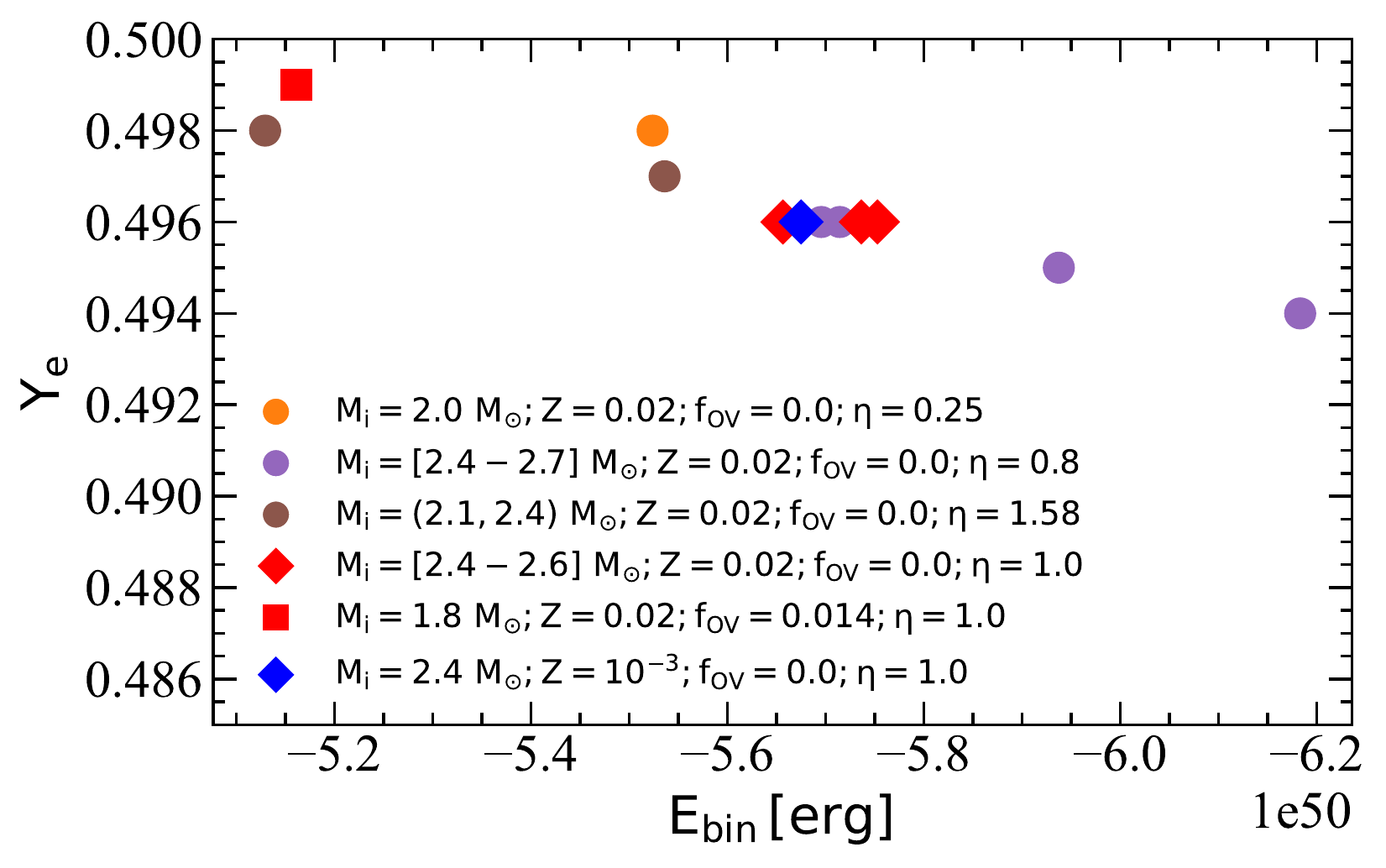}
    \caption{Electron-to-baryon ratio of the core at maximum compactness as a function of central density (top) and binding energy (bottom). Star-marks indicate the electron-to-baryon ratio at the end of the simulation.}
    \label{fig:composition_series2}
\end{figure}

Figure~\ref{fig:composition_series2} summarizes  the relation between $Y_{\rm e}$, central density, and binding energy at the onset of oxygen ignition for the twelve models that evolved past this phase. 
As can be seen, with few exceptions, the binding energy of these models lies within $-5.8 \leq E_{\text{bin}} / 10^{50}\,\text{erg} \leq -5.4$ and the average electron-to-baryon ratio is $0.494 \leq Y_e \leq 0.499$. 
Our simulations indicate that these stars respond to  oxygen burning by expanding. As discussed in \citetalias{antoniadis2020}, if one assumes that the thermonuclear runaway produces a typical SN\,Ia 
composition  ($\sim 0.7\msun$ of iron-peak and $\sim 0.8\msun$ of Si-group elements), 
then the nuclear energy yield suffices to completely unbind the star and produce ejecta with kinetic energies  
$0.77 \leq E_{\text{ej}} / 10^{51}\,\text{erg} \leq 1.26$. 
Consequently, these thermonuclear runaways would belong to the first category (ONe SNIa explosions). 
The exact \iso{Ni}{56} yield depends on  when the deflagration transitions to a detonation, 
and cannot be directly predicted by our simulations.  
If nuclear statistical equilibrium is achieved when the cores are still around their maximum compactness, 
then based on the core compositions shown in Fig.~\ref{fig:composition_series2}, the resulting explosions would synthesize mostly stable iron-group 
elements and only $\sim \numrange{0.1}{0.3}\msun$ of radioactive iron-group elements. 
Such low nickel yields would translate to underluminous explosions with light curves resembling those of  
SNe\,Iax \citep[e.g.,][]{2017hsnJ}. 
On the other hand, if the cores expand before a detonation occurs, 
then up to $1\msun$ of nickel may be synthesized (see \citetalias{antoniadis2020}).

As can be seen, two models ignited oxygen when $\logrhoc\simeq 9.3$. The model represented by a red square corresponds to the CONe cores described in Sect.~\ref{sec:cone_core_evolution} and produces an explosion that belongs to the second category (CONe SNe\,Ia). The other model (depicted with brown color) is a \seriestwo ONe core with initial mass of $2.1$\msun\ and an enhanced wind efficiency of $\eta = 1.6$ (see Sect.~\ref{sec:eta_1p58_evolution}). The explosion would belong to the first category discussed above, producing ejecta with kinetic energy of $\sim  10^{51}\, \text{erg}$. 
Finally, one \seriestwo model with reduced wind efficiency, ignited oxygen only after surpassing the density threshold for 
$e$-captures on \iso{Ne}{20} and will therefore produce an ECSN. As we discuss in Sect.~\ref{sec:physical_uncertainties}, we identify the decisive property for this behavior to be its low residual carbon abundance. Hence, we expect the same outcome  for all stars with  similar compositions.

\subsection{Core-collapse supernovae}\label{sec:ccsn_evolution}
If $ M_i  \gtrsim 2.8\msun$ (for models without overshooting), then the core reaches a 
sufficiently high temperature to undergo on-center carbon ignition, under nondegenerate conditions, leading to more advanced burning stages. The evolution of these stars is similar to those described in \cite{Jones2013apj} and \cite{Woosley:2019sdf}.
Models that fall within this category, develop a silicon core and evolve toward an iron core-collapse that  ultimately forms a neutron 
star. These models retain a \iso{He}{4}-rich envelope, so assuming that a successful explosion follows the collapse of the core, it would 
probably be observable  as a SN\,Ib. Type Ic SNe are also expected to 
form from stripped-envelope stars, but not in the mass range of our 
models, unless additional mass loss takes place, either via winds,  
pulsations \citep{Aguilera-Dena:2021abc}, or case BB mass transfer, 
which may result in the formation of ultra-stripped SNe  \citep{Tauris:2015xra}.

\section{Physical uncertainties} \label{sec:physical_uncertainties}

\subsection{Wind and mass loss uncertainties}\label{sec:series2_grid}
There are two phases where mass loss through stellar winds  affects the outcome of our model computations. Firstly, 
because of its long duration, mass loss during the core helium burning 
phase may influence both the total mass and the size of the degenerate core. In our  \seriesone models the total mass removed in this phase is $\mathcal{O}(0.1\msun)$. Secondly, our models reach very high 
luminosities toward the end of their evolution, when they expand and 
their atmospheres become cooler (up to $\log_{10} \left(L/{\rm L}_\odot \right) \simeq 6.25$ 
for the example in Fig.~\ref{fig:lum_teff_mdot_rad}). In this short phase, the 
remaining \iso{He}{4}-envelope is likely removed. The competition between envelope removal and core growth is critical for the final fate of the
star: if the envelope is lost before the core reaches the Chandrasekhar-mass limit, then the star  becomes a WD. Contrarily, if enough of the envelope survives, helium might become observable in the resulting explosion.  

In our models, we adopted the \texttt{Dutch} \mesa wind scheme which  extrapolates empirical 
mass loss relations for more massive Wolf-Rayet (WR) stars from \cite{Nugis2000} for $\rm 
T_{\text{eff}} > 10^4$\,K, and uses the prescription of \cite{deJager1988} for cool stars. These 
mass loss rates are  uncertain and may  overestimate the total mass lost during the post main sequence evolution \citep{Beasor2021arx}. Empirically mass loss rates for low mass 
helium stars are also still very uncertain, mainly because they are only formed in binary systems where they are difficult to observe \citep{Smith2017,Zapartas2017}. Recent
theoretical work suggests that  mass loss rates in this regime could lie significantly below the relations for massive Wolf-Rayet stars \citep{Graefener2017,Vink:2017ujd}. 

In late phases, when the degenerate core is near its maximum mass, our models reach extremely high 
super-Eddington luminosities of up to $\log_{10}{L/{\rm L}_\odot = 6.25}$.
This results in a strong stellar wind that eventually removes the \iso{He}{4}-rich envelope. While the mass loss rates of such stars are also extremely uncertain, it is notable that the maximum wind strength achieved in our models lies  near the theoretically expected maximum values for super-Eddington winds \citep[][]{Owocki:2004zz,Smith2006}. 

\begin{figure}
    \centering
    \includegraphics[width=\columnwidth]{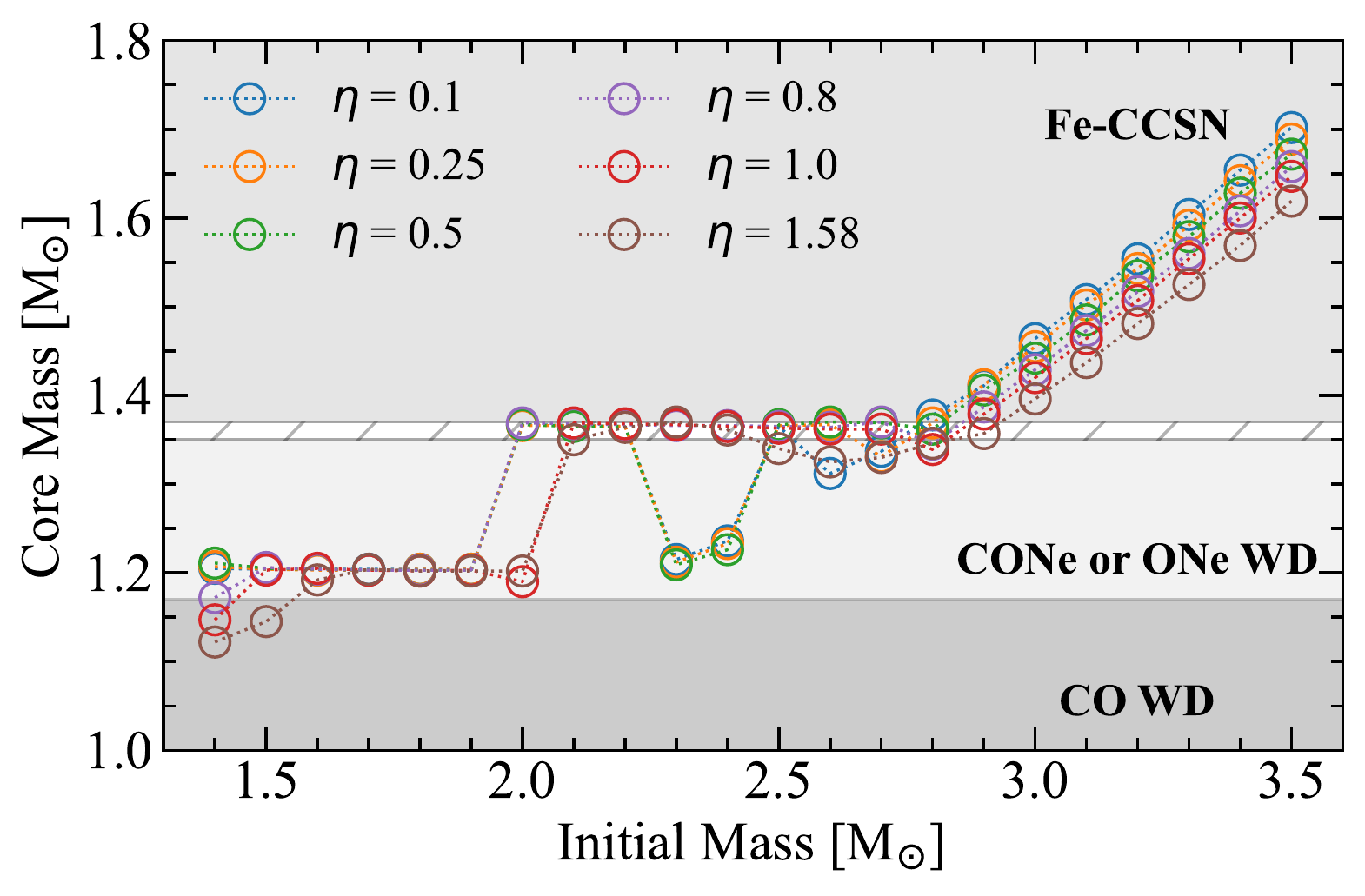}
    \includegraphics[width=\columnwidth]{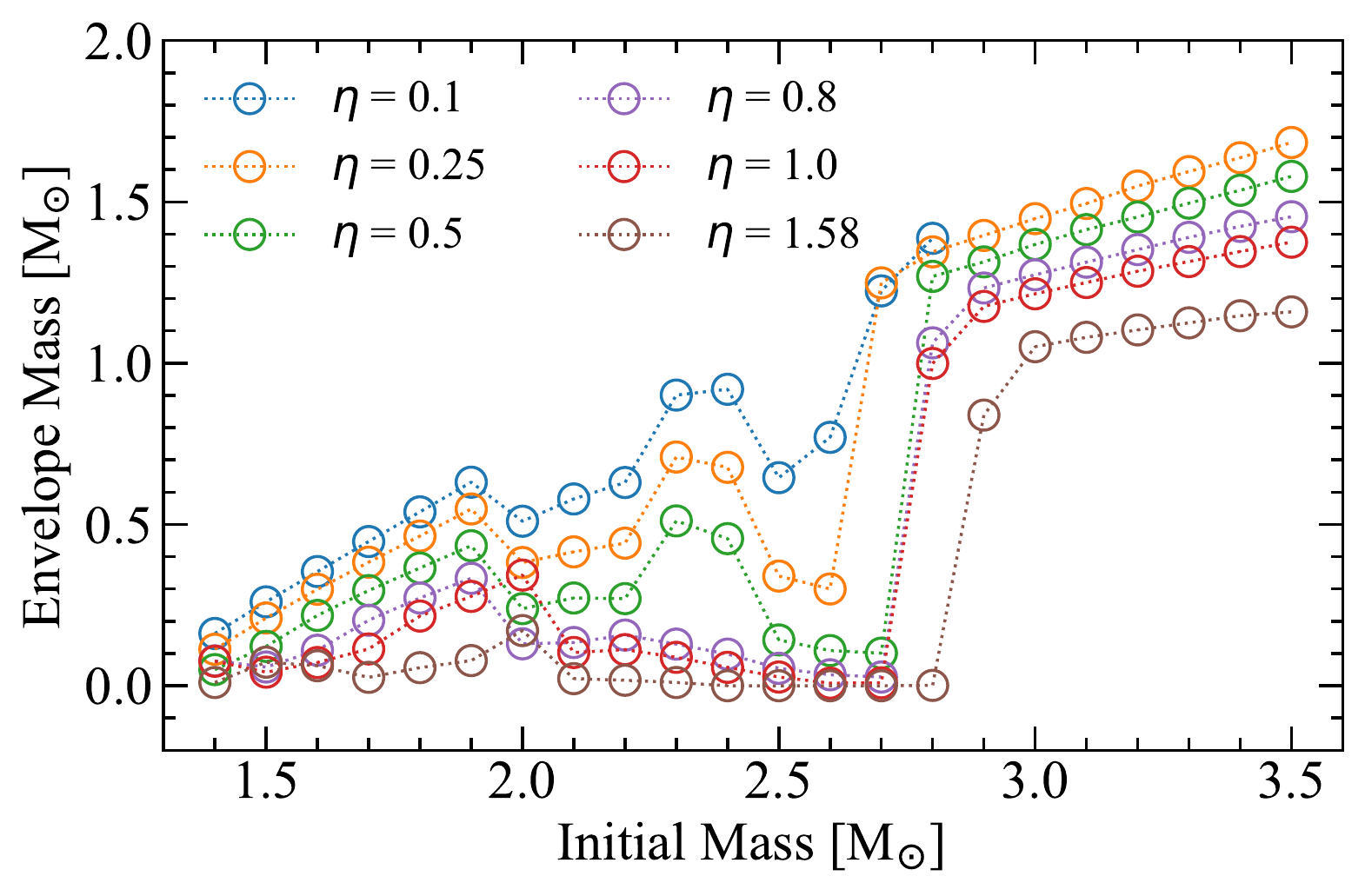}
    \caption{Core mass (top) and envelope mass (bottom) for various values of the wind efficiency parameter ($\eta$) within the ``Dutch'' scheme. These masses represent the value at the end of the simulation. Although the latter does not correspond to the same evolutionary stage for all models, the general trend is evident.
    All models assume solar metallicity and no overshoot mixing.}
    \label{fig:etas}
\end{figure}

To further explore the influence of mass loss uncertainties, we computed an additional grid consisting of 110,  $(Z, f_{\rm OV})=(0.02,0.0)$ models, with varying wind efficiency parameter ($\eta$). 
Figure~\ref{fig:etas} shows the effects of wind on core growth (top panel), and on the retention efficiency of envelope mass (bottom panel). Notwithstanding that most models in the  $\numrange{1.8}{2.7}\msun$ mass range terminated during the Urca reactions stage, we find that the ability of the core to reach a near-Chandrasekhar mass is not hindered by the adopted wind efficiency parameter. Contrarily, the mass of the envelope retained by the star is sensitive to the  $\eta$-parameter. The chemical composition of the envelope is similar to our \seriesone models, where the original helium-rich envelope has been replaced by a metal-rich atmosphere due to intense helium shell burning (see Appendix~\ref{apx:composition}). As all models within the $\eta = 0.1, 0.5$ subsets failed to converge after the $e$-captures on \iso{Mg}{24}, in the remainder of this section we describe some representative evolutionary models from the remaining subsets, namely $\eta = 0.25, 0.8, 1.6$.

\subsubsection{$\eta = 0.25$}\label{sec:eta_0p25_evolution}
In this section, we present the evolution of a \seriestwo model with initial mass of $2.0\msun$, since it is the lowest initial mass model with $\eta < 1.0$ able to achieve oxygen ignition, and simultaneously holds the most massive envelope compared to similar models within this grid.

The off-center ignition of carbon took place at a mass coordinate of $\sim 0.5\msun$ when the total mass of the star was $\sim 1.95\msun$ and the mass of the core was $\sim 1.01\msun$. The main carbon burning phase lasted $\sim 56\,\text{kyr}$, resulting in an ONe core of $M_{\rm c} \approx 1.37\msun$, composed primarily of oxygen and neon with central abundances of $X(\iso{O}{16}) \approx 0.27$ and $X(\iso{Ne}{20}) \approx 0.32$ respectively. The degenerate core also retained some residual carbon mass fraction of $X(\iso{C}{12}) \approx 0.02$. During this time the star grew to $R \simeq 158$\,R$_\odot$ and had an effective temperature of $\log_{10} (T_{\text{eff}} / \text{K}) \simeq 3.78$ while it was losing mass at a rate $\dot{M} \simeq 5.5 \times 10^{-8}\msun \text{yr}^{-1}$. At the end of the simulation it possessed a helium-free, metal-rich envelope of $M_{\text{env}} \simeq 0.38\msun$  (see Fig.~\ref{apx:fig:eta0p25}).

\begin{figure*}
    \centering
    \includegraphics[width=\textwidth]{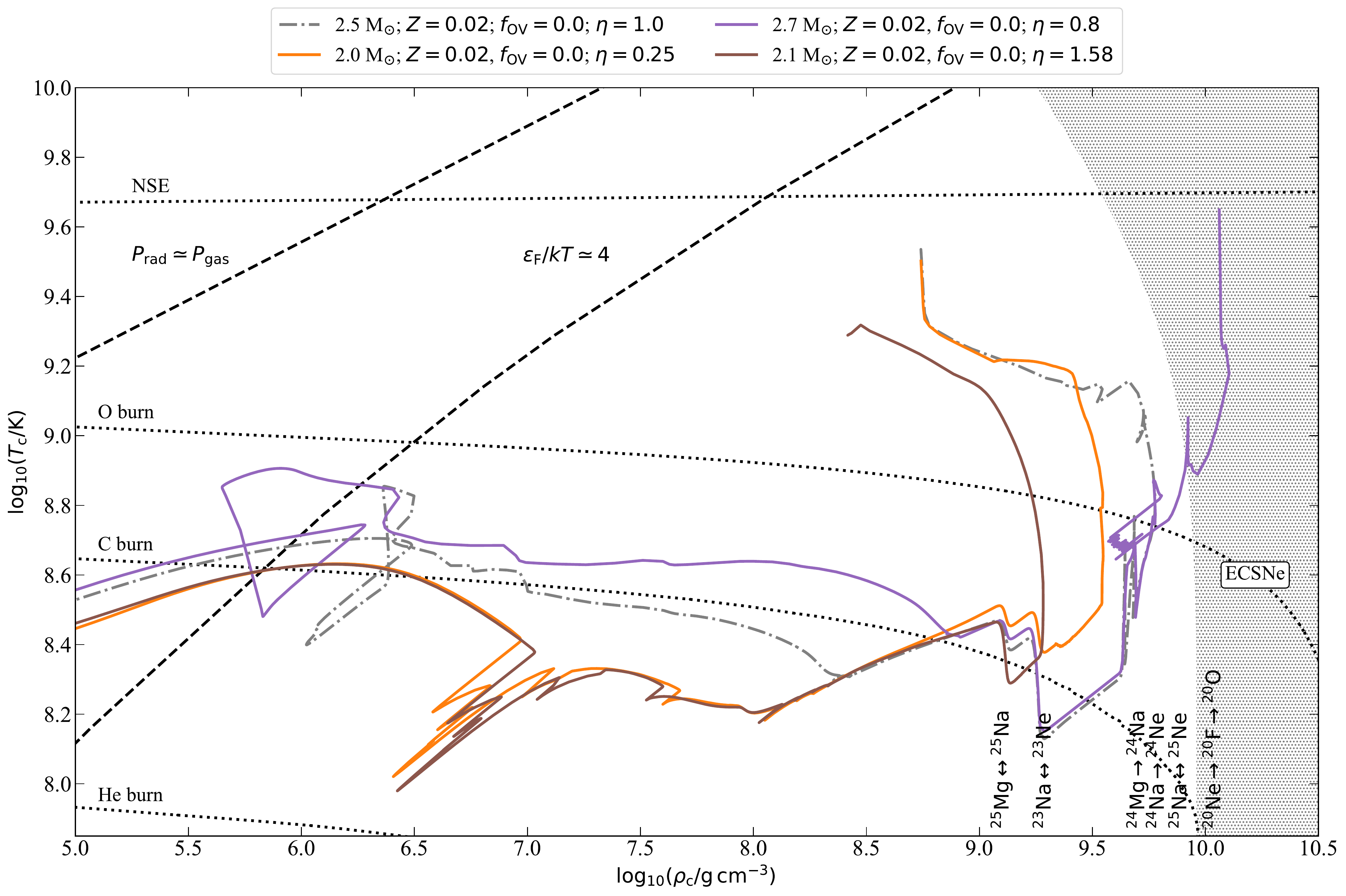}
    \caption{Evolution of central temperature vs central density for three \seriestwo models with different wind efficiencies. Black-dotted lines show approximate ignition thresholds taken from \mesa. Black dashed lines indicate different pressure regimes while the hatched area shows the approximate region for $e$-captures on $\rm ^{20}Ne$ nuclei. An inefficient Urca cooling phase can leave the core with an excess amount of heat igniting the residual carbon before the $e$-captures on \iso{Mg}{24} (orange and brown curves). The grey dash-dotted line shows the---efficient---cooling and subsequent adiabatic evolution of our \seriesone fiducial model, which leads to a thermal runaway caused by the $e$-captures on \iso{Mg}{24}. The model depicted here with purple color ignites oxygen at high densities leading, most likely, to an ECSN.}
    \label{fig:rhot_series2}
\end{figure*}

\begin{figure*}
    \centering
    \begin{subfigure}{.5\textwidth}
        \centering
        \includegraphics[width=\columnwidth]{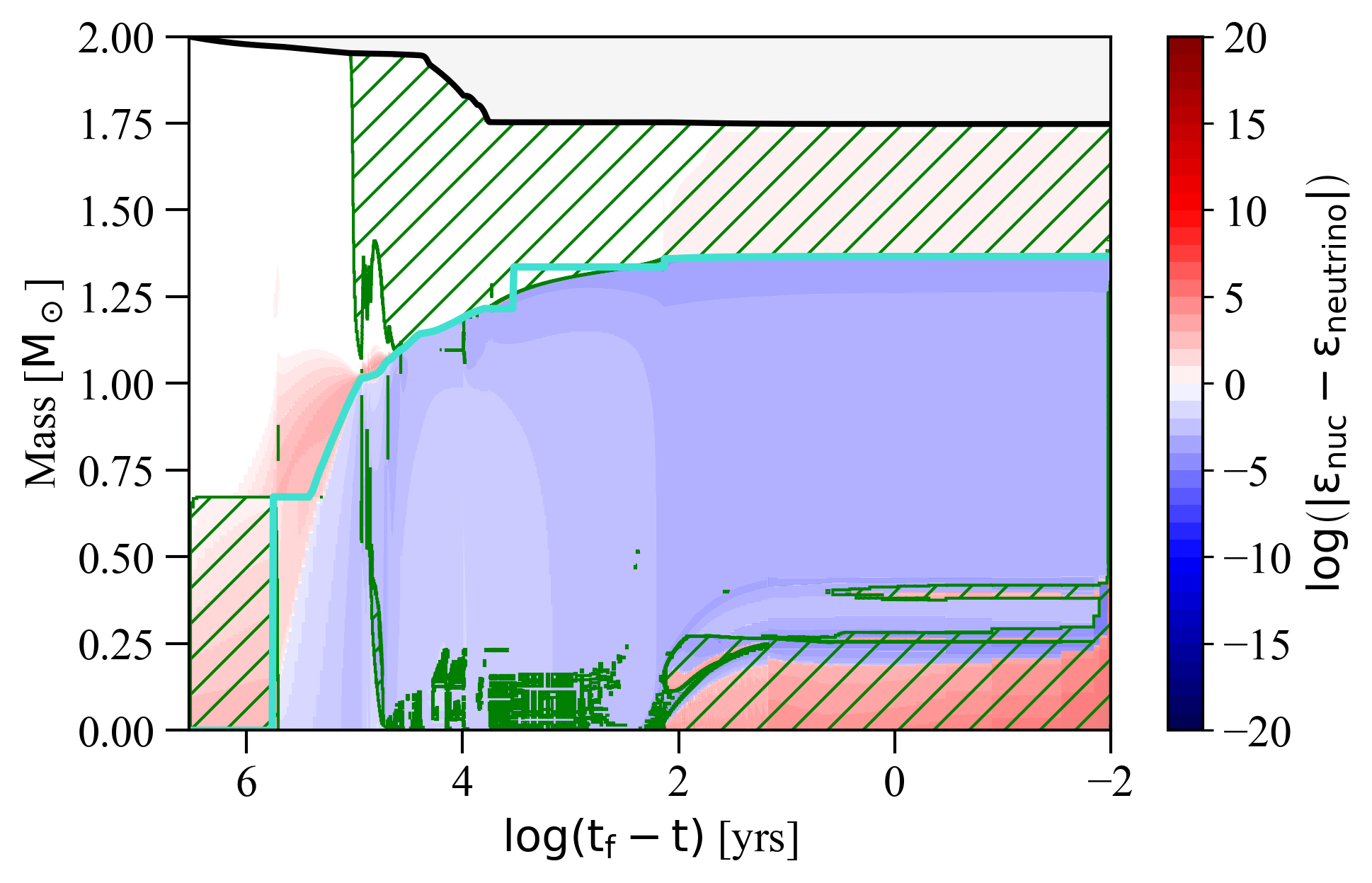}
    \end{subfigure}\hfill
    \begin{subfigure}{.5\textwidth}
        \centering
        \includegraphics[width=\columnwidth]{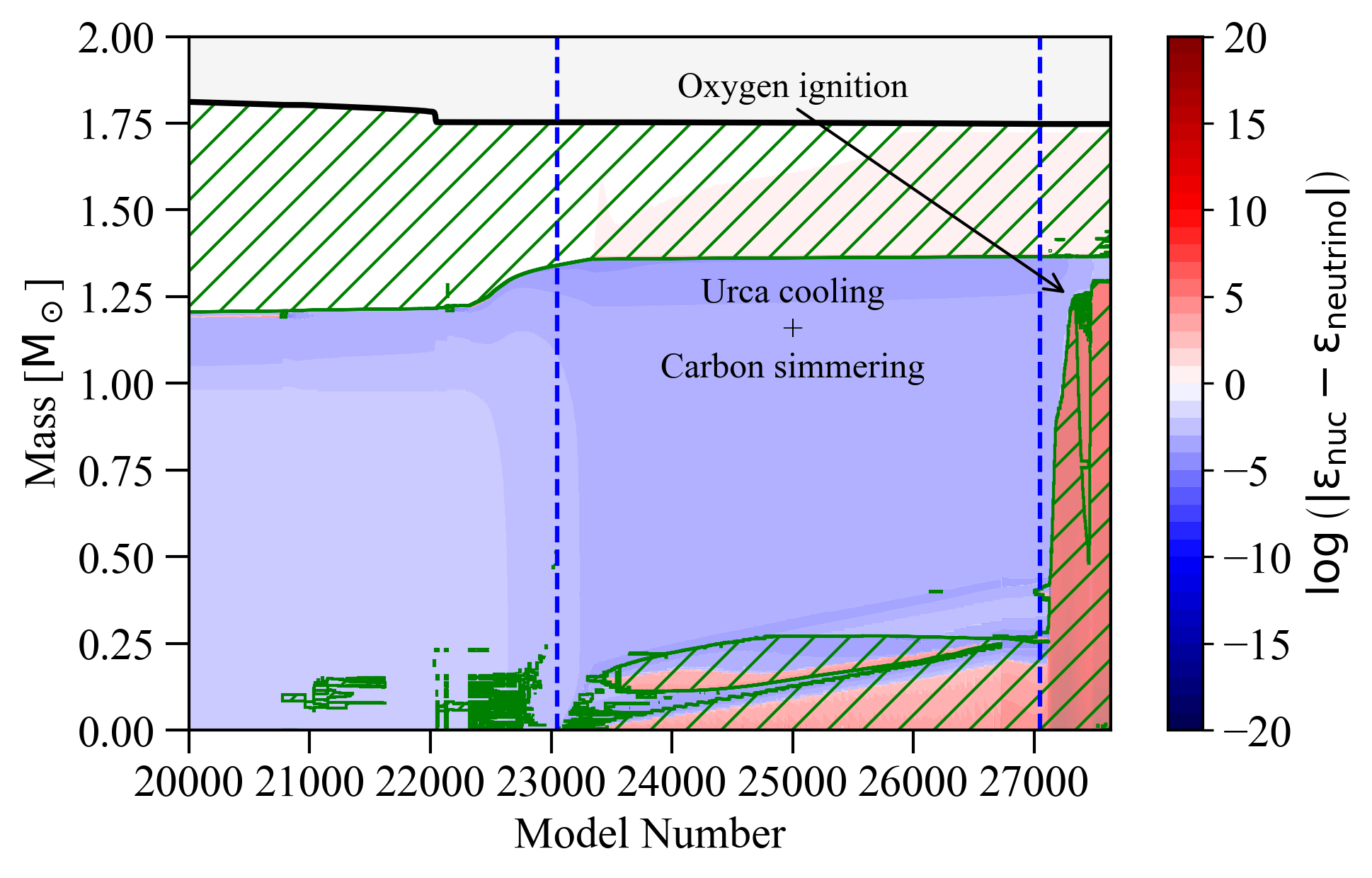}
    \end{subfigure}\hfill
    
    \begin{subfigure}{.5\textwidth}
        \centering
        \includegraphics[width=\columnwidth]{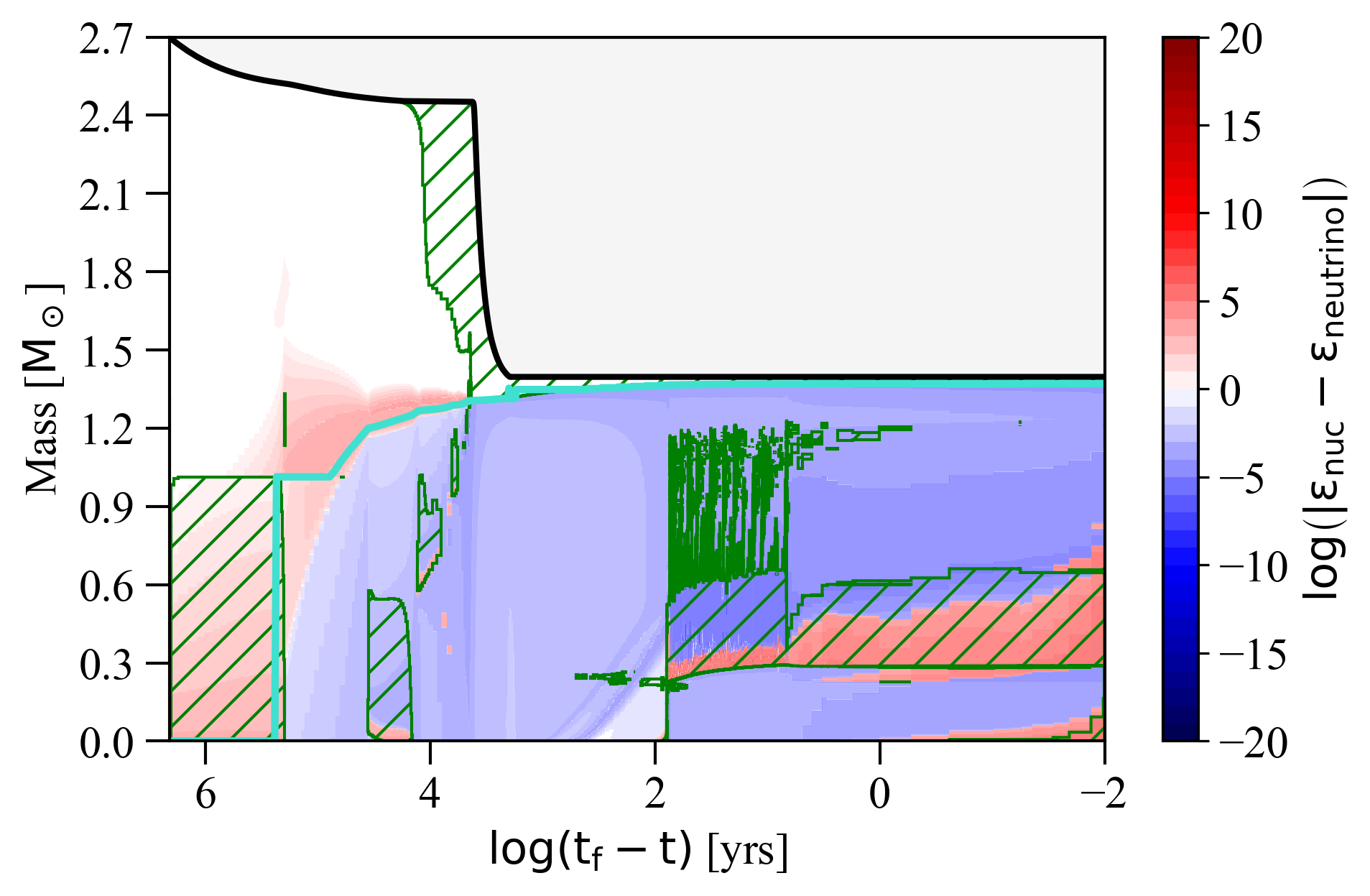}
    \end{subfigure}\hfill
    \begin{subfigure}{.5\textwidth}
        \centering
        \includegraphics[width=\columnwidth]{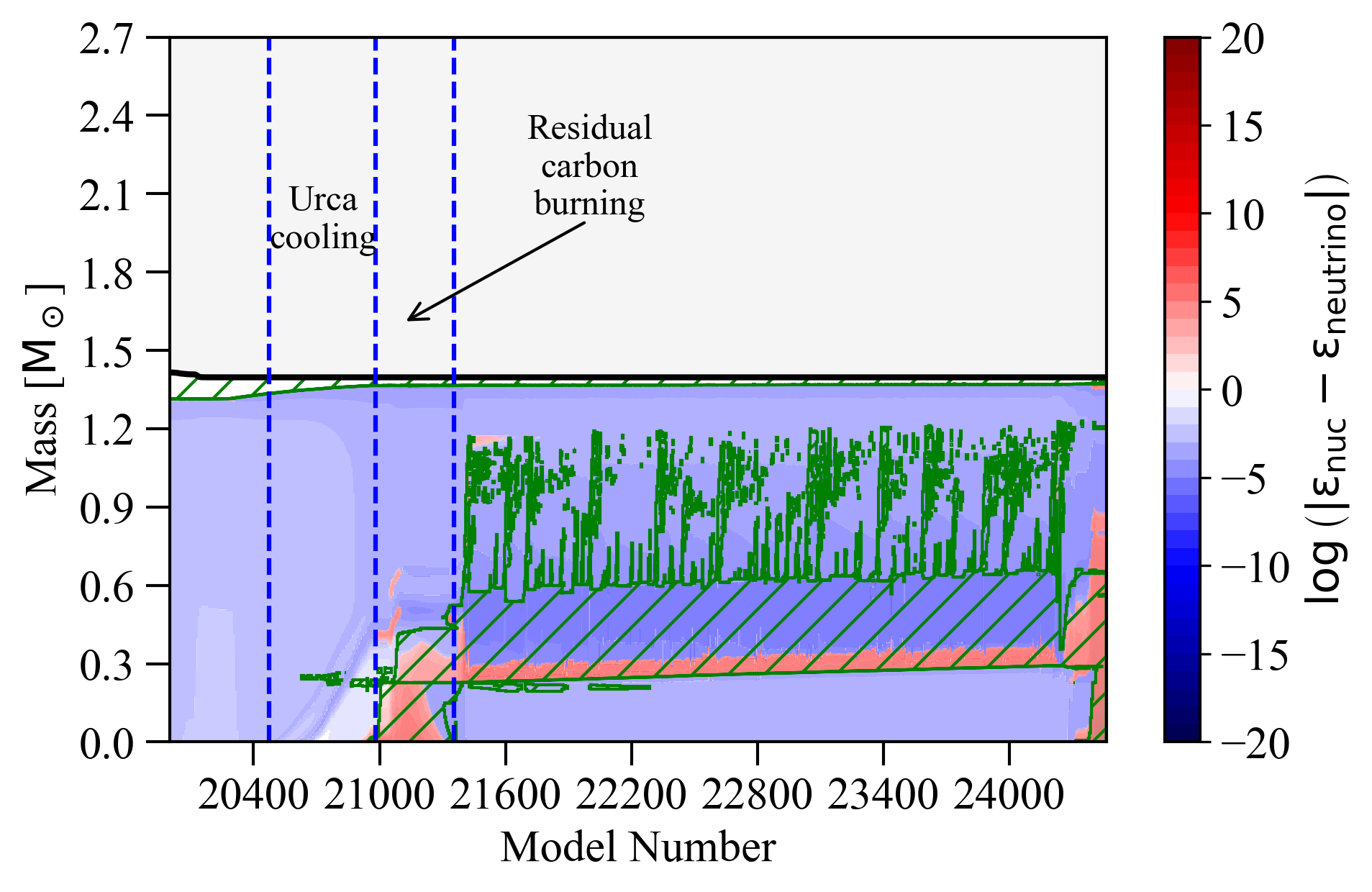}
    \end{subfigure}
    
     \begin{subfigure}{.5\textwidth}
        \centering
        \includegraphics[width=\columnwidth]{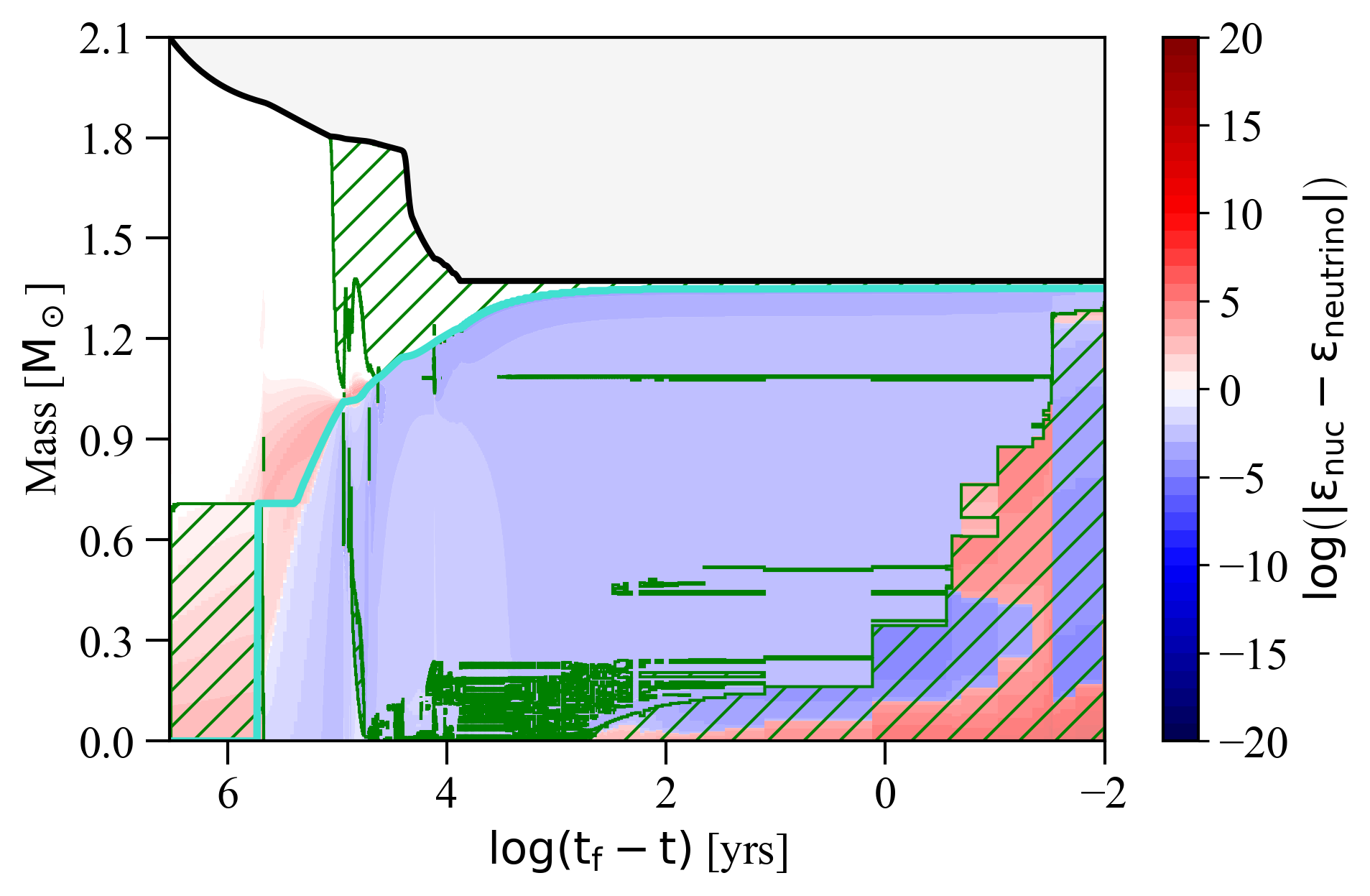}
    \end{subfigure}\hfill
    \begin{subfigure}{.5\textwidth}
        \centering
        \includegraphics[width=\columnwidth]{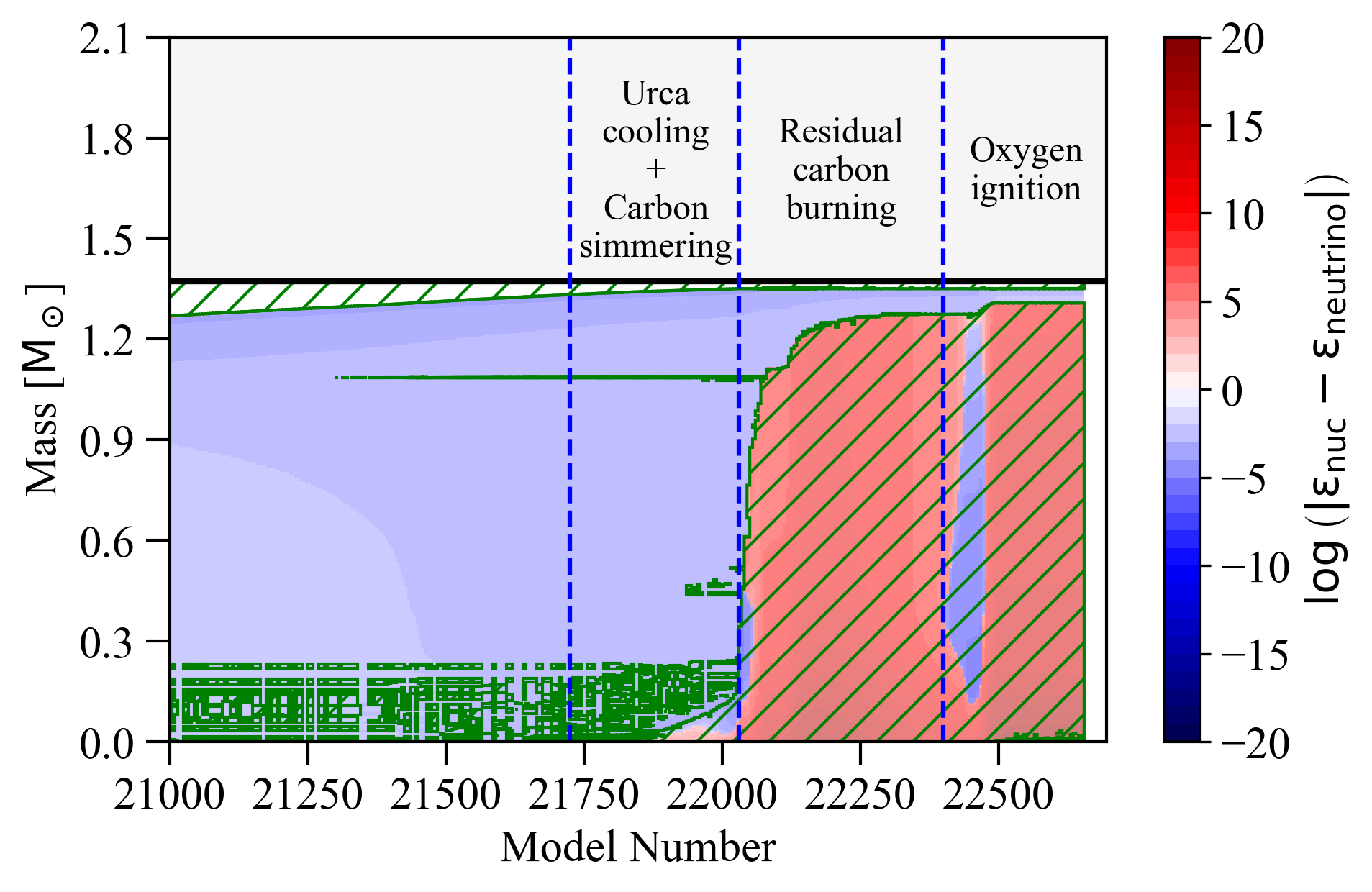}
    \end{subfigure}
    
    \caption{On the left, we present a Kippenhahn diagram of three \seriestwo stellar models as a function of remaining evolution time. On the right, we present a section of the same diagrams but as a function of model number in which the location of some important stages can be seen more accurately. Green hatched areas denote regions of convection. The intensity shown in the colorbar, indicates the net energy rate. Turquoise line shows the CO core. \textbf{Top panel}: Structure of a $2.0 \rm M_{\odot}$ helium star with $Z = 0.02; \eta = 0.25$ and no overshooting. \textbf{Middle panel}: Structure of a $2.7 \rm M_{\odot}$ helium star with $Z = 0.02; \eta = 0.8$ and no overshooting. This model evolves into the ECSN regime. Notice that not all of the core is convective anymore due to intense neutrino emission. \textbf{Bottom panel}: Structure of a $2.1\msun$ helium star with $Z = 0.02; \eta = 1.6$ and no overshooting. The ignition of oxygen occurs during the final few hours  and cannot be discerned on the left plot.}
    \label{fig:Kipp_series2}
\end{figure*}

Figure~\ref{fig:rhot_series2} illustrates the evolution of the core in 
terms of central density and central temperature. By comparing the 
adiabatic tracks of this models (orange line) with the \seriesone 
fiducial model (grey dash-dotted line) one sees that the core is 
significantly hotter after the end of the Urca process. The 
inefficiency of Urca cooling is most likely related to the smaller amount of \iso{Na}{23} distributed in the core, compared to the \seriesone fiducial model $(X(\iso{Na}{24}) \simeq 0.035$ versus 0.057,  respectively).
As the core compresses adiabatically, the star enters a short carbon simmering phase. 
The structure of the star during this period is depicted in the top 
right panel of Fig.~\ref{fig:Kipp_series2}. After $146\,\text{yr}$ 
the temperature has increased up to $\log_{10} (T_c / \text{K}) \approx 8.66$ causing  residual carbon to ignite before the density reaches the threshold value for $e$-captures on \iso{Mg}{24} nuclei. This results in an even lower ignition density for the deflagration front at $\log_{10} (\rho_c / \text{g cm}^{-3}) \approx 9.55$ which, most likely,  affects the transition time from deflagration to detonation and consequently the whole explosion dynamics. At the onset of oxygen ignition, the model has a binding energy of $E_{\text{bin}} = - 5.5 \times 10^{50}\,\text{erg}$ with an average electron-to-baryon ratio of the core $Y_e = 0.498$. Hence, this star is also prone to complete disruption. 

As we discuss above, mass loss rates for Wolf-Rayet stars currently employed by many stellar 
evolution codes probably overestimate the strength of the wind in low mass helium stars 
\citep[e.g.,][]{Vink:2017ujd, Sanders:2020:mnras}. The formation of thermonuclear supernovae 
progenitors through our evolutionary channel in models with such a small wind efficiency 
demonstrates the robustness of our conclusions to assumptions in the wind.

\subsubsection{$\eta = 0.8$}\label{sec:eta_0p8_evolution}

For the subset of models with $\eta=0.8$, only four in the mass range $\numrange{2.4}{2.7}\msun$, reached the residual carbon ignition stage. All of these models develop a degenerate ONe core of $\sim 1.37\msun$ with an average residual carbon abundance of $0.002 \leq X(\iso{C}{12}) \leq 0.008$. Three of them ignited oxygen at low densities, with the subsequent evolution being  almost identical to their \seriesone counterparts. 

One stellar model however, ignited oxygen at higher densities, within the ECSN regime (purple line in Fig.~\ref{fig:rhot_series2}). This model had an initial mass of $2.7\msun$ and retained only $X(\iso{C}{12}) \approx 0.002$ of carbon mass fraction. Shortly after the $e$-captures on \iso{Mg}{24} nuclei, the residual carbon ignited  when $\log_{10} (\rho_c / \text{g cm}^{-3}) \approx 9.69$. After the depletion of carbon, the core expanded, and the burning region moved into a thin shell. This is illustrated in the middle-right panel of Fig.~\ref{fig:Kipp_series2}.
After $\sim 66\,\text{yr}$ the core contracted  again, reaching the threshold density for $e$-captures on \iso{Ne}{20} ($\log_{10} (\rho_c / \text{g cm}^{-3}) \approx 9.92$). The thermal runaway occurred when $\logrhoc \approx 10.1$ and the average electron-to-baryon ratio of the degenerate core was  $Y_e = 0.495$. The core at this point was primarily composed of oxygen and neon with $X(\iso{O}{16}) \approx 0.40$ and $X(\iso{Ne}{20}) \approx 0.42$, respectively.

\subsubsection{$\eta = 1.6$} \label{sec:eta_1p58_evolution}
We also investigated the evolution of models with an enhanced wind efficiency. We were able to follow the evolution of only two models, after oxygen ignition. The first stellar model had an initial mass of $2.1\msun$ and ignited carbon off-center when the total mass is $M_\text{tot} \simeq 1.8\msun$. The characteristics of carbon burning (ignition location, flame propagation) and the evolution of the envelope are similar to the ones we describe in Sect.~\ref{sec:eta_0p25_evolution}. As a result of the enhanced wind efficiency, the star lost mass during this time at a rate of $\dot{M} \approx 3.9 \times 10^{-7}\msun \text{yr}^{-1}$ allowing the degenerate ONe core to grow to $\sim 1.35\msun$ with a central abundance of residual carbon circa $X(\iso{C}{12}) \approx 0.025$. 

After the onset of Urca cooling, the star entered a simmering phase that lasted $\sim 737\,\text{yr}$, until the core reached its maximum compactness when $\log_{10} (\rho_c / \text{g cm}^{-3}) \approx 9.28 $. When this happened, the temperature reached a value of $\log_{10} (T_c / \text{K}) \approx 8.51$, triggering residual carbon ignition. The next $\sim 12.5\,\text{yr}$ were  characterized by vigorous burning which led to a thermal runaway. At the time of oxygen ignition the core had a central abundance of  $X(\iso{O}{16}) \approx 0.35$ and $X(\iso{Ne}{20}) \approx 0.43$, and an average electron-to-baryon ratio of $Y_e = 0.498$.

The second stellar model had an initial mass of $2.4\msun$, an ONe core of $\sim 1.36\msun$, and a central abundance of leftover carbon of $X(\iso{C}{12}) \approx 0.01$. Its evolution followed the same steps already described above. The main difference was that after the depletion of residual carbon, the core expanded significantly and oxygen was ignited off-center. This resulted to  residual oxygen being distributed within a silicon core $(X(\iso{O}{16}) = 0.007)$, as seen in the bottom-right panel of Fig.~\ref{apx:fig:eta1p58}.
Although we could not follow the evolution past this stage, the model most likely evolves toward a core-collapse SN, similar to the ultra-stripped SN scenario described by \cite{Tauris:2015xra}.

\subsection{Residual carbon: (C)ONe \ias vs ECSNe}\label{sec:carbon}
In Sects.~\ref{sec:one_core_evolution} and \ref{sec:cone_core_evolution} we discussed how in a subset of our (C)ONe models, 
the ignition of residual  carbon leads to explosive oxygen burning. Besides its critical role in enabling thermonuclear 
explosions, the amount of residual carbon may also influence the explosion energetics and \iso{Ni}{56} 
yields \citep[see Sect.~\ref{sec:explosion_properties} and also][]{Willcox:2016yyp}. 
The carbon abundance in (C)ONe WDs also influences their ability to trigger \ias via mass accretion   
 \citep[e.g., see][]{Dominguez1993,  Garcia1997, Gutierrez2005, Waldman2007, Schwab:2018cnb}.

To further gauge the importance of residual carbon in determining the final fate of our models, 
we also computed a 2.5\msun\ model in which the rate of all 
carbon-participating nuclear reactions was set to zero, once the
central density became larger than $\rm \log_{10}(\rho_c / \rm g \ cm^{-3}) \simeq 9.0$ for
the first time (black cross-mark in Fig.~\ref{fig:RhoT}).  
The evolution of this model is represented by the blue dashed line in Fig.~\ref{fig:RhoT}. 
As can be seen, in the virtual absence of carbon, the evolution of 
the core is almost identical to our fiducial model (blue line) up to the point 
where $e$-captures on \iso{Mg}{24} occur. 
From there on, instead of igniting oxygen, the model evolves toward higher densities 
until it crosses the threshold for $e$-captures  
on \iso{Ne}{20}, thereby triggering an ECSN. 
A similar outcome was found for the \seriestwo model represented by the purple line in 
Fig.~\ref{fig:rhot_series2}, which had an average residual carbon mass fraction of $\rm X(\iso{C}{12}) = 0.002$. 
This behavior confirms that the final outcome (ONe \ia vs ECSN) of near-\mch\ cores 
depends critically on the residual carbon mass fraction.   
In Fig.~\ref{fig:carbon_mass_fractions}, one sees that the latter varies significantly, 
both with initial mass and metallicity. 
More specifically, one sees that lower-mass models generally retain more carbon. 
Variations are also seen as a function of metallicity, especially for $f_{\rm OV}>0$. For example, for $M_{\rm He, i}=2.4\msun$ and $f_{\rm OV}=0.016$, one sees that the $Z=10^{-3}$ 
model has more than 10 times less carbon than 
the $Z=0.02$ one.  
Finally, one also sees some small variations with wind efficiency, which imply that the 
core evolution is not completely decoupled from the envelope during carbon burning 
(lower panel of Fig.~\ref{fig:carbon_mass_fractions}). 

The  model with the lowest amount of residual carbon that was able to initiate an ONe \ia, 
had an initial mass of $M_{\rm He, i}=2.6\msun$ and a mass averaged carbon mass fraction of $X(\iso{C}{12})\simeq 0.004$ within the degenerate core. 
Hence for an ONe \ia to occur, the core must have a minimum carbon mass fraction of $0.002 < X_{\rm min}(\iso{C}{12}) < 0.004$ (see Appendix~\ref{apx:composition} for distribution profiles).  
Based on these constrains, we conclude (rather conservatively) that near-\mch  ONe cores with carbon mass fractions $X_{\rm min}(\iso{C}{12}) \geq 0.004$ experience an ONe \ia, while those with a smaller carbon content undergo an ECSN (thermonuclear or core-collapse).
Given that the carbon abundance in the core changes with initial metallicity, so does the relative rates of ONe \ias and ECSNe, with the former being more likely for higher metallicities.

\subsection{Binary interactions}

A key feature in the evolution of hydrogen-rich SAGB stars is the occurrence of the 
2DU episode, during which the mass of the helium core is being reduced. Moreover, the existence of two 
burning shells in the structure of these stars leads to periodic thermonuclear instabilities 
that reduce the accretion of material onto the core \citep{Denissenkov:2013qaa}. 
These two effects make it harder for the 
core to grow, leaving only a very narrow window, in the mass range, for which the cores of SAGB 
stars can finally reach the Chandrasekhar-mass limit. Owing to the absence of a hydrogen envelope, 
our models are not subject to any of those effects, although our models do experience a shallower dredge up which it does not seem to have the same impact as the 2DU to the \iso{H}{1}-rich counterparts. However, since helium stars in our considered mass range are expected to be produced in binaries, some residual hydrogen may be present in the envelope when these stars detach from their Roche lobes \citep[][and references therein]{Sen:2022abc}. This thin layer of hydrogen might  
influence the evolution, for instance by increasing the surface luminosity (and therefore mass loss rate) due to additional shell burning, or by inducing  instabilities such as the ones described above. This could potentially inhibit the growth of the core and therefore reduce the expected mass range for core-collapse.  

Similarly, binary interaction, both during and after core helium burning, can 
potentially influence the final core mass. As can be seen in Fig.~\ref{fig:mass_radius}, our models remain relatively compact ($\lesssim 0.5\rsun$) during this phase. Hence, additional mass stripping via Case\,BB mass transfer is only expected to play a significant role in very compact binaries. 
Contrarily, we find significant radius expansion for all models with initial masses above $\sim 1.5\msun$ (see lower panel of Fig.~\ref{fig:mass_radius}), meaning that late mass transfer likely affects the most extended SAGB stars, independently of their final fate ((C)ONe \ias vs ECSNe vs iron core-collapse). While such interaction is not considered in our models, it will most likely influence our results by effectively shifting  the mass boundaries shown in Fig.~\ref{fig:parameterSpace}. 
Bearing in mind the importance of these caveats, we plan to investigate this further by performing detailed binary calculations in a future project.

\subsection{Other sources of uncertainty}\label{sec:uncertainties}

Besides the effects discussed previously in this section, there are several additional physical uncertainties that may influence the final fate of stars in this regime. 
One of these uncertain factors is the propagation of the flame during the main  carbon burning phase. 
Both the ignition location of carbon and the flame propagation speed are sensitive to several physical mechanisms such as 
thermohaline, convective and rotational mixing, the neutrino cooling timescale, and other uncertainties related to winds 
\citep[][and references therein]{Farmer:2015afs}. 
Here, we only investigated one small component of this issue that is related to convective overshooting. 
Our models confirm that the carbon flame propagation is indeed sensitive to assumptions related to mixing, 
suggesting that other mixing mechanisms may  play a significant role as well. 

The accuracy of numerical calculations during the late stages of the evolution also contributes some uncertainty. It is unclear if our simulations can accurately reproduce the structure and properties of the outermost layers of stars with convective envelopes. For instance, our models reach a short phase of super-Eddington luminosities causing convection to become insufficient to transport the energy, due to the necessary high superadiabatic temperature gradient. This leads to increasingly short timesteps when one attempts to resolve this evolutionary stage. 
In our calculations, we circumvented  these issues by artificially reducing the superadiabatic temperature gradient using the \texttt{MLT++} option, which increases the efficiency of convective energy transport. 
However, this approach also influences radii  estimates, as it suppresses envelope inflation, thereby also affecting other properties such as the effective temperature and surface luminosity. 

During the aforementioned super-Eddington phase, some of our models also achieve effective temperatures 
that are below the lowest temperature covered by the \texttt{OPAL} opacity tables included in 
\mesa $\left( \log_{10} (T/\rm K) = 3.75 \right)$. This forces \mesa to use other low-temperature opacity 
tables that in principle assume different composition. This means that there might be significant 
changes in opacity at the location of the blend due to the change in the assumed composition 
\citep[see][for details]{Schwab:2019:RCrB}. These issues motivate further investigations to 
understand the complex underlying physics of stellar winds and envelope ejection. 

Finally, we did not investigate in detail the impact of uncertainties in  nuclear reaction rates. 
For instance, the uncertain  $\iso{C}{12}({\rm \alpha},\iso{O}{16})\gamma$,  $\iso{C}{12}(\iso{C}{12},{\rm \alpha})\iso{Ne}{20}$ and 
$\iso{C}{12}(\iso{C}{12},{\rm p})\iso{Na}{23}$ rates  
may substantially influence the pre-SN \iso{C}{12} and \iso{Na}{23} abundances. 
Similarly, as \cite{Stromberg:nl2021} point out, the details of $e$-capture processes and the associated 
weak reaction rates can also introduce substantial uncertainties. They argue that taking into consideration the second forbidden transition $4^{+}_{\text{g.s.}} \rightarrow 2^{+}_{1}$ lowers the threshold density for \iso{Na}{24}$(\electron,\neutrinoe)$\iso{Ne}{24}, allowing it 
to occur soon after the \iso{Mg}{24}$(\electron,\neutrinoe)$\iso{Na}{24}, for temperatures 
$\log_{10}(\rm T / {\rm K}) \lesssim 8.5$. We did not consider this transition in our calculations. 
Instead, the temperature in this regime  increases in two separate steps, at slightly different 
densities, owing to the allowed transition $0^{+}_{\text{g.s.}} \rightarrow 1^{+}_{1}$ between 
\iso{Mg}{24} and \iso{Na}{24}, and the $1^{+}_{1} \rightarrow 0^{+}_{\text{g.s.}}$ between the 
\iso{Na}{24} and \iso{Ne}{24}. For our models that feature residual carbon in their cores, 
the close succession of those two reactions---forced by the forbidden transition---could have a 
non-negligible effect on the overall outcome by reducing the carbon ignition density.

\section{Summary \& discussion} \label{sec:summary}
 \cite{antoniadis2020}  presented a case study of two helium star models that  develop a 
degenerate (C)ONe core  and subsequently experience a 
thermal runaway when their central density is $9.2 \lesssim \logrhoc \lesssim 9.8$. These stars likely avoid $e$-captures on 
\iso{Ne}{20} and produce a (C)ONe \ia. 
In this work, we build on this result by using \mesa to model the evolution of $\numrange{0.8}{3.5}\msun$ helium stars, exploring the influence of initial composition, stellar-wind efficiency, and mixing assumptions. 
Our models confirm that (C)ONe \ias may occur for a sizeable fraction of  helium stars in the $\numrange{1.7}{2.6}\msun$ range. The salient properties of our models which we identify as responsible for enabling this outcome can be summarized as follows: firstly, unlike hydrogen-rich SAGB stars, our models do not experience a 2DU episode, nor undergo unstable shell burning. Consequently, their cores grow fast via stable \iso{He}{4}-shell burning. This significantly broadens the mass range for which stars develop degenerate Chandrasekhar-mass cores. Our findings (Fig.~\ref{fig:final_fates}) are consistent with previous studies  \citep[e.g.,][]{Podsiadlowski:2003py,Poelarends:2007ip,Poelarends:2017dua}, confirming that ECSNe are mostly relevant to interacting binary systems. In Sects.~\ref{sec:one_core_evolution} and \ref{sec:cone_core_evolution} we argue that all (C)ONe cores originating from helium stars with initial masses $M_{\rm He, i}\gtrsim 2.1\msun$ are able to reach the Chandrasekhar limit, independently of overshooting efficiency and initial metallicity. At $Z=0.02$ and $f_{\rm OV}>0$, near-\mch\ cores become possible for all stars with  $M_{\rm He, i}\gtrsim 1.9\msun$. This includes the entire mass range for which ONe cores are produced. Consequently, ONe WDs are not expected for these initial conditions.

Secondly, (C)ONe cores are formed via carbon burning that initiates off-center and propagates inward. This leaves a certain amount of unburned carbon distributed throughout the core.  
In Sects.~\ref{sec:one_core_evolution} and \ref{sec:carbon} we demonstrate that leftover carbon plays a critical role in determining the final fate of near-\mch\ (C)ONe cores. For hybrid CONe compositions, residual carbon ignites when $\logrhoc \simeq 9.3$ leading to a thermonuclear runaway. Similarly, in ONe cores a carbon simmering phase is triggered by exothermic $e$-captures on \iso{Mg}{24} and \iso{Na}{24}. This may raise the temperature sufficiently to initiate explosive oxygen burning when $\logrho \lesssim 9.8$. 

Our models suggest that degenerate cores with a mass averaged mass fraction of $X(\iso{C}{12})\gtrsim 0.004$ produce ONe \ias, whereas those with lower \iso{C}{12} mass fractions yield ECSNe (Section~\ref{sec:carbon}). This is approximately equal to 0.005\msun\ of residual carbon distributed in the degenerate core. Figure~\ref{fig:carbon_mass_fractions} illustrates how the residual \iso{C}{12} mass fraction varies with initial mass, composition, and overshooting efficiency. Larger residual carbon mass fractions are favored for lower masses and high metallicities, especially for $f_{\rm OV}>0$. This means that (C)ONe \ias are mostly relevant to high-metallicity environments: for $Z=0.02$, the initial mass range for which near-\mch\ degenerate cores have $X(\iso{C}{12})\leq 0.004$ is $(M^{\rm min}_{\rm He, i},M^{\rm max}_{\rm He, i})  \simeq (2.1, 2.7)$ or $(1.8, 2.3)$\msun\ for $f_{\rm OV}=0.0$ and 0.016, respectively. For $Z\leq10^{-3}$, the corresponding range becomes $\sim(2.1, 2.5)$ and $(2.0, 2.1)$\msun\ for $f_{\rm OV}=0.0$ and 0.016.

Thirdly, prior to collapse, near-\mch\ models undergo a period of vigorous shell burning, during which the envelope reaches super-Eddington luminosities. In Sect.~\ref{sec:results} we demonstrate that this ``hot-bottom \iso{He}{4}-burning'' episode leads to helium-free, metal-rich envelopes. In Sect.~\ref{sec:physical_uncertainties} we explore uncertainties related to the power of the stellar wind. We find that the final envelope mass depends on the strength of the stellar wind during the core helium burning phase. However, in all cases, no helium is present at the time of core-collapse.

In Sect.~\ref{sec:physical_uncertainties} we discuss several factors that could affect these explosions, including uncertainties in nuclear reaction rates, mixing processes, stellar winds, and binary interaction. If (C)ONe \ias indeed occur in nature, they can significantly influence the properties of compact object populations, as well as the chemical evolution of galaxies. For instance, the formation of neutron stars via ECSNe in binaries would be affected, particularly at high metallicities. Such potentially important consequences raise the question of how (C)ONe \ias and their progenitors could  be identified. In Sect.~\ref{sec:explosion_properties} we estimated that (C)ONe \ias have a sufficiently large reservoir of nuclear energy to appear similar to normal \ias in terms of iron-group elements yields and kinetic energies. Evidently, more robust constraints on the nucleosynthesis and  explosion energies require detailed multidimensional hydrodynamic simulations of the explosive oxygen burning phase. Another distinct characteristic that could help constrain their observational counterparts is the intense mass loss taking place during the few thousand years leading to the explosion. This mass loss history will likely create  a dense metal-rich circumstellar material (CSM) that can subsequently interact with the SN ejecta. This presents some similarities to the ``core-denegerate'' \ia scenario \citep[e.g.,][]{Wang:2016alt,soker2019,Soker:2021dii}. However, binary interactions---involving these stars with the most extended envelopes---may occur leading to a more diverse CSM spectrum.

In terms of explosion sites, we expect (C)ONe \ias to be relevant mostly to high-metallicity star-forming regions. Similarly to ECSNe, we expect a delay time distribution dominated by the main sequence lifetime of their progenitors, that is to be $\mathcal{O}$(200\,Myr), although it might be possible to form similar ONe cores via WD mergers \citep{Schwab:2016lep}.
\cite{antoniadis2020}, argued that the frequency of ONe \ias can be comparable to the observed SNe Ib/c rates. Considering the results presented here, we update the relative ratio between ONe \ias and SNe Ib/c to be $\numrange{0.17}{0.30}$ at $Z=0.02$ and $\numrange{0.03}{0.13}$ at $Z=10^{-4}$ \citep[assuming a][initial mass function]{Chabrier:2004vw}.

Interestingly, these population properties are very similar to what has been inferred for SNe\,Iax \citep{Foley:2012tu,Meng:2017ijx,Jha:2017gwq} but also for some  high-velocity \citep[e.g.,][]{Pan:2020grt,Zeng:2021vsz} and normal \ias \citep{Maoz:2010pz,Soker:2021dii}. 
SNe\,Iax in particular, present some properties that make them particularly attractive as potential ONe \ia counterparts. While they are usually interpreted as CO or (C)ONe WD deflagrations \citep[e.g.,][]{Kromer:2015lda,Yamanaka:2015qpa,Magee:2016vnu},  at least a fraction of them occur inside a helium-rich CSM \citep{Magee:2018aui,Jacobson-Galan:2018fqu}, while none has been observed in an elliptical galaxy host \citep{Foley:2012tu}. 
Lastly, at least one nearby SN\,Iax, SN\,2012Z, has been securely associated with a luminous blue progenitor \citep{McCully:2014jva,Stritzinger:2014lva,Yamanaka:2015qpa,McCully:2021abc}, with properties similar to those inferred for our ONe models prior to collapse (Section~\ref{sec:results}). 

Besides the explosions themselves, the progenitors of (C)ONe \ias might also be easily identifiable, particularly in the few thousand years leading to the collapse. This is  because of their high luminosities (Fig.~\ref{fig:hrd_series1}), but most importantly due to their hydrogen- and helium-free, metal-rich atmospheres (Fig.~\ref{fig:surface_abun_evol}). Such peculiar atmospheric abundances are not easily produced via ordinary stellar evolution 
paths. A unique object that presents some interesting similarities to our progenitor models is WS35 \citep{Gvaramadze:2019abc,Oskinova:2019abc}. 
This star is luminous ($L\simeq 10^{4.5}$\,L$_{\odot}$), hot ($\sim$200,000\,K), as well as hydrogen- and helium-free \citep{Oskinova:2019abc}. It is also
embedded in a circular nebula \citep{Gvaramadze:2019abc}, that was likely produced via a
strong stellar wind. While our progenitor models  are somewhat cooler prior to collapse, this could be due to artifacts related to, for instance, mixing assumptions or surface boundary conditions.   

Other indirect, but distinct observational evidence could be related to the properties of compact-object populations at different metallicities. For instance if neutron star formation via ECSNe is suppressed at high-$Z$ environments (Fig.~\ref{fig:final_fates}), as our results suggest, both the integrated and $Z$-dependent double NS (DNS) merger rates would be affected \citep{Mandel:2021abc}. One would also expect to see differences  between Galactic and extra-galactic DNSs (e.g., in their mass ratio distributions). Similarly, the Galactic pulsar population might also be affected, for example fewer ECSNe could  lead to a decreased number of young radio pulsars in wide binary systems and globular clusters \citep[][and references therein]{Antoniadis:2020gos,Willcox:2021kbg}. However, we note that there might be other core-collapse SN mechanisms that produce low-velocity pulsars \citep[see ][and references therein]{Antoniadis:2021dhe}

Lastly,  models that do not reach the Chandrasekhar limit, but become (C)ONe WDs instead, also 
present some interest as possible thermonuclear SN and NS progenitors. Their high masses (up to 1.35\msun; Section~\ref{sec:wd_evolution}) 
imply that they only need to accrete very small amounts of mass to become unstable. Their collapse would then lead to a thermonuclear deflagration
\citep{Meng:2014qta,Willcox:2016yyp,Bravo:2019fwh}, or the formation of a neutron star in an accretion-induced collapse \citep[e.g.,][]{Nomoto:1991abc,Tauris:2013zna,Ruiter:2018ouw,Wang:2020pzc}. Similarly to what has been discussed in this work, we expect the core composition---particularly the amount of residual carbon and its distribution in the core---to play a critical role in determining their fate \citep{brooks2017}. It is conceivable that if these WDs can achieve oxygen ignition conditions similar to those of (C)ONe \ias, then the resulting explosions would also have similar appearance. In this case, (C)ONe WDs may contribute \ias with much longer delay times.

\begin{acknowledgements}
We thank the referee for the extremely helpful report.
This work was supported by the Stavros Niarchos Foundation (SNF) and the Hellenic Foundation for Research and 
Innovation (H.F.R.I.) under the 2nd Call of ``Science and Society'' Action Always strive for excellence -- ``Theodoros Papazoglou'' (Project Number: 01431). GG acknowledges support from Deutsche Luft- und Raumfahrt (DLR) grant No. 50\,OR\,2009.

Results presented in this work have been produced using the Aristotle University of Thessaloniki (AUTh) Aristotelis cluster.
SC would like to acknowledge the support provided by the Bonn-Cologne Graduate School for Physics and Astronomy (BCGS) and the IT center of the Aristotle University of Thessaloniki throughout the progress of this research work. 

This research made extensive use of NASA's ADS, \texttt{MESA}\footnote{\url{http://mesastar.org}} \citep{Paxton:2010ji,Paxton:2013pj,Paxton:2015jva,Paxton:2017eie}, \texttt{PyMesaReader}\footnote{\url{https://zenodo.org/record/826958}} \citep{pymesareader}, \texttt{mesaPlot}\footnote{\url{https://zenodo.org/record/4729811}} \citep{mesaplot}, \texttt{Kippenhahn plotter for MESA}\footnote{\url{https://zenodo.org/record/2602098}} \citep{pablo_marchant_2019_2602098}, \texttt{Astropy}\footnote{\url{http://www.astropy.org}} \citep{Price-Whelan:2018hus}, \texttt{Matplotlib} \citep{Hunter:2007:matplotlib}, \texttt{NumPy} \citep{harris2020array:numpy}, and \texttt{Jupyter} \citep{Kluyver2016jupyter}.
\end{acknowledgements}

\bibliography{bibfile.bib}

\begin{thebibliography}{123}
\expandafter\ifx\csname natexlab\endcsname\relax\def\natexlab#1{#1}\fi

\bibitem[{{Aguilera-Dena} {et~al.}(2022){Aguilera-Dena}, Langer, Antoniadis,
  Pauli, Dessart, {Vigna-G{\'o}mez}, Gr{\"a}fener, \&
  Yoon}]{Aguilera-Dena:2021abc}
{Aguilera-Dena}, D.~R., Langer, N., Antoniadis, J., {et~al.} 2022, Astronomy \&
  Astrophysics, 661, A60

\bibitem[{Antoniadis(2021)}]{Antoniadis:2020gos}
Antoniadis, J. 2021, Monthly Notices of the Royal Astronomical Society, 501,
  1116

\bibitem[{Antoniadis {et~al.}(2020)Antoniadis, Chanlaridis, Gr{\"a}fener, \&
  Langer}]{antoniadis2020}
Antoniadis, J., Chanlaridis, S., Gr{\"a}fener, G., \& Langer, N. 2020,
  Astronomy and Astrophysics, 635, A72

\bibitem[{{Antoniadis, John} {et~al.}(2022){Antoniadis, John}, {Aguilera-Dena,
  David R.}, {Vigna-G\'omez, Alejandro}, {Kramer, Michael}, {Langer, Norbert},
  {M\"uller, Bernhard}, {Tauris, Thomas M.}, {Wang, Chen}, \& {Xu,
  Xiao-Tian}}]{Antoniadis:2021dhe}
{Antoniadis, John}, {Aguilera-Dena, David R.}, {Vigna-G\'omez, Alejandro},
  {et~al.} 2022, A\&A, 657, L6

\bibitem[{{Astropy Collaboration} \& {Astropy
  Contributors}(2018)}]{Price-Whelan:2018hus}
{Astropy Collaboration} \& {Astropy Contributors}. 2018, The Astronomical
  Journal, 156, 123

\bibitem[{Beasor {et~al.}(2021)Beasor, Davies, \& Smith}]{Beasor2021arx}
Beasor, E.~R., Davies, B., \& Smith, N. 2021, The Astrophysical Journal, 922,
  55

\bibitem[{{Bravo, E.}(2019)}]{Bravo:2019fwh}
{Bravo, E.} 2019, A\&A, 624, A139

\bibitem[{Brooks {et~al.}(2017)Brooks, Schwab, Bildsten, Quataert, \&
  Paxton}]{brooks2017}
Brooks, J., Schwab, J., Bildsten, L., Quataert, E., \& Paxton, B. 2017, The
  Astrophysical Journal, 834, L9

\bibitem[{Brown {et~al.}(2013)Brown, Garaud, \& Stellmach}]{Brown_2013}
Brown, J.~M., Garaud, P., \& Stellmach, S. 2013, The Astrophysical Journal,
  768, 34

\bibitem[{Canal {et~al.}(1992)Canal, Isern, \& Labay}]{canal1992}
Canal, R., Isern, J., \& Labay, J. 1992, The Astrophysical Journal Letters,
  398, L49

\bibitem[{Chabrier(2005)}]{Chabrier:2004vw}
Chabrier, G. 2005, The Initial Mass Function 50 Years Later, 327, 41

\bibitem[{Chen {et~al.}(2014)Chen, Herwig, Denissenkov, \& Paxton}]{chen2014b}
Chen, M.~C., Herwig, F., Denissenkov, P.~A., \& Paxton, B. 2014, Monthly
  Notices of the Royal Astronomical Society, 440, 1274

\bibitem[{{de Jager} {et~al.}(1988){de Jager}, {Nieuwenhuijzen}, \& {van der
  Hucht}}]{deJager1988}
{de Jager}, C., {Nieuwenhuijzen}, H., \& {van der Hucht}, K.~A. 1988, Astronomy
  and Astrophysics Supplement Series, 72, 259

\bibitem[{Denissenkov {et~al.}(2013)Denissenkov, Herwig, Truran, \&
  Paxton}]{Denissenkov:2013qaa}
Denissenkov, P.~A., Herwig, F., Truran, J.~W., \& Paxton, B. 2013, The
  Astrophysical Journal, 772, 37

\bibitem[{Doherty {et~al.}(2015)Doherty, {Gil-Pons}, Siess, Lattanzio, \&
  Lau}]{doherty2015}
Doherty, C.~L., {Gil-Pons}, P., Siess, L., Lattanzio, J.~C., \& Lau, H. H.~B.
  2015, Monthly Notices of the Royal Astronomical Society, 446, 2599

\bibitem[{Doherty {et~al.}(2010)Doherty, Siess, Lattanzio, \&
  Gil-Pons}]{Doherty:2010slg}
Doherty, C.~L., Siess, L., Lattanzio, J.~C., \& Gil-Pons, P. 2010, Monthly
  Notices of the Royal Astronomical Society, 401, 1453

\bibitem[{{Dominguez} {et~al.}(1993){Dominguez}, {Tornambe}, \&
  {Isern}}]{Dominguez1993}
{Dominguez}, I., {Tornambe}, A., \& {Isern}, J. 1993, \apj, 419, 268

\bibitem[{Farmer(2021)}]{mesaplot}
Farmer, R. 2021, rjfarmer/mesaplot: Release v1.1.0

\bibitem[{Farmer {et~al.}(2015)Farmer, Fields, \& Timmes}]{Farmer:2015afs}
Farmer, R., Fields, C.~E., \& Timmes, F.~X. 2015, The Astrophysical Journal,
  807, 184

\bibitem[{Foley {et~al.}(2013)Foley, Challis, Chornock, Ganeshalingam, Li,
  Marion, Morrell, Pignata, Stritzinger, Silverman, Wang, Anderson, Filippenko,
  Freedman, Hamuy, Jha, Kirshner, McCully, Persson, Phillips, Reichart, \&
  Soderberg}]{Foley:2012tu}
Foley, R.~J., Challis, P.~J., Chornock, R., {et~al.} 2013, \apj, 767, 57

\bibitem[{{Garc{\'\i}a-Berro} {et~al.}(1997){Garc{\'\i}a-Berro}, {Ritossa}, \&
  {Iben}}]{Garcia1997}
{Garc{\'\i}a-Berro}, E., {Ritossa}, C., \& {Iben}, Icko, J. 1997, \apj, 485,
  765

\bibitem[{{Gil-Pons} {et~al.}(2003){Gil-Pons}, {Garc{\'\i}a-Berro}, {Jos{\'e}},
  {Hernanz}, \& {Truran}}]{Gil-Pons:2003sep}
{Gil-Pons}, P., {Garc{\'\i}a-Berro}, E., {Jos{\'e}}, J., {Hernanz}, M., \&
  {Truran}, J.~W. 2003, A\&A, 407, 1021

\bibitem[{{Glebbeek} {et~al.}(2009){Glebbeek}, {Gaburov}, {de Mink}, {Pols}, \&
  {Portegies Zwart}}]{Dutch}
{Glebbeek}, E., {Gaburov}, E., {de Mink}, S.~E., {Pols}, O.~R., \& {Portegies
  Zwart}, S.~F. 2009, Astronomy and Astrophysics, 497, 255

\bibitem[{{Gr{\"a}fener} {et~al.}(2017){Gr{\"a}fener}, {Owocki}, {Grassitelli},
  \& {Langer}}]{Graefener2017}
{Gr{\"a}fener}, G., {Owocki}, S.~P., {Grassitelli}, L., \& {Langer}, N. 2017,
  \aap, 608, A34

\bibitem[{Grevesse \& Sauval(1998)}]{grevesse1998}
Grevesse, N. \& Sauval, A.~J. 1998, Space Science Reviews, 85, 161

\bibitem[{Gutierrez {et~al.}(1996)Gutierrez, {Garcia-Berro}, Iben, Isern,
  Labay, \& Canal}]{gutierrez1996a}
Gutierrez, J., {Garcia-Berro}, E., Iben, Jr., I., {et~al.} 1996, The
  Astrophysical Journal, 459, 701

\bibitem[{{Guti\'errez, J.} {et~al.}(2005){Guti\'errez, J.}, {Canal, R.}, \&
  {Garc\'{\i}a-Berro, E.}}]{Gutierrez2005}
{Guti\'errez, J.}, {Canal, R.}, \& {Garc\'{\i}a-Berro, E.} 2005, A\&A, 435, 231

\bibitem[{Gvaramadze {et~al.}(2019)Gvaramadze, Gr{\"a}fener, Langer, Maryeva,
  Kniazev, Moskvitin, \& Spiridonova}]{Gvaramadze:2019abc}
Gvaramadze, V.~V., Gr{\"a}fener, G., Langer, N., {et~al.} 2019, Nature, 569,
  684

\bibitem[{Harris {et~al.}(2020)Harris, Millman, van~der Walt, Gommers,
  Virtanen, Cournapeau, Wieser, Taylor, Berg, Smith, Kern, Picus, Hoyer, van
  Kerkwijk, Brett, Haldane, del R{\'{i}}o, Wiebe, Peterson,
  G{\'{e}}rard-Marchant, Sheppard, Reddy, Weckesser, Abbasi, Gohlke, \&
  Oliphant}]{harris2020array:numpy}
Harris, C.~R., Millman, K.~J., van~der Walt, S.~J., {et~al.} 2020, Nature, 585,
  357

\bibitem[{{Henyey} {et~al.}(1965){Henyey}, {Vardya}, \&
  {Bodenheimer}}]{MLT_Henyey}
{Henyey}, L., {Vardya}, M.~S., \& {Bodenheimer}, P. 1965, The Astrophysical
  Journal, 142, 841

\bibitem[{{Herwig}(2000)}]{Herwig2000}
{Herwig}, F. 2000, Astronomy and Astrophysics, 360, 952

\bibitem[{Hunter(2007)}]{Hunter:2007:matplotlib}
Hunter, J.~D. 2007, Computing in Science \& Engineering, 9, 90

\bibitem[{Iben \& Renzini(1983)}]{Iben:1983ts}
Iben, Jr., I. \& Renzini, A. 1983, Annual Review of Astronomy and Astrophysics,
  21, 271

\bibitem[{Iben {et~al.}(1997)Iben, Ritossa, \&
  {Garc{\'i}a-Berro}}]{Iben:1997abc}
Iben, Jr., I., Ritossa, C., \& {Garc{\'i}a-Berro}, E. 1997, The Astrophysical
  Journal, 489, 772

\bibitem[{{Iglesias} \& {Rogers}(1996)}]{OPAL}
{Iglesias}, C.~A. \& {Rogers}, F.~J. 1996, The Astrophysical Journal, 464, 943

\bibitem[{{Itoh} {et~al.}(2002){Itoh}, {Tomizawa}, {Tamamura}, {Wanajo}, \&
  {Nozawa}}]{Itoh2002}
{Itoh}, N., {Tomizawa}, N., {Tamamura}, M., {Wanajo}, S., \& {Nozawa}, S. 2002,
  The Astrophysical Journal, 579, 380

\bibitem[{{Jacobson-Gal{\'a}n} {et~al.}(2019){Jacobson-Gal{\'a}n}, Foley,
  Schwab, Dimitriadis, Dong, Jha, Kasen, Kilpatrick, \&
  Thomas}]{Jacobson-Galan:2018fqu}
{Jacobson-Gal{\'a}n}, W.~V., Foley, R.~J., Schwab, J., {et~al.} 2019, Monthly
  Notices of the Royal Astronomical Society, 487, 2538

\bibitem[{{Jha}(2017)}]{2017hsnJ}
{Jha}, S.~W. 2017, {Type Iax Supernovae}, ed. A.~W. {Alsabti} \& P.~{Murdin},
  375

\bibitem[{Jha(2017)}]{Jha:2017gwq}
Jha, S.~W. 2017, Handbook of Supernovae, 375

\bibitem[{Jones {et~al.}(2014)Jones, Hirschi, \& Nomoto}]{jones2014}
Jones, S., Hirschi, R., \& Nomoto, K. 2014, \textbackslash{}apj, 797, 83

\bibitem[{Jones {et~al.}(2013)Jones, Hirschi, Nomoto, Fischer, Timmes, Herwig,
  Paxton, Toki, Suzuki, Mart{\'{\i}}nez-Pinedo, Lam, \&
  Bertolli}]{Jones2013apj}
Jones, S., Hirschi, R., Nomoto, K., {et~al.} 2013, The Astrophysical Journal,
  772, 150

\bibitem[{Jones {et~al.}(2019)Jones, R{\"o}pke, Fryer, Ruiter, Seitenzahl,
  Nittler, Ohlmann, Reifarth, Pignatari, \& Belczynski}]{Jones:2018ule}
Jones, S., R{\"o}pke, F.~K., Fryer, C., {et~al.} 2019, Astronomy \&
  Astrophysics, 622, A74

\bibitem[{Jones {et~al.}(2016)Jones, R{\"o}pke, Pakmor, Seitenzahl, Ohlmann, \&
  Edelmann}]{Jones:2016asr}
Jones, S., R{\"o}pke, F.~K., Pakmor, R., {et~al.} 2016, Astronomy \&
  Astrophysics, 593, A72

\bibitem[{Kippenhahn {et~al.}(1980)Kippenhahn, Ruschenplatt, \&
  Thomas}]{kippenhahn1980}
Kippenhahn, R., Ruschenplatt, G., \& Thomas, H.-C. 1980, Astronomy and
  Astrophysics, 91, 175

\bibitem[{Kirsebom {et~al.}(2019{\natexlab{a}})Kirsebom, Hukkanen, Kankainen,
  Trzaska, Str{\"o}mberg, {Mart{\'i}nez-Pinedo}, Andersen, Bodewits, Brown,
  Canete, Cederk{\"a}ll, Enqvist, Eronen, Fynbo, Geldhof, {de Groote}, Jenkins,
  Jokinen, Joshi, Khanam, Kostensalo, Kuusiniemi, Langanke, Moore, Munch,
  Nesterenko, Ovejas, Penttil{\"a}, Pohjalainen, Reponen, {Rinta-Antila},
  Riisager, {de Roubin}, Schotanus, Srivastava, Suhonen, Swartz, Tengblad,
  Vilen, V{\'i}nals, \& {{\'n}yst{\"o}}}]{kirsebom2019b}
Kirsebom, O.~S., Hukkanen, M., Kankainen, A., {et~al.} 2019{\natexlab{a}},
  Physical Review C, 100, 065805

\bibitem[{Kirsebom {et~al.}(2019{\natexlab{b}})Kirsebom, Jones, Str{\"o}mberg,
  {Mart{\'i}nez-Pinedo}, Langanke, R{\"o}pke, Brown, Eronen, Fynbo, Hukkanen,
  Idini, Jokinen, Kankainen, Kostensalo, Moore, M{\"o}ller, Ohlmann,
  Penttil{\"a}, Riisager, {Rinta-Antila}, Srivastava, Suhonen, Trzaska, \&
  {{\'n}yst{\"o}}}]{kirsebom2019a}
Kirsebom, O.~S., Jones, S., Str{\"o}mberg, D.~F., {et~al.} 2019{\natexlab{b}},
  Physical Review Letters, 123, 262701

\bibitem[{Kitaura {et~al.}(2006)Kitaura, Janka, \& Hillebrandt}]{kitaura2006}
Kitaura, F.~S., Janka, H.-T., \& Hillebrandt, W. 2006, Astronomy and
  Astrophysics, 450, 345

\bibitem[{Kluyver {et~al.}(2016)Kluyver, Ragan-Kelley, P{\'e}rez, Granger,
  Bussonnier, Frederic, Kelley, Hamrick, Grout, Corlay, Ivanov, Avila, Abdalla,
  \& Willing}]{Kluyver2016jupyter}
Kluyver, T., Ragan-Kelley, B., P{\'e}rez, F., {et~al.} 2016, in Positioning and
  Power in Academic Publishing: Players, Agents and Agendas, ed. F.~Loizides \&
  B.~Schmidt, IOS Press, 87 -- 90

\bibitem[{Kromer {et~al.}(2015)Kromer, Ohlmann, Pakmor, Ruiter, Hillebrandt,
  Marquardt, R{\"o}pke, Seitenzahl, Sim, \& Taubenberger}]{Kromer:2015lda}
Kromer, M., Ohlmann, S.~T., Pakmor, R., {et~al.} 2015, Monthly Notices of the
  Royal Astronomical Society, 450, 3045

\bibitem[{{Langer}(1991)}]{Langer1991}
{Langer}, N. 1991, Astronomy and Astrophysics, 252, 669

\bibitem[{Langer {et~al.}(1983)Langer, Fricke, \& Sugimoto}]{Langer1983}
Langer, N., Fricke, K.~J., \& Sugimoto, D. 1983, Astronomy and Astrophysics,
  126, 207

\bibitem[{{Laplace} {et~al.}(2020){Laplace}, {G{\"o}tberg}, {de Mink},
  {Justham}, \& {Farmer}}]{Laplace2020aa}
{Laplace}, E., {G{\"o}tberg}, Y., {de Mink}, S.~E., {Justham}, S., \& {Farmer},
  R. 2020, A\&A, 637, A6

\bibitem[{Lecoanet {et~al.}(2016)Lecoanet, Schwab, Quataert, Bildsten, Timmes,
  Burns, Vasil, Oishi, \& Brown}]{Lecoanet:2016abca}
Lecoanet, D., Schwab, J., Quataert, E., {et~al.} 2016, The Astrophysical
  Journal, 832, 71

\bibitem[{Leung {et~al.}(2020)Leung, Nomoto, \& Suzuki}]{Leung:2019phz}
Leung, S.-C., Nomoto, K., \& Suzuki, T. 2020, The Astrophysical Journal, 889,
  34

\bibitem[{Magee {et~al.}(2016)Magee, Kotak, Sim, Kromer, Rabinowitz, Smartt,
  Baltay, Campbell, Chen, Fink, {Gal-Yam}, Galbany, Hillebrandt, Inserra,
  Kankare, Le~Guillou, Lyman, Maguire, Pakmor, R{\"o}pke, Ruiter, Seitenzahl,
  Sullivan, Valenti, \& Young}]{Magee:2016vnu}
Magee, M.~R., Kotak, R., Sim, S.~A., {et~al.} 2016, Astronomy \& Astrophysics,
  589, A89

\bibitem[{Magee {et~al.}(2019)Magee, Sim, Kotak, Maguire, \&
  Boyle}]{Magee:2018aui}
Magee, M.~R., Sim, S.~A., Kotak, R., Maguire, K., \& Boyle, A. 2019, Astronomy
  \& Astrophysics, 622, A102

\bibitem[{Mandel \& Broekgaarden(2022)}]{Mandel:2021abc}
Mandel, I. \& Broekgaarden, F.~S. 2022, Living Reviews in Relativity, 25, 1

\bibitem[{Maoz \& Badenes(2010)}]{Maoz:2010pz}
Maoz, D. \& Badenes, C. 2010, Monthly Notices of the Royal Astronomical
  Society, 407, 1314

\bibitem[{Marchant(2019)}]{pablo_marchant_2019_2602098}
Marchant, P. 2019, Kippenhahn plotter for MESA

\bibitem[{McCully {et~al.}(2014)McCully, Jha, Foley, Bildsten, Fong, Kirshner,
  Marion, Riess, \& Stritzinger}]{McCully:2014jva}
McCully, C., Jha, S.~W., Foley, R.~J., {et~al.} 2014, Nature, 512, 54

\bibitem[{McCully {et~al.}(2022)McCully, Jha, Scalzo, Howell, Foley, Zeng, Liu,
  Hosseinzadeh, Bildsten, Riess, Kirshner, Marion, \&
  Camacho-Neves}]{McCully:2021abc}
McCully, C., Jha, S.~W., Scalzo, R.~A., {et~al.} 2022, The Astrophysical
  Journal, 925, 138

\bibitem[{Meng \& Podsiadlowski(2014)}]{Meng:2014qta}
Meng, X. \& Podsiadlowski, P. 2014, \apj, 789, L45

\bibitem[{Meng \& Podsiadlowski(2018)}]{Meng:2017ijx}
Meng, X. \& Podsiadlowski, P. 2018, The Astrophysical Journal, 861, 127

\bibitem[{Miyaji {et~al.}(1980)Miyaji, Nomoto, Yokoi, \& Sugimoto}]{miyaji1980}
Miyaji, S., Nomoto, K., Yokoi, K., \& Sugimoto, D. 1980, Publications of the
  Astronomical Society of Japan, 32, 303

\bibitem[{Nomoto(1984)}]{nomoto1984}
Nomoto, K. 1984, The Astrophysical Journal, 277, 791

\bibitem[{Nomoto \& Kondo(1991)}]{Nomoto:1991abc}
Nomoto, K. \& Kondo, Y. 1991, The Astrophysical Journal, 367, L19

\bibitem[{{Nugis} \& {Lamers}(2000)}]{Nugis2000}
{Nugis}, T. \& {Lamers}, H.~J.~G.~L.~M. 2000, Astronomy and Astrophysics, 360,
  227

\bibitem[{Oskinova {et~al.}(2019)Oskinova, Gvaramadze, Gr{\"a}fener, Langer, \&
  Todt}]{Oskinova:2019abc}
Oskinova, L.~M., Gvaramadze, V.~V., Gr{\"a}fener, G., Langer, N., \& Todt, H.
  2019, Nature, 569, 684

\bibitem[{Owocki {et~al.}(2004)Owocki, Gayley, \& Shaviv}]{Owocki:2004zz}
Owocki, S.~P., Gayley, K.~G., \& Shaviv, N.~J. 2004, The Astrophysical Journal,
  616, 525

\bibitem[{Pan(2020)}]{Pan:2020grt}
Pan, Y.-C. 2020, The Astrophysical Journal, 895, L5

\bibitem[{Paxton {et~al.}(2011)Paxton, Bildsten, Dotter, Herwig, Lesaffre, \&
  Timmes}]{Paxton:2010ji}
Paxton, B., Bildsten, L., Dotter, A., {et~al.} 2011, The Astrophysical Journal
  Supplement Series, 192, 3

\bibitem[{Paxton {et~al.}(2013)Paxton, Cantiello, Arras, Bildsten, Brown,
  Dotter, Mankovich, Montgomery, Stello, Timmes, \& Townsend}]{Paxton:2013pj}
Paxton, B., Cantiello, M., Arras, P., {et~al.} 2013, The Astrophysical Journal
  Supplement Series, 208, 4

\bibitem[{Paxton {et~al.}(2015)Paxton, Marchant, Schwab, Bauer, Bildsten,
  Cantiello, Dessart, Farmer, Hu, Langer, Townsend, Townsley, \&
  Timmes}]{Paxton:2015jva}
Paxton, B., Marchant, P., Schwab, J., {et~al.} 2015, The Astrophysical Journal
  Supplement Series, 220, 15

\bibitem[{Paxton {et~al.}(2018)Paxton, Schwab, Bauer, Bildsten, Blinnikov,
  Duffell, Farmer, Goldberg, Marchant, Sorokina, Thoul, Townsend, \&
  Timmes}]{Paxton:2017eie}
Paxton, B., Schwab, J., Bauer, E.~B., {et~al.} 2018, The Astrophysical Journal
  Supplement Series, 234, 34

\bibitem[{Podsiadlowski {et~al.}(2004)Podsiadlowski, Langer, Poelarends,
  Rappaport, Heger, \& Pfahl}]{Podsiadlowski:2003py}
Podsiadlowski, P., Langer, N., Poelarends, A. J.~T., {et~al.} 2004, The
  Astrophysical Journal, 612, 1044

\bibitem[{Poelarends {et~al.}(2008)Poelarends, Herwig, Langer, \&
  Heger}]{Poelarends:2007ip}
Poelarends, A. J.~T., Herwig, F., Langer, N., \& Heger, A. 2008, The
  Astrophysical Journal, 675, 614

\bibitem[{Poelarends {et~al.}(2017)Poelarends, Wurtz, Tarka, Cole~Adams, \&
  Hills}]{Poelarends:2017dua}
Poelarends, A. J.~T., Wurtz, S., Tarka, J., Cole~Adams, L., \& Hills, S.~T.
  2017, The Astrophysical Journal, 850, 197

\bibitem[{Potekhin \& Chabrier(2010)}]{PC:eos}
Potekhin, A.~Y. \& Chabrier, G. 2010, Contributions to Plasma Physics, 50, 82

\bibitem[{Potekhin {et~al.}(2009)Potekhin, Chabrier, \& Rogers}]{PCR2009}
Potekhin, A.~Y., Chabrier, G., \& Rogers, F.~J. 2009, Phys. Rev. E, 79, 016411

\bibitem[{Raddi {et~al.}(2019)Raddi, Hollands, Koester, Hermes, G{\"a}nsicke,
  Heber, Shen, Townsley, Pala, Reding, Toloza, Pelisoli, Geier,
  Gentile~Fusillo, Munari, \& Strader}]{raddi2019a}
Raddi, R., Hollands, M.~A., Koester, D., {et~al.} 2019, Monthly Notices of the
  Royal Astronomical Society, 489, 1489

\bibitem[{{Ritossa} {et~al.}(1999){Ritossa}, {Garc{\'\i}a-Berro}, \&
  {Iben}}]{Ritossa:1999ApJ}
{Ritossa}, C., {Garc{\'\i}a-Berro}, E., \& {Iben}, Icko, J. 1999, ApJ, 515, 381

\bibitem[{Rose(1969)}]{rose1969}
Rose, W.~K. 1969, The Astrophysical Journal, 155, 491

\bibitem[{Ruiter {et~al.}(2019)Ruiter, Ferrario, Belczynski, Seitenzahl,
  Crocker, \& Karakas}]{Ruiter:2018ouw}
Ruiter, A.~J., Ferrario, L., Belczynski, K., {et~al.} 2019, Monthly Notices of
  the Royal Astronomical Society, 484, 698

\bibitem[{{Sander} \& {Vink}(2020)}]{Sanders:2020:mnras}
{Sander}, A. A.~C. \& {Vink}, J.~S. 2020, \mnras, 499, 873

\bibitem[{Schwab(2019)}]{Schwab:2019:RCrB}
Schwab, J. 2019, The Astrophysical Journal, 885, 27

\bibitem[{Schwab {et~al.}(2017)Schwab, Bildsten, \& Quataert}]{Schwab:2017epw}
Schwab, J., Bildsten, L., \& Quataert, E. 2017, Monthly Notices of the Royal
  Astronomical Society, 472, 3390

\bibitem[{Schwab {et~al.}(2020)Schwab, Farmer, \& Timmes}]{schwab2020}
Schwab, J., Farmer, R., \& Timmes, F.~X. 2020, The Astrophysical Journal, 891,
  5

\bibitem[{Schwab {et~al.}(2016)Schwab, Quataert, \& Kasen}]{Schwab:2016lep}
Schwab, J., Quataert, E., \& Kasen, D. 2016, Monthly Notices of the Royal
  Astronomical Society, 463, 3461

\bibitem[{Schwab \& Rocha(2019)}]{Schwab:2018cnb}
Schwab, J. \& Rocha, K.~A. 2019, The Astrophysical Journal, 872, 131

\bibitem[{Sen {et~al.}(2022)Sen, Langer, Marchant, Menon, {de Mink},
  Schootemeijer, Sch{\"u}rmann, Mahy, Hastings, Nathaniel, Sana, Wang, \&
  Xu}]{Sen:2022abc}
Sen, K., Langer, N., Marchant, P., {et~al.} 2022, Astronomy \&amp;
  Astrophysics, Volume 659, id.A98, {$<$}NUMPAGES{$>$}33{$<$}/NUMPAGES{$>$}
  pp., 659, A98

\bibitem[{Siess(2006)}]{siess2006}
Siess, L. 2006, Astronomy and Astrophysics, 448, 717

\bibitem[{{Siess}(2007)}]{dec07:siess}
{Siess}, L. 2007, A\&A, 476, 893

\bibitem[{{Siess}(2009)}]{Siess2009}
{Siess}, L. 2009, \aap, 497, 463

\bibitem[{{Siess} \& {Lebreuilly}(2018)}]{Siess2018leb}
{Siess}, L. \& {Lebreuilly}, U. 2018, A\&A, 614, A99

\bibitem[{{Smith} {et~al.}(2018){Smith}, {G{\"o}tberg}, \& {de
  Mink}}]{Smith2017}
{Smith}, N., {G{\"o}tberg}, Y., \& {de Mink}, S.~E. 2018, \mnras, 475, 772

\bibitem[{{Smith} \& {Owocki}(2006)}]{Smith2006}
{Smith}, N. \& {Owocki}, S.~P. 2006, \apjl, 645, L45

\bibitem[{Soker(2019)}]{soker2019}
Soker, N. 2019, New Astronomy Reviews, 87, 101535

\bibitem[{{Soker}(2022)}]{Soker:2021dii}
{Soker}, N. 2022, Research in Astronomy and Astrophysics, 22, 035025

\bibitem[{Stritzinger {et~al.}(2015)Stritzinger, Valenti, Hoeflich, Baron,
  Phillips, Taddia, Foley, Hsiao, Jha, McCully, Pandya, Simon, Benetti, Brown,
  Burns, Campillay, Contreras, F{\"o}rster, Holmbo, Marion, Morrell, \&
  Pignata}]{Stritzinger:2014lva}
Stritzinger, M.~D., Valenti, S., Hoeflich, P., {et~al.} 2015, Astronomy and
  Astrophysics, 573, A2

\bibitem[{Str\"omberg {et~al.}(2022)Str\"omberg, Mart\'{\i}nez-Pinedo, \&
  Nowacki}]{Stromberg:nl2021}
Str\"omberg, D.~F., Mart\'{\i}nez-Pinedo, G., \& Nowacki, F. 2022, Phys. Rev.
  C, 105, 025803

\bibitem[{Suzuki {et~al.}(2016)Suzuki, Toki, \& Nomoto}]{Suzuki2016}
Suzuki, T., Toki, H., \& Nomoto, K. 2016, The Astrophysical Journal, 817, 163

\bibitem[{Tauris \& Janka(2019)}]{tauris2019}
Tauris, T.~M. \& Janka, H.-T. 2019, The Astrophysical Journal Letters, 886, L20

\bibitem[{{Tauris} {et~al.}(2017){Tauris}, {Kramer}, {Freire}, {Wex}, {Janka},
  {Langer}, {Podsiadlowski}, {Bozzo}, {Chaty}, {Kruckow}, {van den Heuvel},
  {Antoniadis}, {Breton}, \& {Champion}}]{Tauris2017ApJ}
{Tauris}, T.~M., {Kramer}, M., {Freire}, P.~C.~C., {et~al.} 2017, \apj, 846,
  170

\bibitem[{Tauris {et~al.}(2015)Tauris, Langer, \&
  Podsiadlowski}]{Tauris:2015xra}
Tauris, T.~M., Langer, N., \& Podsiadlowski, P. 2015, Monthly Notices of the
  Royal Astronomical Society, 451, 2123

\bibitem[{Tauris {et~al.}(2013)Tauris, Sanyal, Yoon, \&
  Langer}]{Tauris:2013zna}
Tauris, T.~M., Sanyal, D., Yoon, S.-C., \& Langer, N. 2013, Astronomy \&
  Astrophysics, 558, A39

\bibitem[{{Timmes} \& {Swesty}(2000)}]{HELM:eos}
{Timmes}, F.~X. \& {Swesty}, F.~D. 2000, \apjs, 126, 501

\bibitem[{Timmes \& Woosley(1992)}]{timmes1992a}
Timmes, F.~X. \& Woosley, S.~E. 1992, The Astrophysical Journal, 396, 649

\bibitem[{Vassiliadis \& Wood(1993)}]{Vassiliadis:1993zz}
Vassiliadis, E. \& Wood, P.~R. 1993, The Astrophysical Journal, 413, 641

\bibitem[{Vink(2017)}]{Vink:2017ujd}
Vink, J.~S. 2017, Astronomy \& Astrophysics, 607, L8

\bibitem[{Waldman \& Barkat(2006)}]{waldman2006a}
Waldman, R. \& Barkat, Z. 2006, Ph.D. Thesis

\bibitem[{Waldman \& Barkat(2007)}]{Waldman2007}
Waldman, R. \& Barkat, Z. 2007, The Astrophysical Journal, 665, 1235

\bibitem[{{Waldman} {et~al.}(2008){Waldman}, {Yungelson}, \&
  {Barkat}}]{waldman2008}
{Waldman}, R., {Yungelson}, L.~R., \& {Barkat}, Z. 2008, in Astronomical
  Society of the Pacific Conference Series, Vol. 391, Hydrogen-Deficient Stars,
  ed. A.~{Werner} \& T.~{Rauch}, 359

\bibitem[{Wang \& Liu(2020)}]{Wang:2020pzc}
Wang, B. \& Liu, D. 2020, Research in Astronomy and Astrophysics, 20, 135

\bibitem[{Wang {et~al.}(2017)Wang, Zhou, Zuo, Li, Luo, Zhang, Liu, \&
  Wu}]{Wang:2016alt}
Wang, B., Zhou, W.-H., Zuo, Z.-Y., {et~al.} 2017, Monthly Notices of the Royal
  Astronomical Society, 464, 3965

\bibitem[{Wheeler(1978)}]{wheeler1978}
Wheeler, J.~C. 1978, The Astrophysical Journal, 225, 212

\bibitem[{Willcox {et~al.}(2016)Willcox, Townsley, Calder, Denissenkov, \&
  Herwig}]{Willcox:2016yyp}
Willcox, D.~E., Townsley, D.~M., Calder, A.~C., Denissenkov, P.~A., \& Herwig,
  F. 2016, The Astrophysical Journal, 832, 13

\bibitem[{Willcox {et~al.}(2021)Willcox, Mandel, Thrane, Deller, Stevenson, \&
  {Vigna-G{\'o}mez}}]{Willcox:2021kbg}
Willcox, R., Mandel, I., Thrane, E., {et~al.} 2021, The Astrophysical Journal,
  920, L37

\bibitem[{Wolf \& Schwab(2017)}]{pymesareader}
Wolf, B. \& Schwab, J. 2017

\bibitem[{Woosley(2019)}]{Woosley:2019sdf}
Woosley, S.~E. 2019, The Astrophysical Journal, 878, 49

\bibitem[{Yamanaka {et~al.}(2015)Yamanaka, Maeda, Kawabata, Tanaka, Tominaga,
  Akitaya, Nagayama, Kuroda, Takahashi, Saito, Yanagisawa, Fukui, Miyanoshita,
  Watanabe, Arai, Isogai, Hattori, Hanayama, Itoh, Ui, Takaki, Ueno, Yoshida,
  Ali, Essam, Ozaki, Nakao, Hamamoto, Nogami, Morokuma, Oasa, Izumiura, \&
  Sekiguchi}]{Yamanaka:2015qpa}
Yamanaka, M., Maeda, K., Kawabata, K.~S., {et~al.} 2015, The Astrophysical
  Journal, 806, 191

\bibitem[{Yan {et~al.}(2016)Yan, Zhu, Wang, \& Lü}]{Yan_2016}
Yan, J.-Z., Zhu, C.-H., Wang, Z.-J., \& Lü, G.-L. 2016, Research in Astronomy
  and Astrophysics, 16, 009

\bibitem[{{Zapartas} {et~al.}(2017){Zapartas}, {de Mink}, {Van Dyk}, {Fox},
  {Smith}, {Bostroem}, {de Koter}, {Filippenko}, {Izzard}, {Kelly}, {Neijssel},
  {Renzo}, \& {Ryder}}]{Zapartas2017}
{Zapartas}, E., {de Mink}, S.~E., {Van Dyk}, S.~D., {et~al.} 2017, \apj, 842,
  125

\bibitem[{Zeng {et~al.}(2021)Zeng, Wang, Esamdin, Pellegrino, Burke, Stahl,
  Zheng, Filippenko, Howell, Sand, Valenti, Mo, Xi, Liu, Zhang, Li, Iskandar,
  Zhang, Lin, Sai, Xiang, Wei, Zhang, Reichart, Brink, McCully, Hiramatsu,
  Hosseinzadeh, Jeffers, Ross, Stegman, Wang, Zhang, \& Ma}]{Zeng:2021vsz}
Zeng, X., Wang, X., Esamdin, A., {et~al.} 2021, The Astrophysical Journal, 919,
  49

\end{thebibliography}

\onecolumn
\begin{appendix}
\section{Composition profiles}\label{apx:composition}

\subsection{The \seriesone Grid}

\begin{figure*}[hbt!]
    \centering
    \includegraphics[height=6cm]{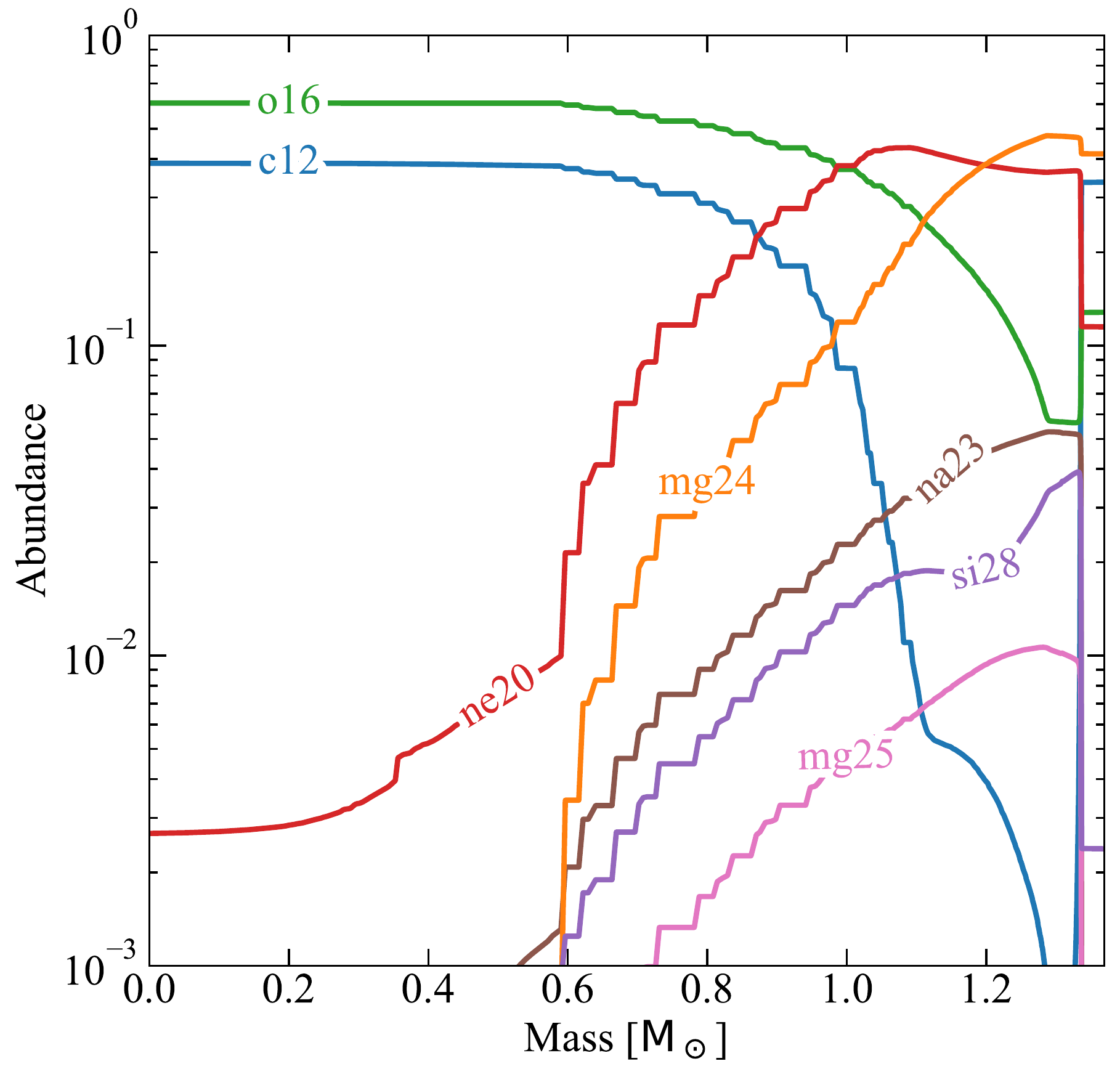}\quad
    \includegraphics[height=6cm]{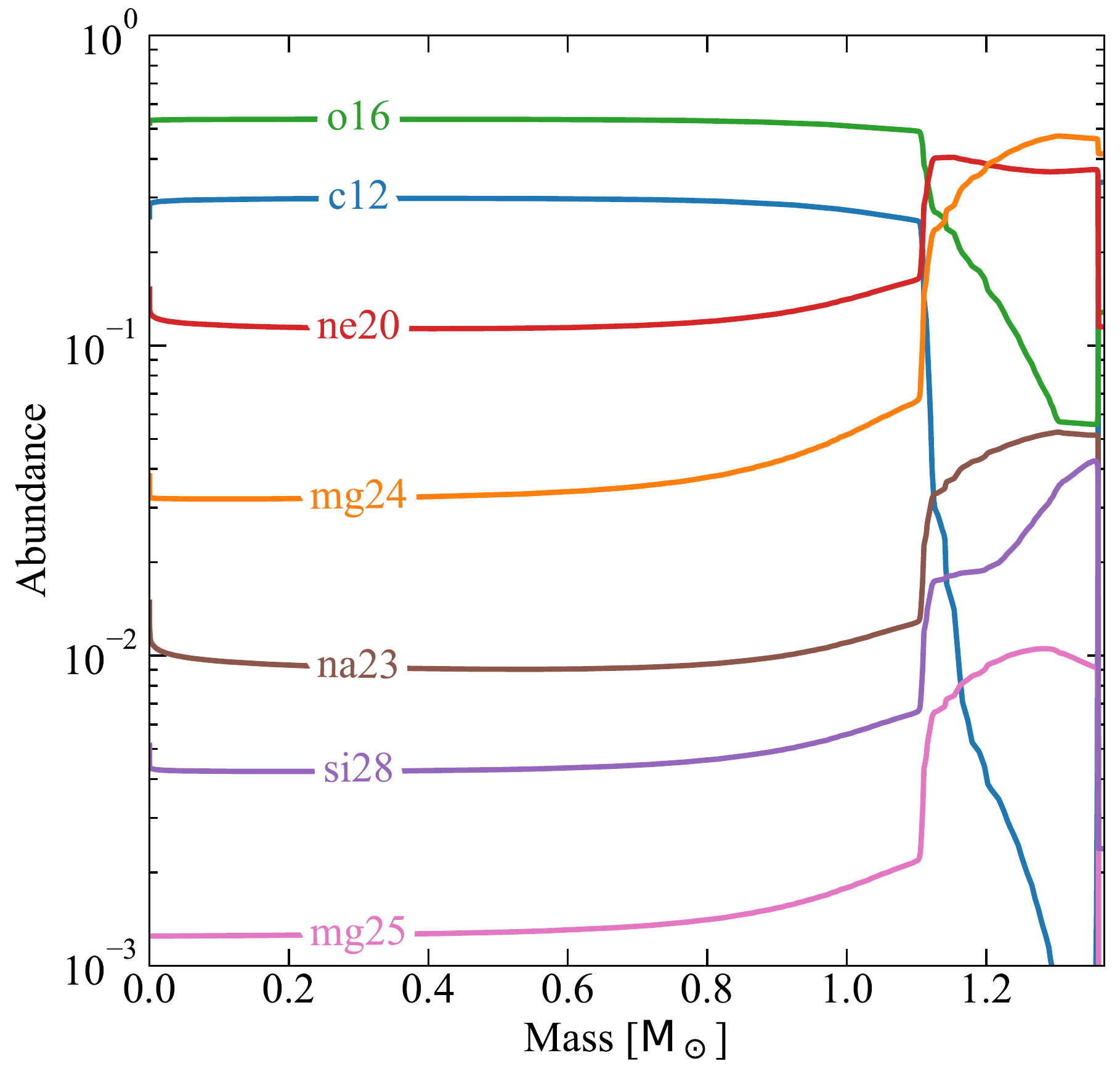}
    \caption{Abundance profiles of the \seriesone hybrid stellar model with $ 1.8\msun,  Z=0.02;f_{OV}=0.014$. The left panel refers to the structure when $\rm \log(\rho_c / g \ cm^{-3}) \simeq 9.0$ (indicated by the black cross-mark in Fig.~\ref{fig:RhoT}). The right panel shows the final chemical composition of the model.} \par\medskip
    \includegraphics[height=6cm]{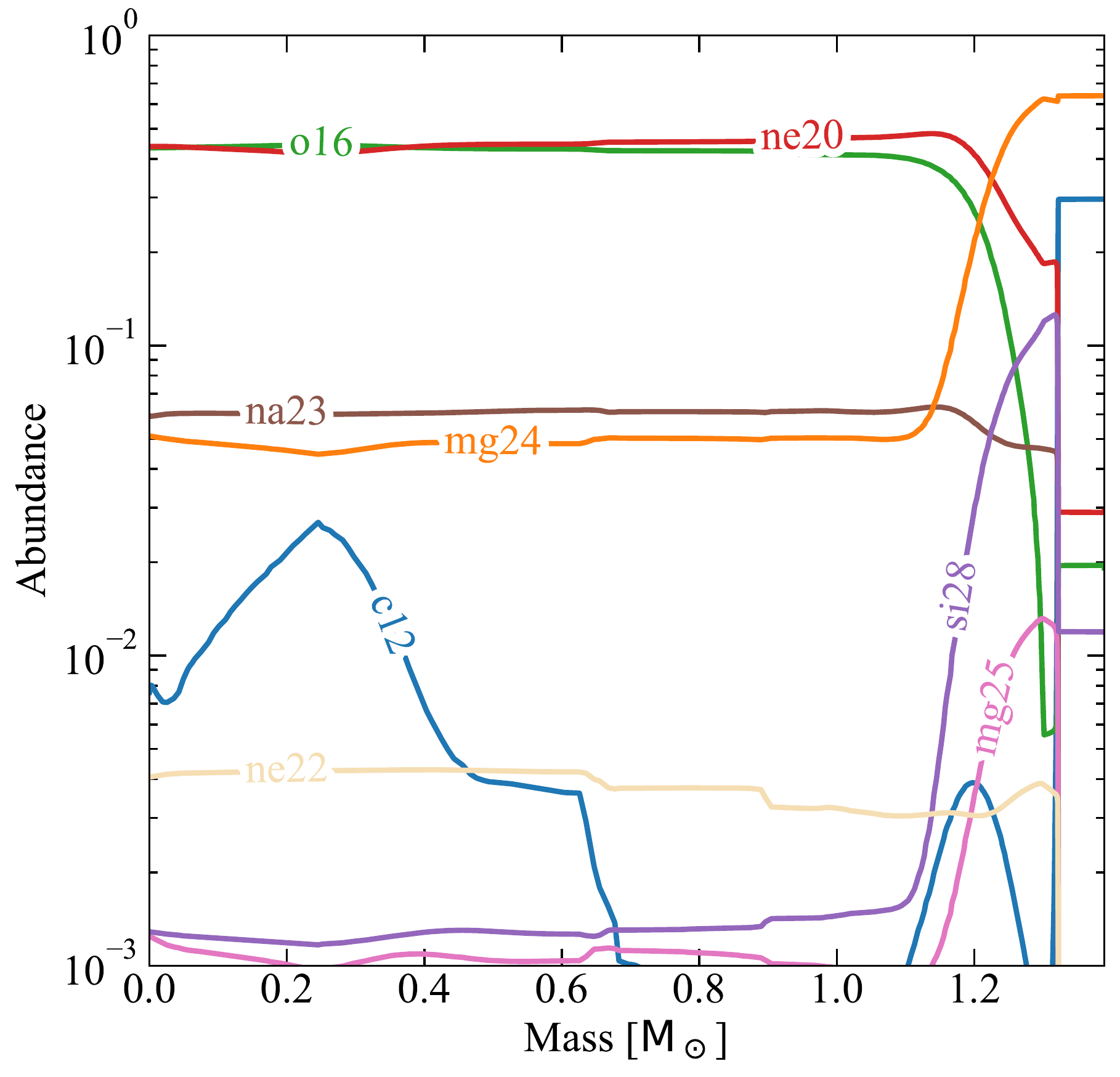}\quad
    \includegraphics[height=6cm]{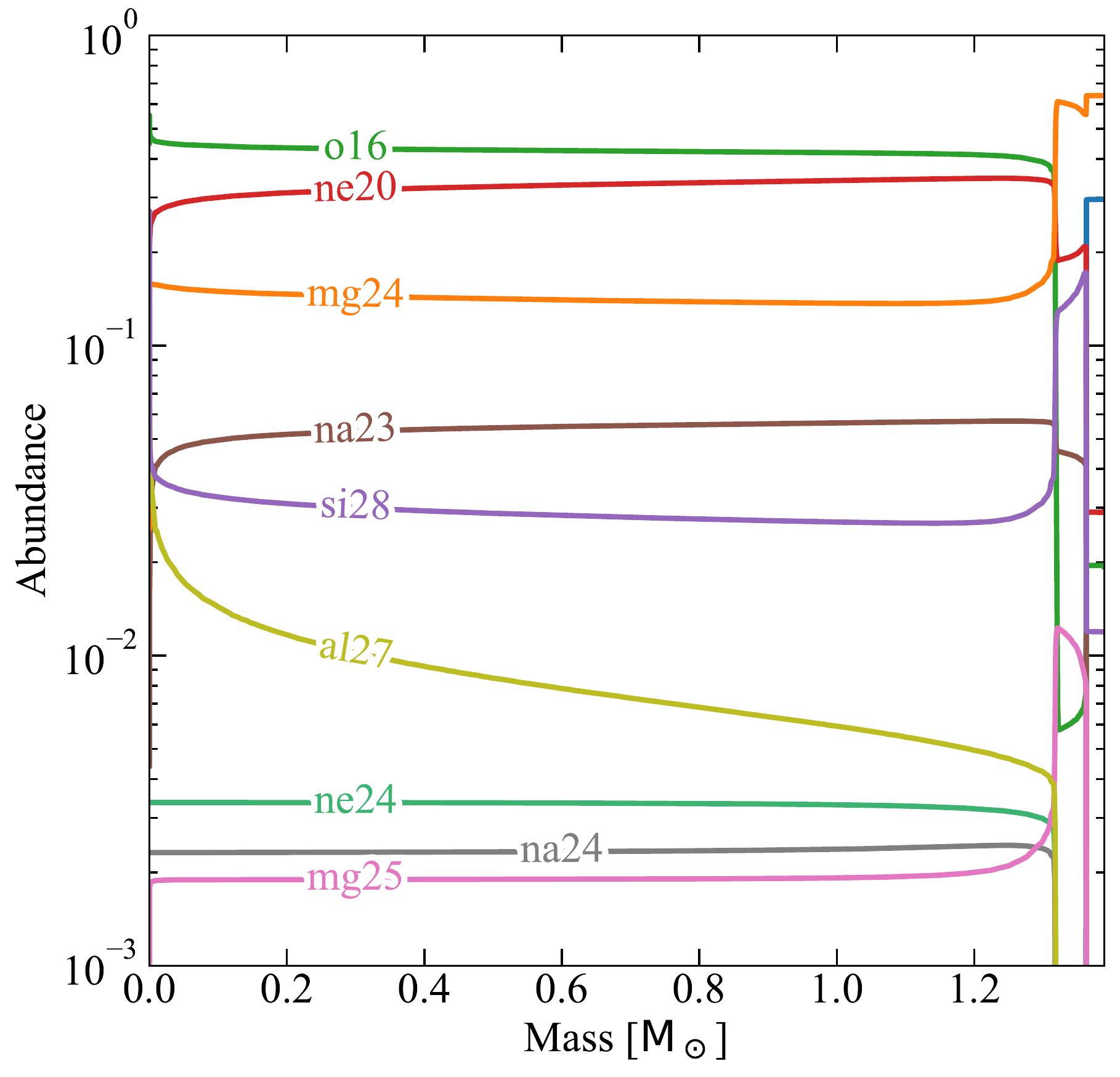}\par\medskip
    \includegraphics[height=6cm]{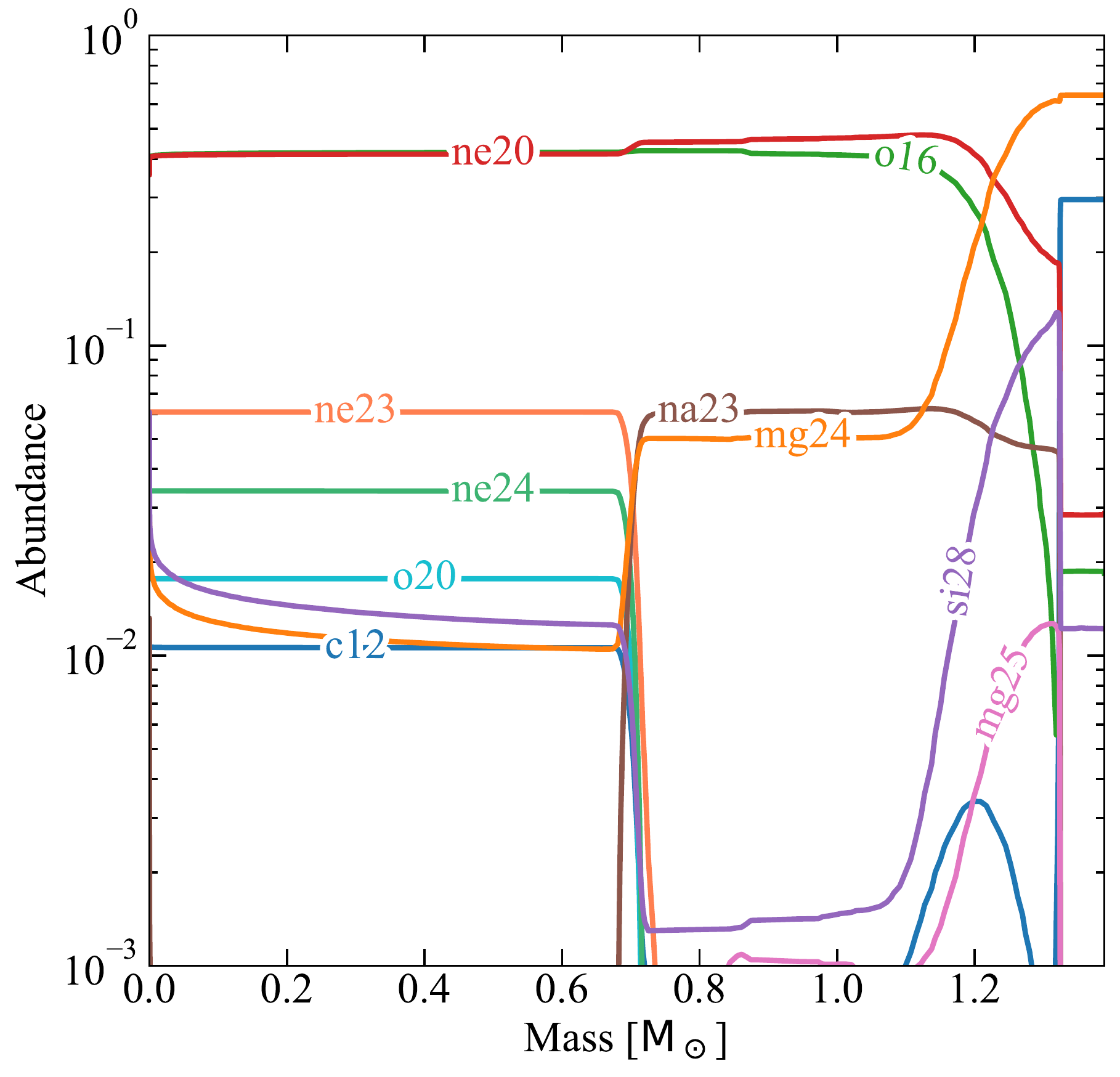}
    \caption{Abundance profiles of the \seriesone $ 2.5\msun,  Z=0.02;f_{OV}=0.0$ stellar model. The top-left panel refers to the structure when $\rm \log(\rho_c / g \ cm^{-3}) \simeq 9.0$. The distribution of residual carbon from previous burning stages is visible. The top-right panel shows the final composition of this model. Ignition of residual carbon leads to thermonuclear explosion.
    The panel at the bottom shows the final structure of the same model, in the case where carbon-consuming nuclear reactions have been turned off.
    In this scenario, the core reaches the density threshold for $e$-captures on Ne nuclei to commence, leading to formation of \iso{O}{20}, and most likely to an ECSN.}
    \label{apx:fig:eta1p0}
\end{figure*}

\clearpage
\subsection{The \seriestwo Grid}

\begin{figure*}[hbt!]
    \centering 
    \includegraphics[height=6cm]{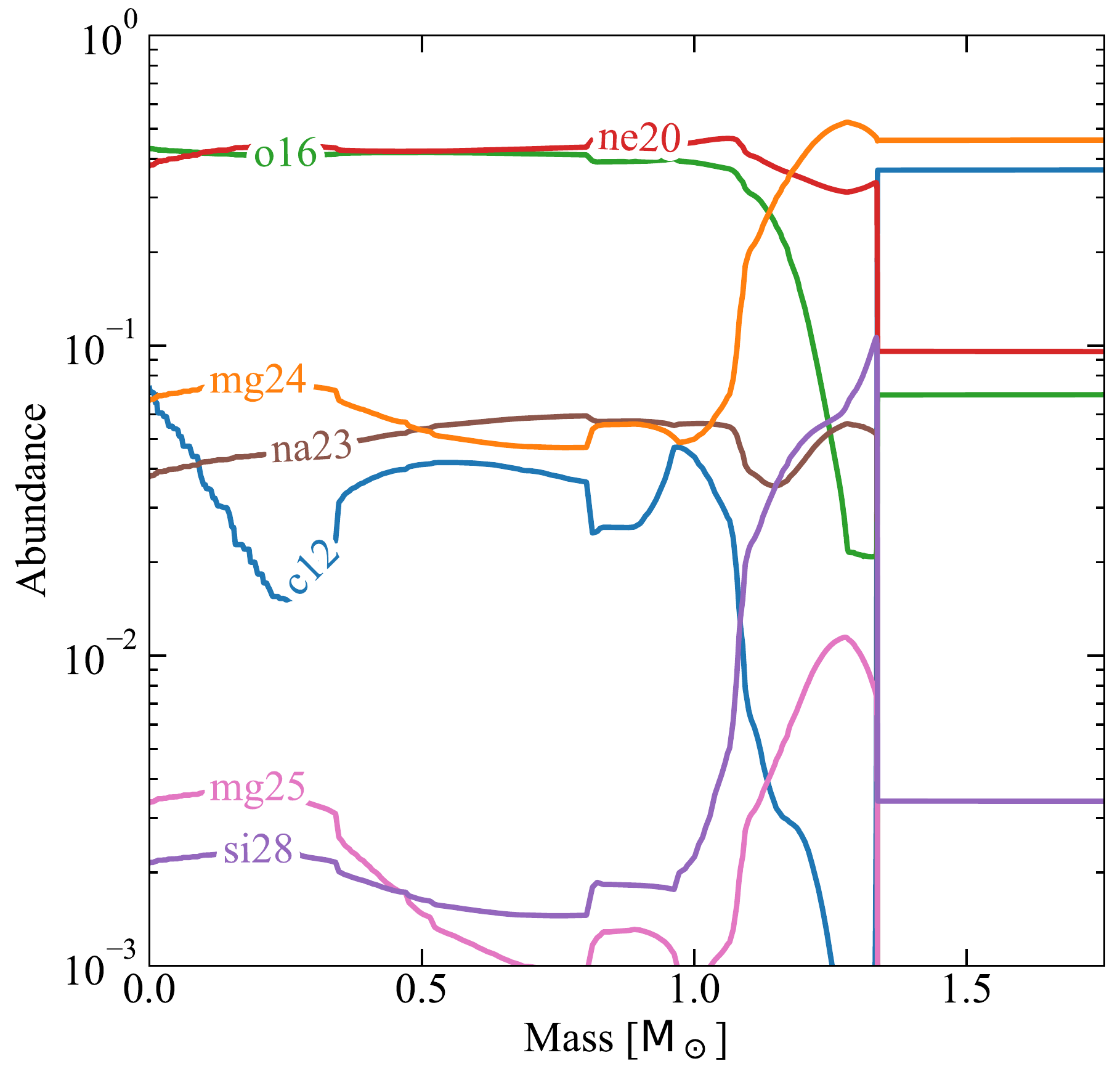}\quad
    \includegraphics[height=6cm]{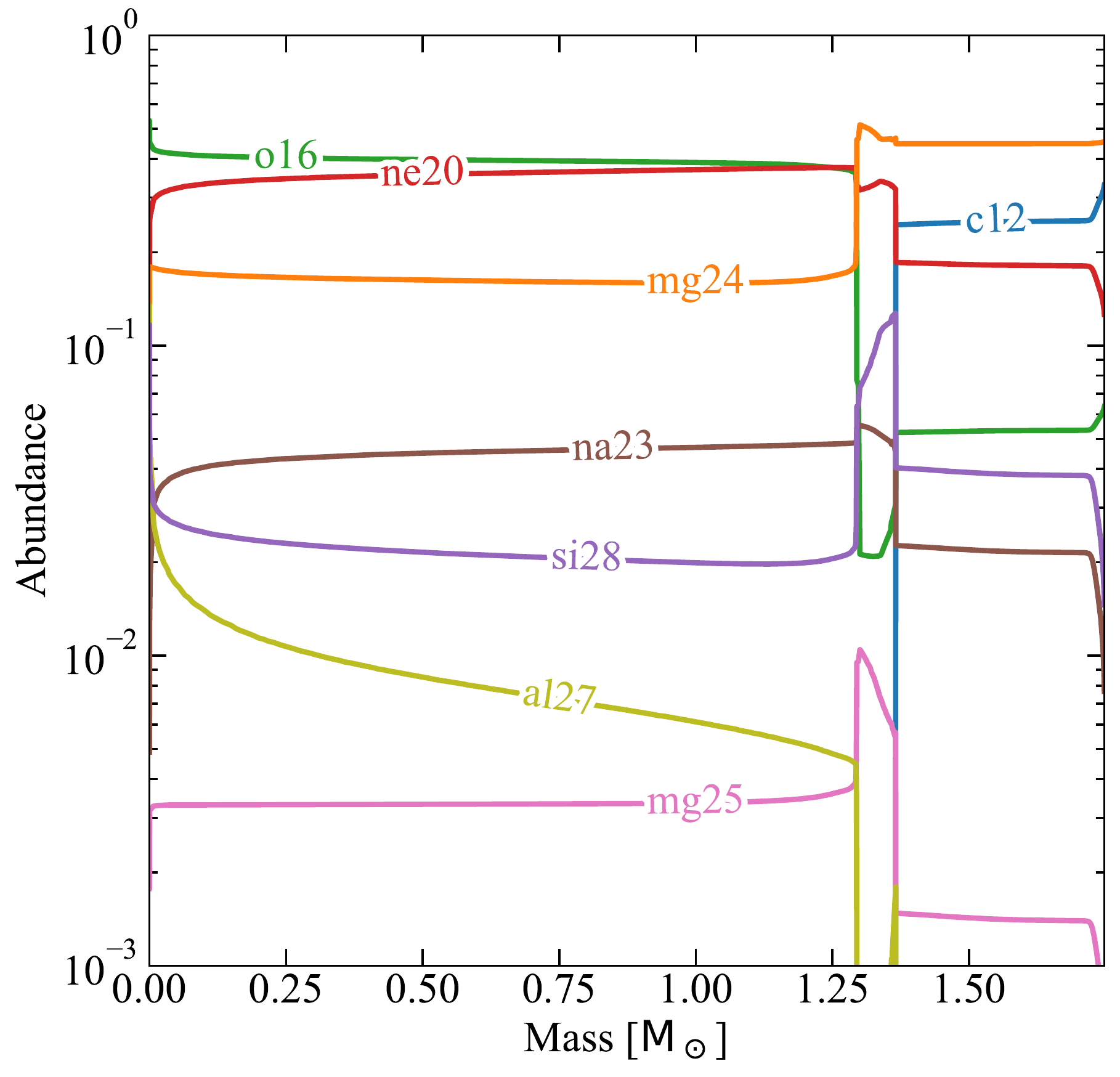}
    \caption{Abundance profiles of the \seriestwo $ 2.0\msun,  Z=0.02;f_{OV}=0.0; \eta=0.25$ stellar model. The left panel refers to the structure when $\rm \log(\rho_c / g \ cm^{-3}) \simeq 9.0$ (shortly before the Urca cooling phase). The right panel shows the final chemical composition of the model after the ignition of neon.}
    \label{apx:fig:eta0p25}
\end{figure*}

\begin{figure*}[hbt!]
    \centering 
    \includegraphics[height=6cm]{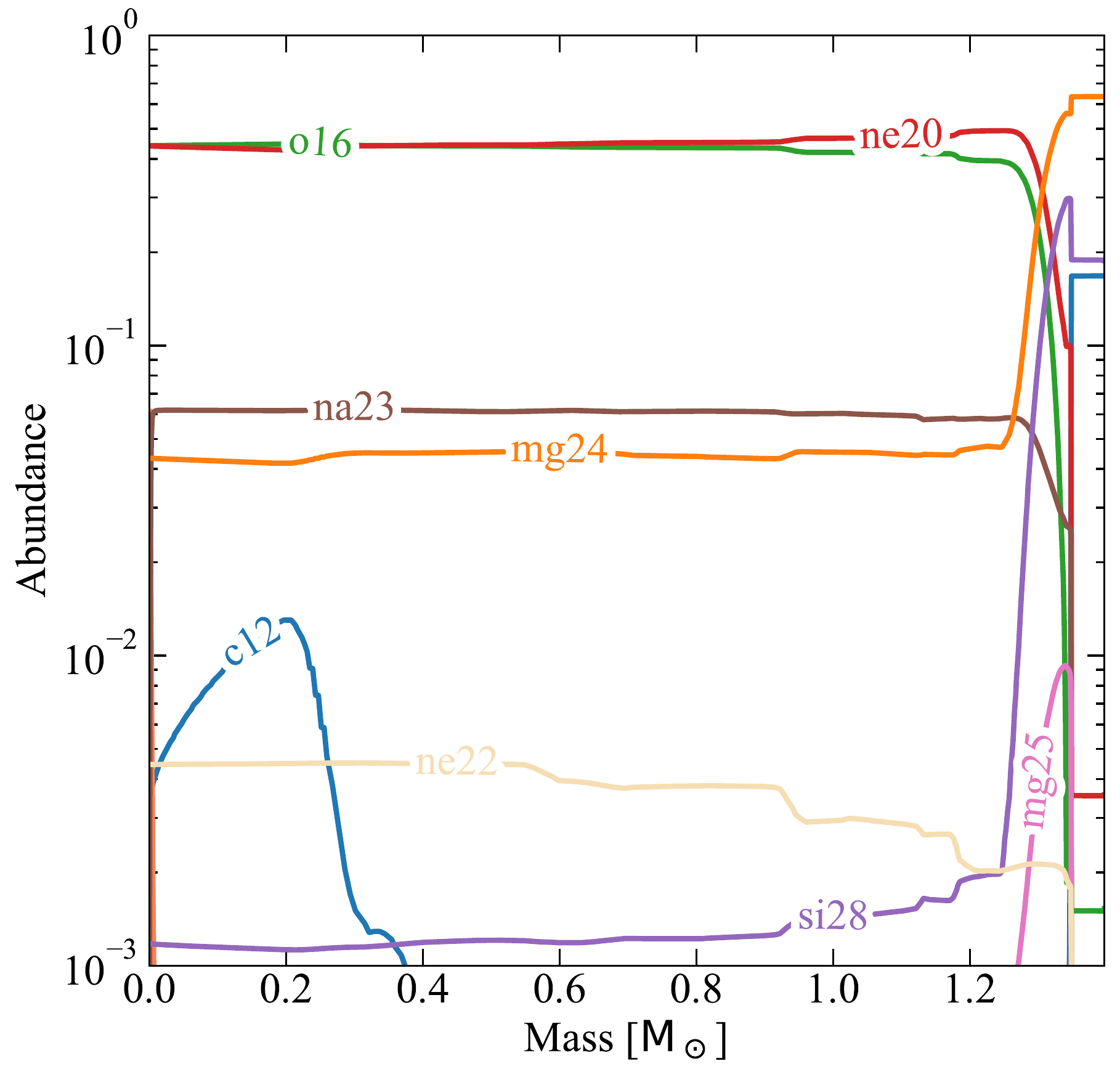}\quad
    \includegraphics[height=6cm]{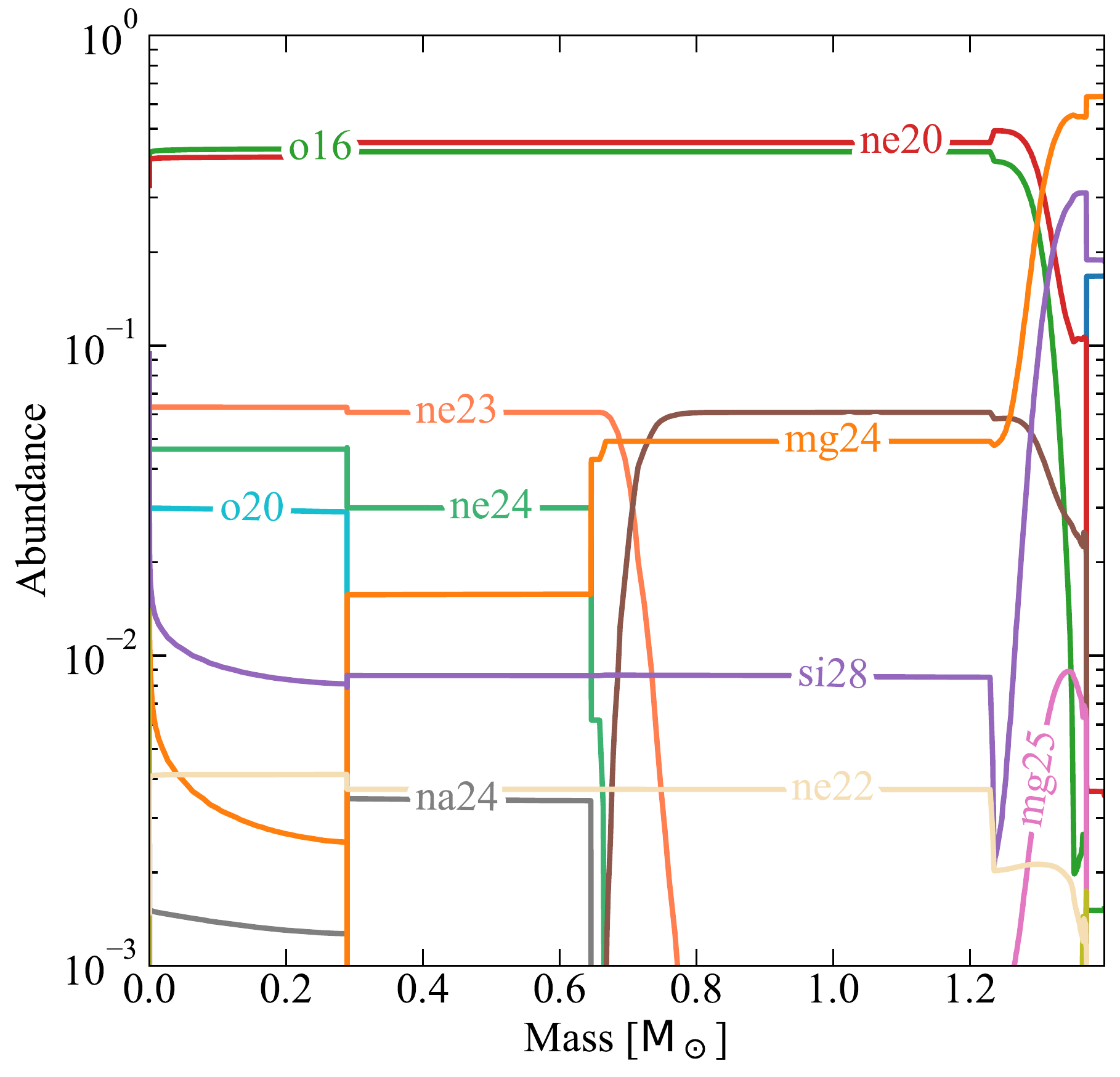}
    \caption{Abundance profiles of the \seriestwo $ 2.7\msun,  Z=0.02;f_{OV}=0.0; \eta=0.8$ stellar model. The left panel refers to the structure when $\rm \log(\rho_c / g \ cm^{-3}) \simeq 9.0$ (shortly before the Urca cooling phase). The right panel shows the final chemical composition of the model after the ignition of neon. This model has reached the critical density for $e$-captures on \iso{Ne}{20} which leads to the formation of \iso{O}{20} (cyan line).}
    \label{apx:fig:eta0p8}
\end{figure*}

\begin{figure*}[hbt!]
    \centering 
    \includegraphics[height=6cm]{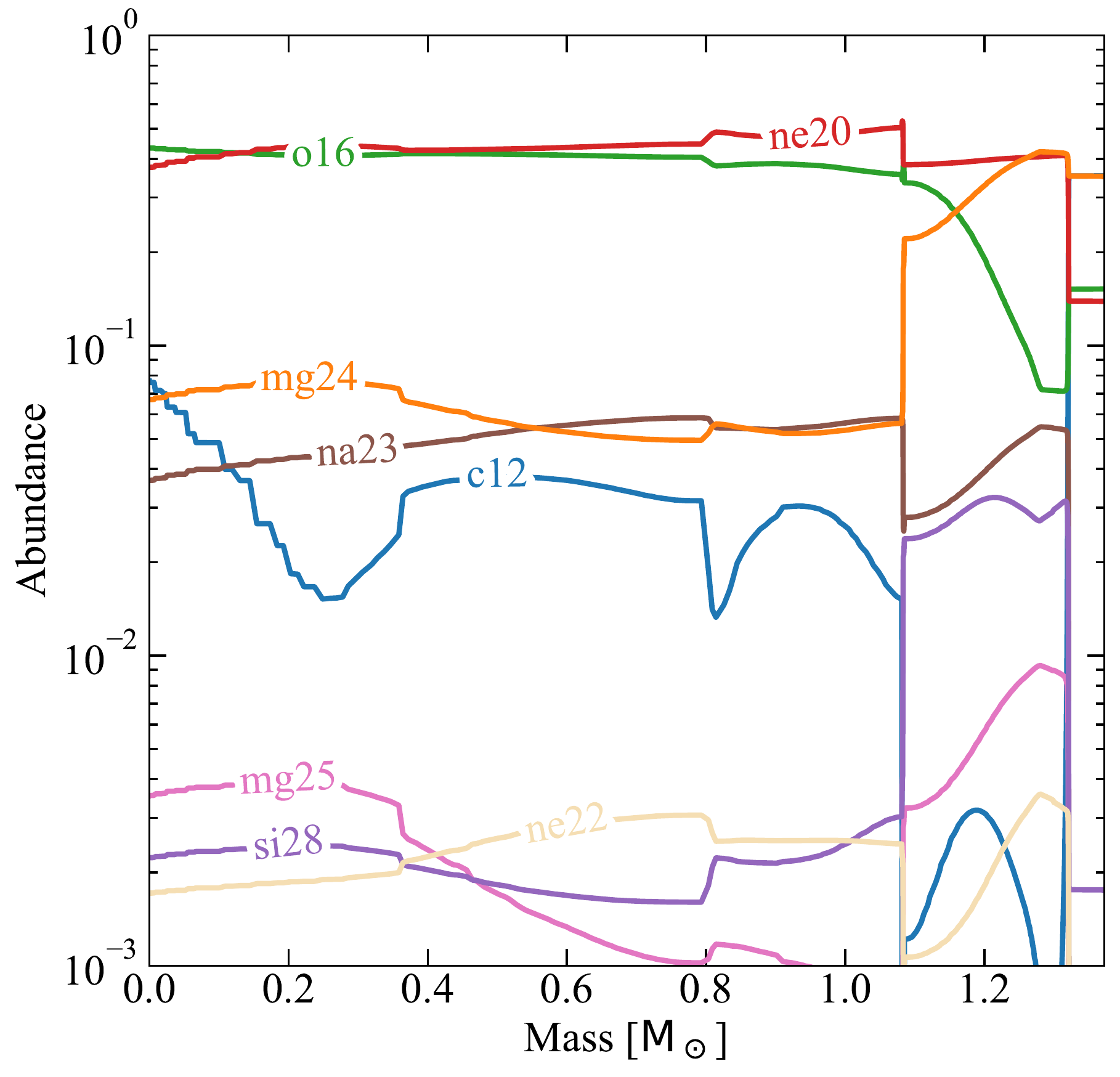}\quad
    \includegraphics[height=6cm]{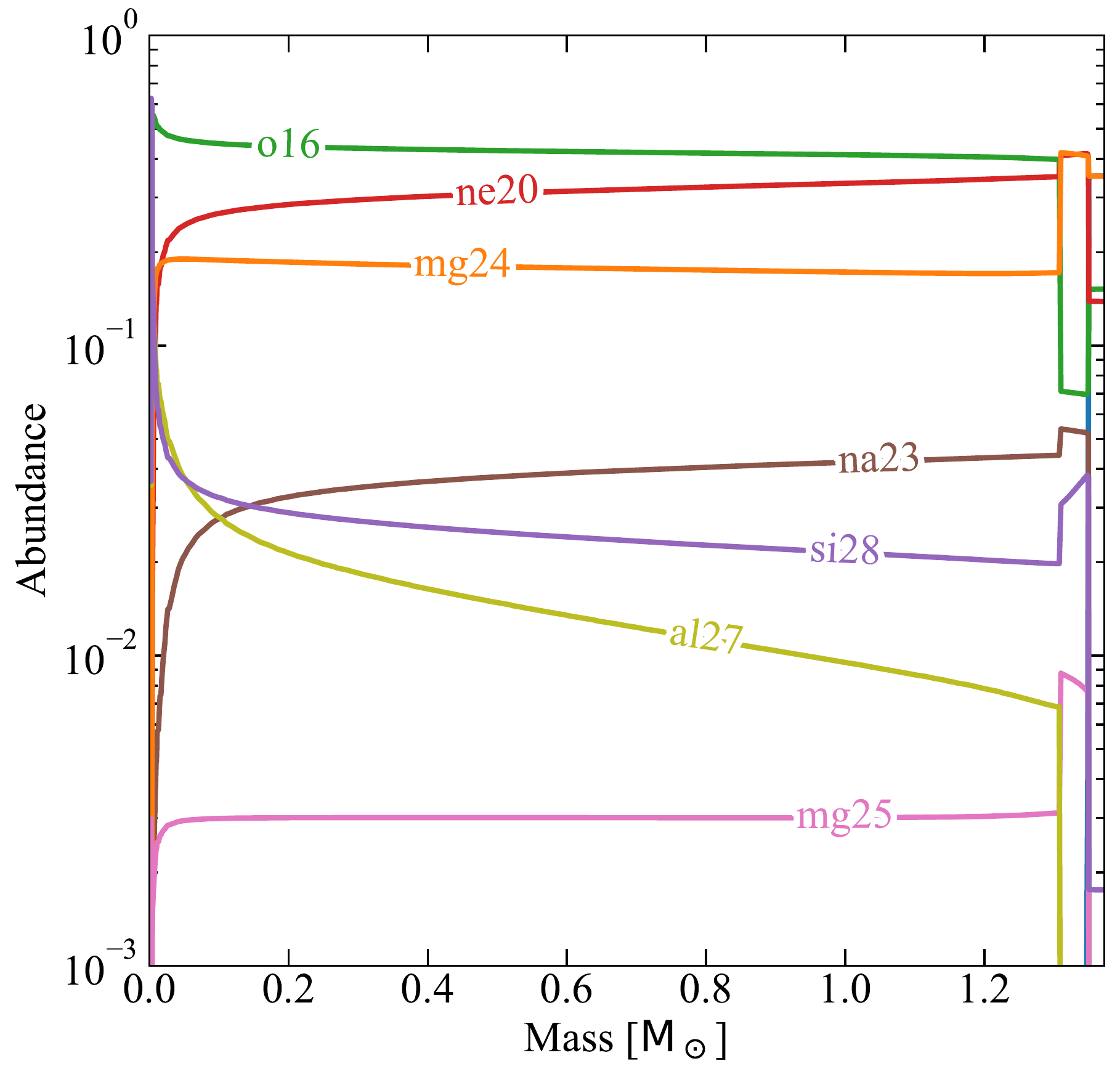}
    \includegraphics[height=6cm]{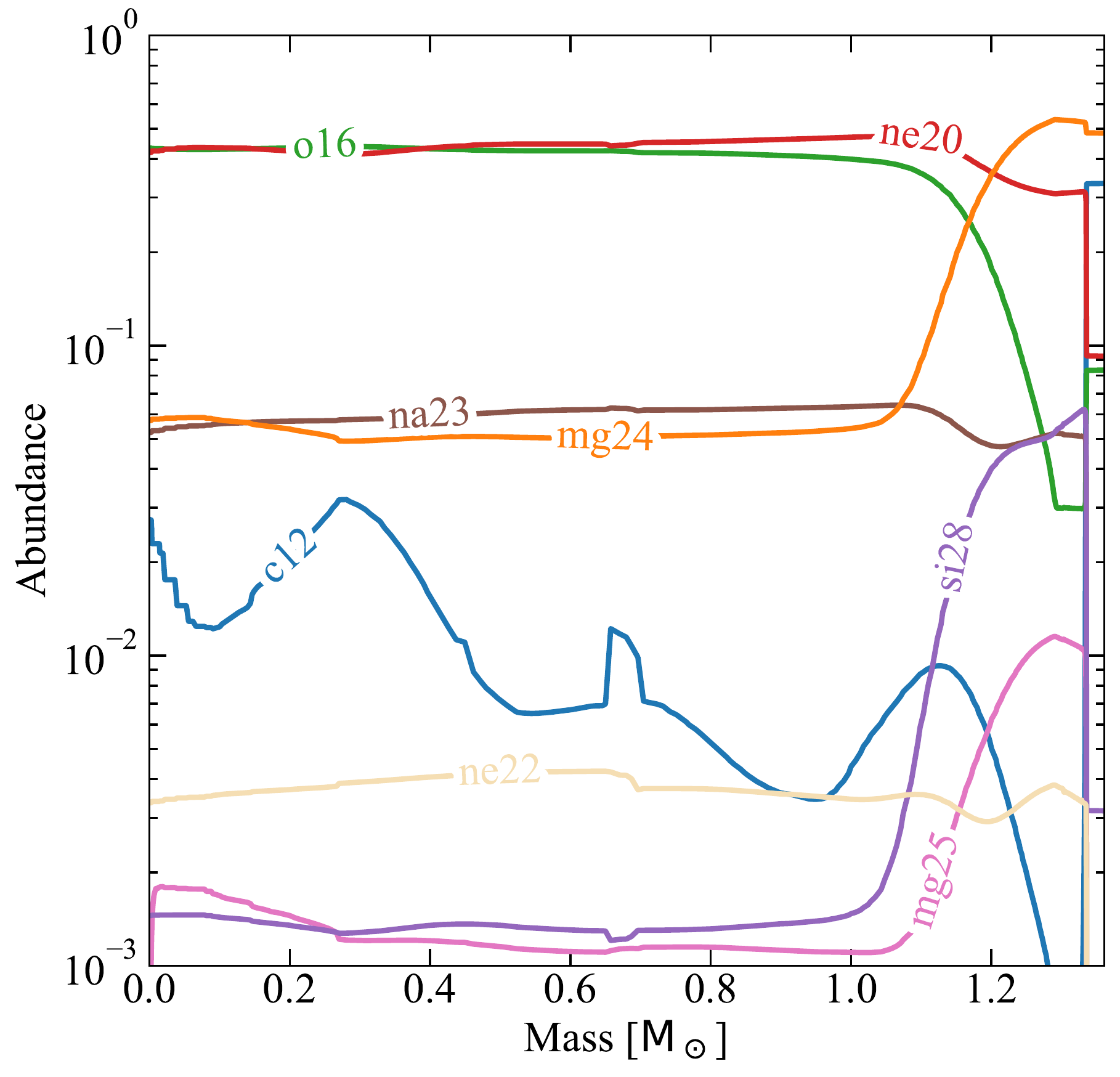}\quad
    \includegraphics[height=6cm]{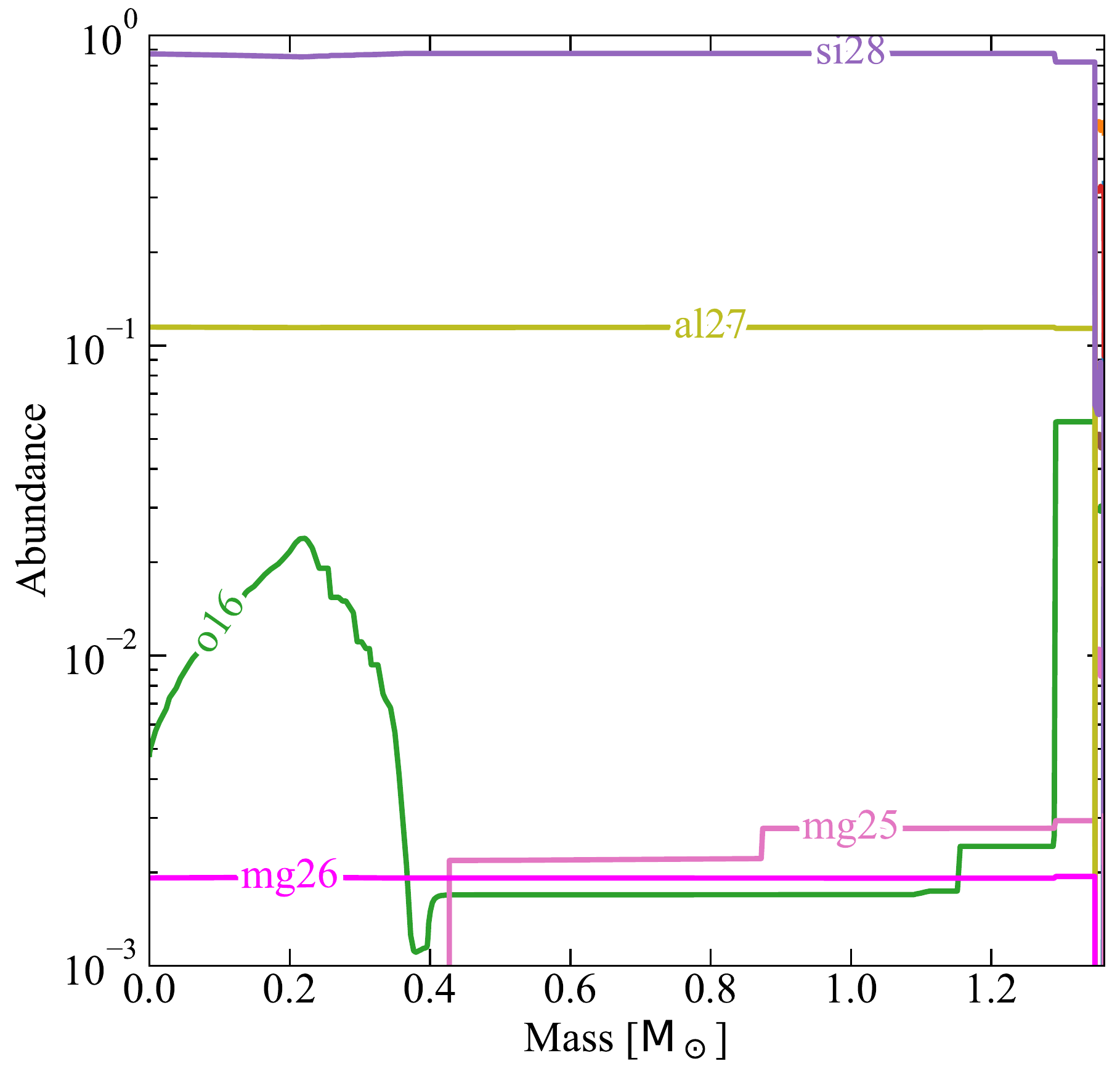}
    \caption{Abundance profiles for two of our \seriestwo models with $2.1\msun,  Z=0.02;f_{OV}=0.0; \eta=1.6$ (top panel) and $2.4\msun,  Z=0.02;f_{OV}=0.0; \eta=1.6$ (bottom panel). On the left, the plots refer to chemical composition when $\rm \log(\rho_c / g \ cm^{-3}) \simeq 9.0$ (shortly before the Urca cooling phase). On the right, the plots refer to the final chemical composition of our models. In the case of the $2.4\msun$ model, off-center ignition of oxygen leads to the formation of a silicon core with amounts of residual oxygen distributed within the core.}
    \label{apx:fig:eta1p58}
\end{figure*}

\end{appendix}

\end{document}